 \documentclass[letterpaper]{JHEP3}
 \usepackage[T1]{fontenc}
 \usepackage[latin9]{inputenc}
 \usepackage{float}
 \usepackage{graphicx}
\usepackage{amsmath}
\usepackage{amssymb}
\usepackage{wasysym}

\usepackage{color}
\let\normalcolor\relax
\usepackage{array}

\usepackage{setspace}

 \makeatletter

 \providecommand{\tabularnewline}{\\}
\newcommand{\twist}{\gamma}

\title{ 
\Large{Spin Chains in ${\cal N} =2$ Superconformal Theories} \\
\vspace{0.3cm}
\large{From the $\mathbb{Z}_2$ Quiver to Superconformal QCD} 
}
\preprint{YITP-SB-10-20}

 \author{Abhijit Gadde\footnote{abhijit@insti.physics.sunysb.edu}, Elli Pomoni\footnote{pomoni@insti.physics.sunysb.edu}, and Leonardo Rastelli\footnote{leonardo.rastelli@stonybrook.edu}
 \\
\\
\it C.N. Yang Institute for Theoretical Physics,\\
\it Stony Brook University, \\
\it Stony Brook, NY 11794-3840, USA
}
\bigskip

 \abstract{
 \vspace{0.2cm}
 
 In this paper we find preliminary evidence that ${\cal N}=2$ superconformal
 QCD, the $SU(N_c)$  SYM theory with $N_f= 2 N_c$ fundamental hypermultiplets,
might be integrable  in the large~$N$ Veneziano limit. 
 We evaluate the one-loop dilation operator in the scalar sector of the ${\cal N}=2$
superconformal quiver with $SU(N_c) \times SU(N_{\check c})$ gauge group, for $N_c \equiv N_{\check c}$.
Both gauge couplings $g$ and $\check g$ are exactly marginal. This theory interpolates
between the $\mathbb{Z}_2$ orbifold of ${\cal N}=4$ SYM,
 which corresponds to $\check g=g$,
and
${\cal N}=2$ superconformal QCD,
which is obtained for $\check g \to 0$.
The planar one-loop dilation operator takes the form of a nearest-neighbor spin-chain Hamiltonian.
For superconformal QCD 
the spin chain is of novel form: besides the color-adjoint fields $\phi^a_{\;b}$, which occupy individual sites of the chain,
there are ``dimers''  $Q^a_{\;i} \bar Q^i_{\;b}$ of  flavor-contracted fundamental fields,
  which occupy two neighboring sites. We solve the two-body scattering problem of magnon excitations and study
the  spectrum of  bound states, for general $\check g/g$.
   The  dimeric excitations of
  superconformal QCD are seen to arise smoothly for $\check g \to 0$ as the limit of bound wavefunctions
  of the interpolating  theory. Finally we check   the Yang-Baxter equation for the two-magnon S-matrix.
  It holds as expected at the orbifold point  $\check g = g$. While  violated for general $\check g \neq g$, 
it holds again in the limit $\check g \to 0$, hinting at one-loop integrability of planar ${\cal N}=2$ superconformal QCD.

 }
 \usepackage{rotating}

 \renewcommand{\[}{\begin{equation}}
 \renewcommand{\]}{\end{equation}}

 \@ifundefined{showcaptionsetup}{}{%
  \PassOptionsToPackage{caption=false}{subfig}}
 \usepackage{subfig}
 \makeatother

\begin{document}
 
\newcommand{\gbar}{{g_+}}
\newcommand{\gdiff}{{g_-}}
\newcommand{\gen}[1]{{\mathfrak{#1}}} 
 \global \long \def \m{\mu}

 \global \long \def \n{\nu}

 \global \long \def \La{\Lambda}

 \global \long \def \s{\sigma}

 \global \long \def \f{\phi}

 \global \long \def \e{\epsilon}

 \global \long \def \del{\partial}

 \global \long \def \D{\Delta}

 \global \long \def \al{\alpha}

 \global \long \def \ad{\dot{\alpha}}

 \global \long \def \bd{\dot{\beta}}

 \global \long \def \la{\lambda}

 \global \long \def \ra{\rightarrow}

 \global \long \def \fbar{\bar{\phi}}

 \global \long \def \p{\partial}

 \global \long \def \bA{{\bf A}}

 \global \long \def \OO{\mathcal{O}}

 \global \long \def \II{\mathcal{I}}

 \global \long \def \JJ{\mathcal{J}}

 \global \long \def \KK{\mathcal{K}}

 \global \long \def \LL{\mathcal{L}}

 \global \long \def \TT{\mathcal{T}}

 \global \long \def \NN{\mathcal{N}}

 \global \long \def \MM{\mathcal{M}}

 \global \long \def \PP{\mathcal{P}}

 \global \long \def \nn{  \mathfrak{n} }

 \global \long \def \qq{  \mathfrak{q} }

 \global \long \def \mm{  \mathfrak{m} }

 \global \long \def \pp{  \mathfrak{p}}

 \global \long \def \Tr{\mbox{Tr}}

 \global \long \def \Q{\mathcal{Q}}

 \global \long \def \TT{\mathcal{T}}

 \global \long \def \SS{\mathcal{S}}

 \global \long \def \RR{\mathcal{R}}

 \global \long \def \TpT{\mathcal{T}^{\prime}}

 \global \long \def \IIh{\hat{\mathcal{I}}}

 \global \long \def \JJh{\hat{\mathcal{J}}}

 \global \long \def \KKh{\hat{\mathcal{K}}}

 \global \long \def \LLh{\hat{\mathcal{L}}}

 \global \long \def \SSh{\hat{\mathcal{S}}}

 \global \long \def \RRh{\hat{\mathcal{R}}}

 \global \long \def \dprime{\prime \prime}

 \global \long \def \topp#1{\check{#1}}

 \global \long \def \fh{\topp{\f}}

 \global \long \def \QQ{\mathcal{Q}}

 \global \long \def \AA{\mathcal{A}}

 \global \long \def \BB{\mathcal{B}}

 \global \long \def \CC{\mathcal{C}}

 \global \long \def \DD{\mathcal{D}}

 \global \long \def \EE{\mathcal{E}}

 \newcommand{\UU}{{\cal U}}
 \newcommand{\VV}{{\cal V}}
 \newcommand{\XX}{{\cal X}}

 \newcommand{\ssu}[1]{\mathfrak{#1}}

%My commands
\newcommand{\apm}{\alpha'}
\def\d{\partial}
\newcommand{\rb}[1]{\raisebox{3.1ex}[0pt]{#1}}
\newcommand{\rbr}[1]{\raisebox{1.5ex}[0pt]{#1}}

\newcommand{\ir}[1]{\ensuremath{\boldsymbol{#1}}}
\def\bea{\begin{eqnarray}}
\def\eea{\end{eqnarray}}
\def\be{\begin{equation}}
\def\ee{\end{equation}}
\def\ea{\end{align}}
\def\bse{\begin{subequations}}
\def\ese{\end{subequations}}
\def\1F1{{}_1\!F_1}
\def\2F0{{}_2\!F_0}

\def\qeq{\, {? \atop ~} \hskip-4mm =}

\def\bP{$\bar{\rm P}$\,}

\newcommand{\Torb}{{\cal T}}

 \tableofcontents{}

 \section{Introduction}

The gauge/gravity duality has given crucial insights into the dynamics of four-dimensional gauge theories.
The long-standing  hope is to find a precise string theory description of realistic field theories such as QCD. At present however
we lack a systematic procedure to find the string dual of a given gauge theory, and
all well-understood dual pairs  fall into the ``universality class'' of the original example, the duality between ${\cal N}=4$ super Yang-Mills and IIB on $AdS_5 \times S^5$. 
These dualities are motivated by taking the decoupling limit of brane configurations in critical string theory. Field theories
in this class share a few  common features, for instance: all  fields are in bifundamental representations of the gauge group;
 the $a$ and $c$ conformal anomaly coefficients are equal at large $N$; there is an exactly marginal coupling $\lambda$
 such that for $\lambda$ large the dual worldsheet sigma-model is weakly coupled and the gravity approximation is valid.

To break outside the ${\cal N}=4$ universality class, an important case study is  ${\cal N}=2$ superconformal QCD, namely the
${\cal N}=2$ super Yang-Mills theory with gauge group $SU(N_c)$ and $N_f = 2N_c$ fundamental hyper multiplets.
There is a large number of fundamental flavors, and  $a \neq c$ at large $N_c$. Nevertheless the theory
 shares with ${\cal N}=4$ SYM the crucial simplifying feature of an exactly marginal gauge coupling.
In a recent paper \cite{Gadde:2009dj} we  made some progress towards the AdS dual of ${\cal N}=2$ SCQCD. We attacked the problem
from two fronts: from the bottom-up, we performed a systematic analysis of the protected spectrum using superconformal representation
theory; from the top-down, we considered the decoupling limit of known brane constructions in string theory.
We concluded that the string dual is a sub-critical string background with seven geometric dimensions,
containing both and $AdS_5$ and an $S^1$ factor. In this paper we take the next step of the bottom-up (=field theory) analysis,
by evaluating the one-loop dilation operator in the scalar sector of the theory.

Perturbative calculations of anomalous dimensions have given important
clues into the nature of ${\cal N}=4$ SYM. They gave the first hint
for integrability of the planar theory: the one-loop dilation operator in the scalar sector is the Hamiltonian of the integrable $SO(6)$
spin chain \cite{Minahan:2002ve} -- a result later generalized to the full theory and to
higher loops, using the formalism of the asymptotic Bethe ansatz (see {\it e.g.} \cite{Beisert:2003yb, Staudacher:2004tk,Beisert:2005fw,Beisert:2006ez,Gromov:2009tv} for a very incomplete list
of references.)
Remarkably, the asymptotic S-matrix of magnon excitations in the field theory
spin chain can be exactly matched with the analogous S-matrix for the
dual string sigma-model. 
Thus perturbative calculations open a window
into the structure of the dual string theory.\footnote{The calculation of the circular  Wilson loop
by localization techniques
\cite{Rey:2010ry} is another interesting probe of the dual theory.}
 It is  natural to attempt the same strategy for ${\cal N}=2$ SCQCD.
The theory admits a large $N$ expansion 
in the Veneziano sense \cite{Veneziano:1976wm}: the number of colors
$N_c$ and the number of fundamental flavors $N_f$ are both sent to infinity 
keeping fixed their ratio $(N_f/N_c \equiv 2$ in our case) and the combination
$\lambda = g^2_{YM} N_c$.  We focus on the flavor-singlet sector of the theory,
which is a consistent truncation since flavor singlets close under operator product expansion.
Let us denote a generic color-adjoint field by
$\phi^a_b$, with $a,b =1, \dots N_c$, and a generic color-fundamental and flavor-fundamental field by
  $Q_{\; i}^{a}$, where $i=1, \dots N_f$; 
we are suppressing all other quantum numbers.
In the Veneziano limit, single-trace
``glueball'' operators, of the schematic form ${\rm Tr} \, \phi^\ell$, are {\it  not} closed under the action
of the dilation operator -- this is a major difference with respect to the  the standard 't Hooft limit of large $N_c$ with $N_f$ fixed \cite{'tHooft:1973jz}.
Rather, glueball operators mix at order one (in the large $N$ counting) with flavor-singlet meson operators
of the  form $\sum_i \,\bar Q^i \phi^k Q_i$. The simplest example is the mixing of  ${\rm Tr} (\phi \bar \phi)$ with the singlet meson $\sum_i \bar Q^i Q_i$,
which occurs  at one-loop in planar perturbation theory (order $O(\lambda)$).
The basic ``elementary'' operators are thus what we call
 {} {\it generalized single-trace} operators, of the schematic form 
\begin{equation}
{\rm Tr}\left(\phi^{k_{1}}{\cal M}^{\ell_{1}}\phi^{k_{2}}\dots\phi^{k_{n}}{\cal M}^{\ell_{n}}\right)\,,\qquad{\cal M}_{\; b}^{a}\equiv\sum_{i=1}^{N_{f}}Q_{\; i}^{a}\,\bar{Q}_{\, b}^{i}\, ,
\label{generalizedsingletrace}
\end{equation}
where   ${\rm Tr}$ is a color trace.
 We have  introduced a flavor-contracted combination of a fundamental and an antifundamental
field, ${\cal M}_{\;\, b}^{a}$, which for the purpose of the large $N$ expansion
plays the role of just another color-adjoint field. 
The usual large $N$ factorization theorems apply: correlators
of generalized multi-traces factorize into correlators of 
generalized single-traces. In particular, acting with the dilation operator
on a generalized single-trace operator yields (at leading order in $N$)
 another generalized single-trace operator, so we may consistently diagonalize the
dilation operator in the space of generalized single-traces. 
The dilation operator
acting on generalized single-traces can then be interpreted, in the usual fashion,
as the Hamiltonian of a closed spin chain. Just as in the 't Hooft limit, planarity of the perturbative diagrams
translates into locality of the spin chain:
at one-loop   the spin chain has only
nearest neighbor interactions, at two two-loops there are next-to-nearest neighbors interactions,
and each higher loop spreads the range interaction one site further.

More insight is gained by viewing ${\cal N} = 2$ SCQCD as part of an ``interpolating'' ${\cal N}=2$ superconformal field theory (SCFT)
that has a product gauge group  $SU(N_c)  \times SU(N_{\check c})$, with $N_{\check c} \equiv N_c$,
 and correspondingly two exactly marginal couplings $g$ and $\check g$. 
For $ \check g \to 0$ one recovers ${\cal N} = 2$ SCQCD {\it plus} a decoupled free vector multiplet, 
 while for  $ \check g= g $ one finds  the familiar
$ \mathbb{Z}_2$ orbifold  of ${\cal N} = 4$ SYM. We have evaluated the one-loop
dilation operator for the whole interpolating theory, in the sector of operators made out of scalar fields.
The magnon excitations of the spin chain and their bound states
undergo an interesting evolution as a function of $\kappa = \check g/g$. 
For $\kappa = 0$ (that is, for ${\cal N}=2$ SCQCD itself), the basic
asymptotic  excitations of the spin chain are linear combinations of the  
 the adjoint impurity $\bar \phi$ and of ``dimer'' impurities ${\cal M}_{\;\, b}^{a}$
(we refer to them as dimers since they occupy two sites of the chain). 
 From the point of view of the interpolating theory with $\kappa >0$, these dimeric asymptotic states of ${\cal N}=2$
SCQCD are {\it bound states} of two elementary magnons; the bound-state wavefunction localizes
in the limit $\kappa \to 0$, giving an impurity that occupies two sites.

 Armed with the one-loop Hamiltonian in the scalar sector, we can easily determine
 the complete spectrum of one-loop protected composite operators made of scalar fields.
  It is instructive to follow the evolution of the protected eigenstates as a function of $\kappa$,
 from the orbifold point to ${\cal N}=2$ SCQCD.
 Some of these results were quoted with no derivation in our previous paper \cite{Gadde:2009dj}, where they served as  input to the analysis of the full protected
 spectrum, carried out with the help of the superconformal index \cite{Kinney:2005ej}.

An important question is whether the one-loop spin chain of ${\cal N}=2$ SCQCD is integrable. The spin chain
for the $\mathbb{Z}_2$ orbifold of ${\cal N}=4$ SYM (which by definition has $\check g = g$)
is known to be integrable \cite{Beisert:2005he,Solovyov:2007pw}.  We find that as we move away from the orbifold point
 integrability is broken, indeed for general $\kappa = \check g/g$  the Yang-Baxter equation for the two-magnon S-matrix does not hold.
Remarkably however the Yang-Baxter equation is satisfied again in the ${\cal N}=2$ SCQCD  limit $\kappa \to 0$. 
Ordinarily a check of the Yang-Baxter equation is  strong evidence in favor of integrability. In our case things are more subtle:
the
elementary $Q$ excitations freeze in the limit $\kappa \to 0$
(their dispersion relation becomes constant), 
 while some (but not all) of their dimeric bound states retain non-trivial dynamics.
 Nevertheless,   for infinitesimal $\kappa$ the elementary $Q$s {\it are} propagating excitations, and the
  Yang-Baxter equation fails only infinitesimally, so
  it seems plausible that one can define consistent 
Bethe equations by taking small $\kappa$ as a regulator, to be removed at the end of the calculation.

In section 2 we review the Lagrangian and symmetries of ${\cal N}=2$ SCQCD and of the interpolating superconformal field theory.
In section
3.1 we evaluate the one-loop dilation operator of SCQCD (in the scalar sector), 
and write it as a spin-chain Hamiltonian. In section 3.2 we find the spectrum of magnon excitations of this spin chain.
These  calculations are repeated in sections 3.3 and 3.4 for the
 the interpolating SCFT. A simplified derivation of the Hamiltonians is presented in appendix A, while appendix B contains
 an equivalent way to write the Hamiltonian for ${\cal N}=2$ SCQCD in terms of composite (dimeric) impurities.
 In section 4 we study the spectrum of protected operators of the interpolating theory,
 and follow its evolution in the limit $\kappa \to 0$. In section 5 we solve the two-magnon scattering problem 
 and analyze the spectrum of bound states in the spin chain of the interpolating SCFT, with particular attention
 to the $\kappa \to 0$ limit. In section 5 we check the Yang-Baxter equation for the two-body S-matrix of the interpolating
 theory, finding that it holds for $\kappa =1$ and $\kappa \to 0$.
 We conclude in section 6 with a brief discussion of  integrability and of future directions of research.

 \section{Lagrangian and Symmetries}

 \subsection{${\cal N} =2$ SCQCD} 
\label{Lagrangians}

Our main interest is
${\cal N}=2$  SYM theory with gauge group $SU(N_{c})$ and $N_{f} = 2 N_c$ fundamental hypermultiplets. 
We refer to this theory as ${\cal N} = 2$ superconformal QCD (SCQCD). 
Its global symmetry group is $U(N_f)  \times SU(2)_R  \times U(1)_r$,
where  $SU(2)_R  \times U(1)_r$  is the R-symmetry subgroup of the superconformal group. 
We use indices $ \II,  \JJ = \pm$  for $SU(2)_R$, $i, j=1,  \dots N_f$ for the flavor group $U(N_f)$
and $a,b=1,  \dots N_c$ for the color group $SU(N_c)$.

Table  \ref{charges} summarizes the field content and quantum numbers  of the model:
The Poincar\'e supercharges  ${\cal Q}^{\II}_ \alpha$, $ \bar {\cal Q}_{\II  \,  \dot  \alpha}$ and the conformal
supercharges ${\cal S}_{\II  \,  \alpha}$, $ \bar {\cal S}^{\II}_ {\dot  \alpha}$
 are $SU(2)_R$ doublets with charges $ \pm 1/2$ under $U(1)_r$.
The $ \NN=2$ vector multiplet consists of a gauge field $A_{\mu}$, two Weyl spinors
$ \lambda_{\alpha}^{\II}$, $ \II= \pm$, which form a doublet under $SU(2)_{R}$,
and one complex scalar $ \phi$, all in the adjoint representation of $SU(N_{c})$. 
Each $ \NN=2$ hypermultiplet consists of 
 an $SU(2)_{R}$ doublet $Q_{\II}$ of complex scalars
and of two Weyl spinors $ \psi_{\alpha}$ and $ \tilde{\psi}_{\alpha}$, $SU(2)_{R}$ singlets.
It is convenient to define the flavor contracted mesonic operators 
 \[ \label{meson}
 \MM_{\JJ  \,  \,  \, b}^{\,  \,  \II a}  \equiv  \frac{1}{\sqrt{2}} Q_{\JJ  \mbox{ }i}^{\mbox{ }a} \, \bar{Q}_{\mbox{ }b}^{\II  \mbox{ }i}  \, ,
 \]  
which may be  decomposed into
into the $SU(2)_{R}$ singlet and triplet combinations
   \begin{equation}  \label{M1M3}
 \MM_{{\bf {1}}}  \equiv  \MM^{\,  \,   \II}_{\II} \quad \mbox{and} \quad \MM_{ {\bf {3} }  \JJ  }^{\quad  \II}   \equiv
 \MM^{\,  \,  \II}_{\JJ}- \frac{1}{2} \MM^{\,  \,  \KK}_{\KK} \,  \delta^{\II}_{\JJ}  \, .
 \end{equation}
The operators ${\cal M}$ decompose  into the adjoint plus the singlet representations of the color group $SU(N_c)$;
the singlet piece is however subleading in the large $N_c$ limit.

 \begin{table}
 \begin{centering}
 \begin{tabular}{|c||c|c|c|c|}
 \hline 
 & $SU(N_{c})$  & $U(N_{f})$  & $SU(2)_{R}$  & $U(1)_{r}$ \tabularnewline
 \hline
 \hline 
$ \mathcal{Q}_{\alpha}^{\II}$  &  \textbf{$ \mathbf{1}$}  &  \textbf{$ \mathbf{1}$}  &  \textbf{$ \mathbf{{2}}$}  & $+1/2$ \tabularnewline
 \hline 
$ \SS_{\II  \, \alpha}$ &  \textbf{$ \mathbf{1}$}  &  \textbf{$ \mathbf{1}$}  &  \textbf{$ \mathbf{2}$}  & $-1/2$ \tabularnewline
 \hline
 \hline 
$A_{\m}$  & Adj  &  \textbf{$ \mathbf{1}$}  &  \textbf{$ \mathbf{1}$}  & $0$ \tabularnewline
 \hline 
$ \f$  & Adj  &  \textbf{$ \mathbf{1}$}  &  \textbf{$ \mathbf{1}$}  & $-1$ \tabularnewline
 \hline 
$ \la_{\alpha}^{\II}$  & Adj  &  \textbf{$ \mathbf{1}$}  &  \textbf{$ \mathbf{2}$}  & $-1/2$ \tabularnewline
 \hline 
$Q_{\II}$  & $ \Box$  & $ \Box$  &  \textbf{$ \mathbf{2}$}  & $0$ \tabularnewline
 \hline 
$ \psi_{\alpha}$  & $ \Box$  & $ \Box$  &  \textbf{$ \mathbf{1}$}  & $+1/2$ \tabularnewline
 \hline 
$ \tilde{\psi}_{\alpha}$  & $ \overline{\Box}$  & $ \overline{\Box}$  &  \textbf{$ \mathbf{1}$}  & $+1/2$ \tabularnewline
 \hline
 \hline 
$ \MM_{\bf 1}$ & Adj + {\bf1}  &  \textbf{$ \mathbf{1}$}  &  \textbf{$ \mathbf{1}$}  & $0$ \tabularnewline
 \hline 
$ \MM_{\bf 3}$ & Adj  + {\bf 1}&  \textbf{$ \mathbf{1}$}  &  \textbf{$ \mathbf{3}$}  & $0$ \tabularnewline
 \hline
 \end{tabular}
 \par \end{centering}
 \caption{\label{charges} Symmetries of  $ \NN=2$ SCQCD.
We show the quantum numbers of  the supercharges ${\cal Q}^ \II$, ${\cal S}_ \II$,
of the elementary components fields and of the mesonic operators ${\cal M}$. 
Conjugate objects (such as $ \bar {\cal Q}_{\II  \dot{\alpha}}  $ and $ \bar  \phi$) are not written explicitly.}
 \end{table}

 The Lagrangian  is
  \[
 \LL= \LL_{V}+ \LL_{H}\,,
 \]
where $ \LL_V$ stands for the Lagrangian of the $\NN=2$ vector multiplet and the $ \LL_H$ for the Lagrangian of $\NN=2$ hypermultiplet. Explicitly\footnote{In our conventions,
$D_\mu \equiv \partial_\mu  + i g_{YM} A_\mu$. We raise and lower $SU(2)_R$ indices with the antisymmetric symbols $\epsilon_{\II \JJ}$  and $\epsilon^{\II \JJ}$, which obey
$ \epsilon_{\II \JJ} \, \epsilon^{\JJ \KK}= \delta_{\II}^{\KK}$.}
\begin{eqnarray}
 \LL_{V} & = & - \Tr \Big[ \frac{1}{4}F^{\mu \nu}F_{\mu \nu}+i \, \bar{\lambda}_{\II} \bar{\sigma}^{\mu}D_{\mu} \lambda^{\II}+(D^{\mu} \phi)(D_{\mu} \phi)^{\dagger} \nonumber  \\
 &  & +i \sqrt{2} \, (g_{YM} \, \epsilon_{\II \JJ} \lambda^{\II} \lambda^{\JJ} \phi^{\dagger}- g_{YM} \, \epsilon^{\II \JJ} \bar{\lambda}_{\II} \bar{\lambda}_{\JJ} \phi)+ \frac{g_{YM}^{2}}{2} \big[ \phi \,, \, \phi^{\dagger} \big]^{2} \Big]\,.\label{eq:vectorlag}
  \end{eqnarray}
 \begin{eqnarray}
 \LL_{H} & = & - \Big[(D^{\mu} \bar{Q}^{\II})(D_{\mu}Q_{\II})+i \, \bar{\psi} \bar{\sigma}^{\mu}D_{\mu} \psi+i \, \tilde{\psi} \bar{\sigma}^{\mu}D_{\mu} \bar{\tilde{\psi}}   \\
 &  & +i  \sqrt{2} \, (g_{YM} \, \epsilon^{\II \JJ} \bar{\psi} \bar{\lambda}_{\II}Q_{\JJ}- g_{YM} \, \epsilon_{\II \JJ} \bar{Q}^{\II} \lambda^{\JJ} \psi \nonumber \\
& & + g_{YM} \, \tilde{\psi} \lambda^{\II}Q_{\II}- g_{YM} \, \bar{Q}^{\II} \bar{\lambda}_{\II} \bar{\tilde{\psi}} \nonumber  \\
 &  & + g_{YM} \, \tilde{\psi} \phi \psi- g_{YM} \, \bar{\psi} \bar{\phi} \bar{\tilde{\psi}}) \nonumber \\
 &  & +g_{YM}^{2} \bar{Q}_{\II}( \phi^{\dagger} \phi+ \f \f^{\dagger})Q^{\II}+g_{YM}^{2} \mathcal{V} (Q)\Big] \, ,\nonumber
  \label{eq:hyperlag}
  \end{eqnarray}
 where the potential for the squarks is
  \begin{eqnarray}
 \mathcal{V}(Q) & = & ( \bar{Q}_{\mbox{ }a}^{\II \mbox{ }i}Q_{\II \mbox{ }j}^{\mbox{ }a})( \bar{Q}_{\mbox{ }b}^{\JJ \mbox{ }j}Q_{\JJ \mbox{ }i}^{\mbox{ }b})- \frac{1}{2}( \bar{Q}_{\mbox{ }a}^{\II \mbox{ }i}Q_{\JJ \mbox{ }j}^{\mbox{ }a})( \bar{Q}_{\mbox{ }b}^{\JJ \mbox{ }j}Q_{\II \mbox{ }i}^{\mbox{ }b}) \nonumber  \\
 &  & + \frac{1}{N_{c}}( \frac{1}{2}( \bar{Q}_{\mbox{ }a}^{\II \mbox{ }i}Q_{\II \mbox{ }i}^{\mbox{ }a})( \bar{Q}_{\mbox{ }b}^{\JJ \mbox{ }j}Q_{\JJ \mbox{ }j}^{\mbox{ }b})-( \bar{Q}_{\mbox{ }a}^{\II \mbox{ }i}Q_{\JJ \mbox{ }i}^{\mbox{ }a})( \bar{Q}_{\mbox{ }b}^{\JJ \mbox{ }j}Q_{\II \mbox{ }j}^{\mbox{ }b}))\,. \label{eq:SQCDpotential}
  \end{eqnarray}
Using the flavor contracted mesonic operator ($ \ref{meson}$), $ \mathcal{V}$ can be written more compactly as 
 \begin{eqnarray*}
 \mathcal{V} & = &  \Tr[ \MM^{\JJ} \,_{\II} \MM^{\II} \,_{\JJ}]- \frac{1}{2} \Tr[ \MM^{\II} \,_{\II} \MM^{\JJ} \,_{\JJ}] \\
 &  & - \frac{1}{N_{c}} \Tr[ \MM^{\JJ} \,_{\II}] \Tr[ \MM^{\II} \,_{\JJ}]+ \frac{1}{2} \frac{1}{N_{c}} \Tr[ \MM^{\II} \,_{\II}] \Tr[ \MM^{\JJ} \,_{\JJ}] \\
& = &  \Tr[ \MM_{{\bf {3}}} \MM_{{\bf {3}}}]- \frac{1}{N_{c}} \Tr[ \MM_{{\bf {3}}}] \Tr[ \MM_{{\bf {3}}}]\,.
  \end{eqnarray*}

\subsection{\label{sec:orbifold} $\mathbb{Z}_2$ orbifold of ${\cal N} = 4$ and interpolating family of SCFTs}

${\cal N} = 2$ SCQCD can be viewed as a limit of a family of superconformal theories;
 in the opposite limit the family reduces to a $\mathbb{Z}_2$ orbifold of ${\cal N} =4$ SYM.  In this subsection we first describe
the orbifold theory and then its connection to ${\cal N} = 2$ SCQCD.

As familiar, the field content of ${\cal N}=4$ SYM 
 comprises the gauge field $A_\mu$, four Weyl fermions
$\lambda^A_\alpha$ and six real scalars $X_{AB}$, where $A,B=1,\dots 4$ are indices of the $SU(4)_R$
R-symmetry group. Under $SU(4)_R$, the fermions are in the ${\bf 4}$ representation, while
the scalars are in ${\bf 6}$ (antisymmetric self-dual)  and  obey the reality condition\footnote{The $\dagger$ indicates
hermitian conjugation of the matrix in color space. We choose hermitian generators for the color group.}
\begin{equation}
X_{AB}^\dagger = \frac{1}{2} \epsilon^{ABCD} X_{CD}  \,. 
\end{equation}
We may parametrize $X_{AB}$ in terms of six real scalars $X_k$, $k=4,\dots 9$,
 \begin{equation}
X_{AB}=\frac{1}{\sqrt{2}}\left(\begin{array}{cc|cc}
0 & X_{4}+iX_{5} & X_{7}+iX_{6} & X_{8}+iX_{9}\\
-X_{4}-iX_{5} & 0 & X_{8}-iX_{9} & -X_{7}+iX_{6}\\
\hline -X_{7}-iX_{6} & -X_{8}+iX_{9} & 0 & X_{4}-iX_{5}\\
-X_{8}-iX_{9} & X_{7}-iX_{6} & -X_{4}+iX_{5} & 0\end{array}\right)\label{Xmatrix}\end{equation}
Next, we pick an $SU(2)_L \times SU(2)_R \times U(1)_r$ subgroup of $SU(4)_R$,
\begin{equation}
\begin{array}{cc}
1 & +\\
2 & -\\
3 &  \hat + \\
4 &  \hat -\end{array}
\left(\begin{array}{cc|cc}
SU(2)_{R}  \times U(1)_r&  & \,\\
 &  & \,\\
\hline  &  & \,\\
 &  & \, & SU(2)_{L} \times U(1)_r^* \end{array}\right) \, .\end{equation}
We use indices $\II, \JJ = \pm$ for $SU(2)_R$ (corresponding to $A,B =1,2$) and indices
$\hat \II, \hat \JJ = \hat \pm$  for $SU(2)_L$ (corresponding to $A,B =3,4$). To make more manifest
their transformation properties, the scalars are rewritten as the $SU(2)_L \times SU(2)_R$ singlet
$Z$ (with  charge $-1$ under $U(1)_r$) and as the bifundamental  $\mathcal{X}_{\II\hat{\II}}$ (neutral under $U(1)_r$),
\begin{equation}
Z \equiv \frac{X_{4}+iX_{5}}{\sqrt{2}} \, ,
\qquad \mathcal{X}_{\II\hat{\II}}\equiv \frac{1}{\sqrt{2}}\left(\begin{array}{cc}
X_{7}+iX_{6} & X_{8}+iX_{9}\\
X_{8}-iX_{9} & -X_{7}+iX_{6}\end{array}\right) \,. \end{equation}
Note the reality condition $\mathcal{X}_{\II\hat{\II}}^\dagger = -\epsilon^{\II \JJ } \epsilon^{\hat \II \hat \JJ } \mathcal{X}_{\JJ \hat{\JJ}}$.
Geometrically,  $SU(2)_{L}\times SU(2)_{R}\cong SO(4)$
is the group of $6789$ rotations  and $U(1)_{R}\cong SO(2)$
the group of $45$ rotations.  Diagonal 
$SU(2)$ transformations $\mathcal{X} \rightarrow U\mathcal{X} U^{-1}$ ($U_R = U, U_L = U^*$)
preserve the trace, $\Tr[\mathcal{X}]=2iX_{6}$,  and thus
 correspond to $789$ rotations.

We  are now ready to discuss the orbifold projection. 
In  R-symmetry space, the orbifold group is  chosen to be $\mathbb{Z}_2 \subset SU(2)_L$ 
with elements  $\pm \mathbb{I}_{2 \times 2}$. This is the well-known
quiver theory \cite{Douglas:1996sw}
obtained by placing $N_c$ D3 branes at the $A_1$ singularity
$ \mathbb{R}^{2}\times\mathbb{R}^{4}/\mathbb{Z}_{2}$,  with $(X_6, X_7, X_8, X_9) \to \pm (X_6, X_7, X_8, X_9)$ and $X_4$ and $X_5$ invariant. Supersymmetry
is broken to $\NN =2$, since the supercharges with $SU(2)_L$ indices are projected out.
The $SU(4)_R$ symmetry is broken to $SU(2)_L \times SU(2)_R \times U(1)_r $, or more precisely
to $SO(3)_L \times SU(2)_R \times U(1)_r$ since only objects with integer $SU(2)_L$ spin survive.
The $SU(2)_R \times U(1)_r$ factors are the R-symmetry of the unbroken ${\cal N} = 2$ superconformal group,
while $SO(3)_L$ is an extra global symmetry under which the unbroken supercharges are neutral.

In color  space, we start with gauge group $SU(2N_c)$, and declare
 the non-trivial element of the orbifold to be 
 \begin{equation} \label{tau}
 \twist\equiv\left(\begin{array}{cc}
\mathbb{I}_{N_{c}\times N_{c}} & 0\\
0 & -\mathbb{I}_{N_{c}\times N_{c}}\end{array}\right) \, .
\end{equation} 
All in all the $\mathbb{Z}_2$ action on the ${\cal N} =4$ fields is
\begin{equation}
A_\mu  \rightarrow  \twist A_\mu \twist \, , \quad
Z_{\II\JJ}  \rightarrow  \twist Z_{\II\JJ}\twist \, ,\quad   \lambda_{\II} \to \twist \lambda_{\II} \twist \, , \quad 
\mathcal{X}_{\II\IIh}  \rightarrow  -\twist\mathcal{X}_{\II\IIh}\twist\, ,\quad\lambda_{\hat \II} \to -\twist \lambda_{\hat \II} \twist \,.
\end{equation}
The components that  survive the projection are
 \begin{eqnarray}  \label{survive}
A_\mu & = &   \left( \begin{array}{cc}
A_{\mu b}^{a} & 0 \\
0 &  \topp A_{\mu \topp b}^{\topp a} \end{array} \right) \quad
Z  =   \left( \begin{array}{cc}
 \f_{\mbox{ } \mbox{ }b}^{a} & 0 \\
0 &  \topp{\f}_{\mbox{ } \mbox{ } \topp b}^{\topp a} \end{array} \right) 
\\
 \lambda_{\II}& =&  \left( \begin{array}{cc}
 \lambda_{\II b}^{a} & 0 \\
0 &  \topp{\la}_{\II \topp b}^{\topp a} \end{array} \right)\quad
 \lambda_{\hat{\II}}  =   \left( \begin{array}{cc}
0 &  \psi_{\hat{\II} \topp a}^{a} \\
 \tilde{\psi}_{\hat{\II}b}^{\topp b} & 0 \end{array} \right)
 \label{fermident}
\\
 \mathcal{X}_{\II \IIh} & = &  \left( \begin{array}{cc}
0 & Q_{\II \IIh \topp a}^{\mbox{ }a} \\
- \epsilon_{\II \JJ} \epsilon_{\hat{\II} \hat{\JJ}} \bar{Q}_{\mbox{ } \mbox{ }b}^{\topp b \hat{\JJ} \JJ} & 0 \end{array} \right) 
  \,.
 \end{eqnarray}
The gauge group  is broken to $SU(N_c) \times SU(N_{\check c}) \times U(1)$,
where the $U(1)$ factor is the {\it relative}\footnote{Had we started with $U(2 N_c)$ group,
we would also have an extra {\it diagonal} $U(1)$, which would completely decouple since no fields are charged under it.}
 $U(1)$ generated by $\twist$ (equ.(\ref{tau})): it
must be removed by hand, since its beta function is non-vanishing. The process of removing the relative $U(1)$ modifies the scalar potential
by double-trace terms, which arise from the fact that the auxiliary fields (in ${\cal N}=1$ superspace) are now missing the $U(1)$ component. Equivalently
we can evaluate the beta function for the double-trace couplings, and tune them to their fixed point \cite{Dymarsky:2005nc}.

Supersymmetry organizes the component fields  into the $\NN =2$ vector multiplets of each factor of the gauge group,
 $(\phi, \lambda_\II, A_\mu)$ and  $(\check \phi, \check \lambda_\II,\check A_\mu)$,
 and into two bifundamental hypermultiplets,  $(Q_{\II, \hat +}, \psi_{\hat +}, \tilde \psi_{\hat +})$
 and  $(Q_{\II, \hat -}, \psi_{\hat -}, \tilde \psi_{\hat -})$. Table 2 summarizes the field
 content and quantum numbers of the orbifold theory.

\begin{table}
\begin{centering}
\begin{tabular}{|c||c|c|c|c|c|}
\hline 
 & $SU(N_{c})$  & $SU(N_{\check c})$  & $SU(2)_{R}$  & $SU(2)_{L}$  & $U(1)_{R}$\tabularnewline
\hline
\hline 
$\QQ_{\alpha}^{\II}$ & \textbf{${\bf 1}$}  & \textbf{${\bf 1}$}  & \textbf{${\bf 2}$}  & \textbf{${\bf 1}$}  & +1/2\tabularnewline
\hline 
$\SS_{\II \, \alpha}$ & \textbf{${\bf 1}$}  & \textbf{${\bf 1}$}  & \textbf{${\bf 2}$}  & \textbf{${\bf 1}$}  & --1/2\tabularnewline
\hline
\hline 
$A_{\mu}$  & Adj  & \textbf{${\bf 1}$}  & \textbf{${\bf 1}$}  & \textbf{${\bf 1}$}  & 0\tabularnewline
\hline 
$\topp A_{\mu}$  & \textbf{${\bf 1}$}  & Adj  & \textbf{${\bf 1}$}  & \textbf{${\bf 1}$}  & 0\tabularnewline
\hline 
$\f$  & Adj  & \textbf{${\bf 1}$}  & \textbf{${\bf 1}$}  & \textbf{${\bf 1}$}  & --1\tabularnewline
\hline 
$\topp{\f}$  & \textbf{${\bf 1}$}  & Adj  & \textbf{${\bf 1}$}  & \textbf{${\bf 1}$}  & --1\tabularnewline
\hline 
$\la^{\II}$  & Adj  & \textbf{${\bf 1}$}  & \textbf{${\bf 2}$}  & \textbf{${\bf 1}$}  & --1/2\tabularnewline
\hline 
$\topp{\la}^{\II}$  & \textbf{${\bf 1}$}  & Adj  & \textbf{${\bf 2}$}  & \textbf{${\bf 1}$}  & --1/2\tabularnewline
\hline 
$Q_{\II\hat{\II}}$  & $\Box$  & $\overline{\Box}$  & \textbf{${\bf 2}$}  & \textbf{${\bf 2}$}  & 0\tabularnewline
\hline 
$\psi_{\hat{\II}}$  & $\Box$  & $\overline{\Box}$  & \textbf{${\bf 1}$}  & \textbf{${\bf 2}$}  & +1/2\tabularnewline
\hline 
$\tilde{\psi}_{\hat{\II}}$  & $\overline{\Box}$  & $\Box$  & \textbf{${\bf 1}$}  & \textbf{${\bf 2}$}  & +1/2\tabularnewline
\hline
\end{tabular}
\par\end{centering}
\caption{\label{orbifoldcharges}
Symmetries of  the $\mathbb{Z}_2$ orbifold of ${\cal N} = 4$ SYM and of the interpolating family of ${\cal N} = 2$ SCFTs.
}
\end{table}

The two gauge-couplings $g_{YM}$ and $\check g_{YM}$ can be independently varied
while preserving ${\cal N} =2$ superconformal invariance, thus defining a two-parameter family
of ${\cal N} = 2$ SCFTs. Some care is needed in adjusting the
 Yukawa and scalar potential terms so that ${\cal N} =2$ supersymmetry is preserved.
We find 
\begin{eqnarray} 
\label{yukawa}
\LL_{Yukawa}(g_{YM},\check{g}_{YM}) & = & i\sqrt{2}\mbox{Tr}\big[-g_{YM}\epsilon^{\II\JJ}\bar{\lambda}_{\II}\bar{\lambda}_{\JJ}\f-\check{g}_{YM}\epsilon^{\II\JJ}\bar{\topp{\la}}_{\II}\bar{\topp{\la}}_{\JJ}\topp{\f}\nonumber \\
 &  & +g_{YM}\epsilon^{\hat{\II}\hat{\JJ}}\tilde{\psi}_{\hat{\II}}\f\psi_{\hat{\JJ}}+\check{g}_{YM}\epsilon^{\hat{\II}\hat{\JJ}}\psi_{\hat{\JJ}}\topp{\f}\tilde{\psi}_{\hat{\II}}\nonumber \\
 &  & +g_{YM}\e^{\IIh\JJh}\tilde{\psi}_{\hat{\JJ}}\lambda^{\II}Q_{\II\IIh}+\check{g}_{YM}\e^{\IIh\JJh}Q_{\II\IIh}\topp{\la}^{\II}\tilde{\psi}_{\hat{\JJ}}\nonumber \\
 &  & -g_{YM}\e_{\II\JJ}\bar{Q}^{\hat{\JJ}\II}\lambda^{\JJ}\psi_{\hat{\JJ}}-\check{g}_{YM}\e_{\II\JJ}\psi_{\hat{\JJ}}\topp{\la}^{\II}\bar{Q}^{\hat{\JJ}\JJ}\big]+h.c.
 \end{eqnarray}
\begin{eqnarray}
 \label{potential}
\mathcal{V}(g_{YM},\check{g}_{YM}) & = & g_{YM}^{2}\mbox{Tr}\big[\frac{1}{2}[\fbar,\f]^{2}+\MM_{\II}^{\:\:\II}(\f\fbar+\fbar\f)+\MM_{\II}^{\:\:\JJ}\MM_{\JJ}^{\:\:\II}-\frac{1}{2}\MM_{\II}^{\:\:\II}\MM_{\JJ}^{\:\:\JJ}\big]\nonumber \\
 &  & +\topp g_{YM}^{2}\mbox{Tr}\big[\frac{1}{2}[\bar{\fh},\fh]^{2}+\topp{\MM}_{\:\:\II}^{\II}(\fh\bar{\fh}+\bar{\fh}\fh)+\topp{\MM}_{\:\:\JJ}^{\II}\topp{\MM}_{\:\:\II}^{\JJ}-\frac{1}{2}\topp{\MM}_{\:\:\II}^{\II}\topp{\MM}_{\:\:\JJ}^{\JJ}\big]\nonumber \\
 &  & +g_{YM}\topp g_{YM}\mbox{Tr}\big[-2Q_{\II\IIh}\fh\bar{Q}^{\IIh\II}\bar{\f}+h.c.\big]-\frac{1}{N_{c}}\mathcal{V}_{d.t.}  \, ,
 \end{eqnarray}
 where  the mesonic operators $\MM$ are defined as\footnote{Note that $\mbox{Tr}[\MM_{\II}^{\:\:\JJ}]=\mbox{Tr}[\topp{\MM}_{\:\:\II}^{\JJ}]$.}
\begin{equation} 
\label{mesonic}
\MM_{\JJ\:\:\: b}^{\:\:\II a}  \equiv  \frac{1}{\sqrt{2}}Q_{\JJ\hat{\JJ}\topp a}^{a}\bar{Q}_{\quad\:\:\:\: b}^{\JJh\II\topp a}\, \, , \qquad
\topp{\MM}_{\:\:\JJ\topp b}^{\II\topp a}  \equiv  \frac{1}{\sqrt{2}}\bar{Q}_{\quad\:\:\:\: a}^{\JJh\II\topp a}Q_{\JJ\hat{\JJ}\topp b}^{a} \, ,
\end{equation}
 and the double-trace terms in the potential are
\begin{eqnarray} \label{doubletraces}
\mathcal{V}_{d.t.} & = & g_{YM}^{2}\big(\mbox{Tr}[\MM_{\II}^{\:\:\JJ}]\mbox{Tr}[\MM_{\JJ}^{\:\:\II}]-\frac{1}{2}\mbox{Tr}[\MM_{\II}^{\:\:\II}]\mbox{Tr}[\MM_{\JJ}^{\:\:\JJ}]\big)\\
 &  & +\topp g_{YM}^{2}\big(\mbox{Tr}[\topp{\MM}_{\:\:\JJ}^{\II}]\mbox{Tr}[\topp{\MM}_{\:\:\II}^{\JJ}]-\frac{1}{2}\mbox{Tr}[\topp{\MM}_{\:\:\II}^{\II}]\mbox{Tr}[\topp{\MM}_{\:\:\JJ}^{\JJ}]\big)  \nonumber \\
 & = & \big(g_{YM}^{2}+\topp g_{YM}^{2}\big)\big(\mbox{Tr}[\MM_{\II}^{\:\:\JJ}]\mbox{Tr}[\MM_{\JJ}^{\:\:\II}]-\frac{1}{2}\mbox{Tr}[\MM_{\II}^{\:\:\II}]\mbox{Tr}[\MM_{\JJ}^{\:\:\JJ}]\big) \,. \nonumber
 \end{eqnarray}

The $SU(2)_L$ symmetry is present for all values of the couplings (and so is the $SU(2)_R \times U(1)_r$ R-symmetry, of course).
At the orbifold point $g_{YM} = \check g_{YM}$
there is an extra $\mathbb{Z}_2$ symmetry (the quantum symmetry of the orbifold) acting as 
\begin{equation} \label{quantumZ2}
\phi \leftrightarrow \check \phi \,, \quad
\lambda_\II \leftrightarrow \check \lambda_\II \, , \quad  A_\mu \leftrightarrow \check A_\mu \, ,\quad \psi_{\hat \II} \leftrightarrow \tilde \psi_{\hat \II} \, ,\quad
Q_{\II \hat \II} \leftrightarrow  -\epsilon_{\II\JJ}\epsilon_{\hat{\II}\hat{\JJ}}\bar{Q}^{ \JJ \hat \JJ } \, .
\end{equation}

Setting $\check g_{YM} = 0$, the second vector multiplet $(\check \phi, \check \lambda_\II, \check A_\mu)$ becomes free and completely decouples 
from the rest of theory, which happens to coincide 
with ${\cal N} = 2$ SCQCD (indeed the field content is the same and ${\cal N} = 2$ susy does the rest). 
The $SU(N_{\check{c}})$
symmetry can now be interpreted as a global flavor symmetry. In fact there is a symmetry enhancement
$SU(N_{\check c}) \times SU(2)_L \to U(N_f = 2 N_c)$: one sees in (\ref{yukawa}, \ref{potential}) that for $\check g_{YM} = 0$
 the $SU(N_{\check c})$ index $\check a$
and the $SU(2)_L$ index $\hat \II$
 can  be combined  into a single flavor index $i \equiv (\check a, \hat I) =1, \dots 2 N_c$.

In the rest of the paper, unless otherwise stated, we will work in the large $N_c \equiv N_{\check c}$ limit, keeping fixed
 the `t Hooft  couplings
\be
\lambda \equiv g_{YM}^2 N_c  \equiv 8 \pi^2 g^2 \, , \qquad  \check \lambda \equiv \check g_{YM}^2 N_{\check c}  \equiv 8 \pi^2 \check g^2 \, .
\ee
We will refer to the theory with arbitrary $g$ and $\check g$ as the ``interpolating SCFT'', thinking of keeping $g$ fixed
as we vary $\check g$ from  $\check g = g$  (orbifold theory) to  $\check g = 0$ 
 (${\cal N} =2$ SCQCD  $\oplus$ extra  $N_{\check c}^2-1$ free vector multiplets).

\section{One-loop Dilation Operator in the Scalar Sector}

\label{preview}

At large $N_c  \sim N_f$, the dilation operator 
of $\NN =2$ SCQCD  can be diagonalized in the sector of
 generalized single-trace operators, of the form (\ref{generalizedsingletrace}), indeed the mixing with generalized multi-traces is subleading.
Motivated by the success of the analogous calculation in  ${\cal N} =4$ SYM \cite{Minahan:2002ve},
we have evaluated the one-loop dilation operator on generalized single-trace operators
made out of scalar fields.  An example of such an operator is 
\begin{equation}
\label{local-operator}
\mbox{Tr}[\bar{\phi}\phi\phi Q_{\II}\bar{Q}^{\JJ}\bar{\phi}] =    \bar \phi^a_{\, \, \,b} \phi^b_{\, \, \,c}  \phi^c_{\, \, \,d}  Q^{\, d}_{\II \, \, i}  \bar Q^{\JJ  i}_{\quad e} \bar \phi^e_{\, \, \, a}    \, , \quad a,b,c,d,e=1,\dots N_c\, ,\quad i = 1, \dots N_f \,.
\end{equation} 
Since the  color or flavor indices of consecutive elementary fields
are contracted, we can assign each field to a definite ``lattice site''\footnote{Up to cyclic re-ordering of course,
under which the trace is invariant.} and think of a generalized single-trace operator
as a state in a periodic spin chain. In the scalar sector, the state space $V_l$ at each lattice site 
is six-dimensional, spanned by  $\{ \phi, \bar \phi,  Q_{\mathcal{I}}, \bar{Q}^{\mathcal{J}} \}$.
However the index structure of the fields imposes  restrictions on the total space $\otimes_{l=1}^L V_l$: not
all states in the tensor product are allowed. Indeed a  $Q$ at site $l$ must always be followed by  a $\bar Q$ at site $l+1$, and viceversa a $\bar Q$
must always be preceded by  a $Q$. Equivalently, as in appendix \ref{composite}, we may use instead
the color-adjoint objects $\phi$, $\bar \phi$,  ${\cal M}_{\bf 1}$ and ${\cal M}_{\bf 3}$
(recall the definitions (\ref{M1M3})), where the ${\cal M}$'s are  viewed as ``dimers'' occupying two sites of the chain.
 
  As usual, we may interpret the perturbative dilation operator as the Hamiltonian of the spin chain. It is convenient
to factor out the overall coupling from the definition of the  Hamiltonian $H$,
 \begin{equation}
 \Gamma^{(1)}  \equiv g^2  H  \, ,  \qquad  g^2  \equiv  \frac{\la}{8  \pi^2} \, , \quad  \lambda  \equiv g_{YM}^2 N_c  \, ,
 \end{equation}
where $ \Gamma^{(1)}$ is the one-loop anomalous dimension matrix. By a simple extension of the usual arguments, the 
Veneziano double-line   notation (see figure \ref{quartic} for an example) makes it clear that for large  $N_c  \sim N_f$ (with $ \lambda$ fixed)
the  dominant contribution comes from planar diagrams. Planarity implies that the 
 one-loop Hamiltonian is of nearest-neighbor type,
 $H =  \sum_{l=1}^L H_{k k+1}$ (with $k  \equiv k + L$), where $H_{k,k+1}:  \,V_{k} \otimes V_{k+1}  \to V_{k} \otimes V_{k+1}$.
 The two-loop correction is next-to-nearest-neighbor and so on. In section \ref{SCQCDHamiltonian} we present our results for the one-loop Hamiltonian of the spin chain for SCQCD. We then derive (section 3.2) the one-particle ``magnon''
excitations of the infinite chain above the BPS vacuum $\dots \phi \phi \phi \dots$. The one-particle eigenstates are  interesting admixtures of the adjoint  $\bar \phi$ impurity and
of the ``dimeric'' $Q \bar Q$ impurities. 
 
 The generalization to the full interpolating SCFT is straightforward and is carried out in sections 3.3 and 3.4. The structure of this more general spin chain is in a sense more conventional,
 and it is somewhat reminiscent of the spin chain \cite{Minahan:2008hf,Bak:2008cp,Gaiotto:2008cg,Bak:2008vd} for the ABJM \cite{Aharony:2008ug} and ABJ \cite{Aharony:2008gk} theories.\footnote{An important difference is that our spin chain has an exact parity symmetry, whereas
 the spin chain of the ABJ theory is expected to violate parity at sufficiently high perturbative order (though somewhat surprisingly 
 the ABJ planar theory appears to be parity invariant to low perturbative order \cite{Bak:2008vd, Minahan:2009te, Minahan:2009aq}.)}

  There are two types of color indices,
for the two gauge groups $SU(N_c)$ and $SU(N_{\check c})$, with adjoint fields $\phi^{a}_{\; b}$ and $\check \phi^{\check a}_{\; \check b}$ carrying two indices of the same type,
and bifundamental fields $Q^{a}_{\; \check b}$ and $\bar Q^{\check a}_{\;  b}$ carrying two indices of opposite type. Of course one must contract
neighboring indices of the same type. 
Now a $Q$ and a $\bar Q$ need not be adjacent since they can be separated for $\check \phi$ fields.
The infinite chain admits two BPS vacua, the state with all $\phi$s and the state with all $\check \phi$s. The magnons  are momentum
eigenstates containing a single $Q$ or $\bar Q$ impurity, separating one BPS vacuum on the left from the other vacuum on the right.
We will see in section 5 how the ``dimeric'' $Q\bar Q$ impurities of the SCQCD chain arise in the limit $\check g \to 0$ from the localization of the  bound state wavefunctions
of the interpolating chain.

 \begin{figure}[t]
 \begin{centering}
 \includegraphics[scale=0.6]{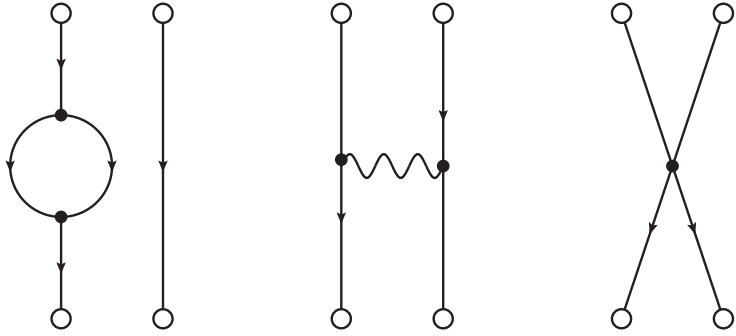} 
 \par \end{centering}
 \caption{\label{feyn-dia}Various types of Feynman diagrams that
contribute, at one loop, to anomalous dimension. The first diagram
is the self-energy contribution. The second diagram represents the
gluon exchange contribution whereas the third one stands for the quartic
interaction between the fields. The first and the second diagrams
are proportional to the identity in the R symmetry space while the
third one carries a nontrivial R symmetry index structure.}
\end{figure}
 
\subsection{\label{SCQCDHamiltonian}Hamiltonian for ${\cal N}=2$ super QCD}

 We have determined the one-loop dilation operator in the scalar sector
 by explicit evaluation of the divergent part of all the relevant Feynman diagrams, 
  which can be classified as self energy diagrams, gluon interaction diagrams
and quartic vertex diagrams and are schematically shown in figure  \ref{feyn-dia}. The calculation is
straightforward and its details will not be reproduced here. 
In appendix \ref{trick} we present a shortcut derivation that bypasses
the explicit evaluation of the self-energy and gluon exchange diagrams,
whose contribution can be fixed by requiring the vanishing of the anomalous dimension of certain protected operators.

As we are at it, we may as well consider the case of arbitrary $N_f$, though we are ultimately
interested in the conformal case $N_f = 2 N_c$. In the non-conformal case, it is more useful
to normalize the fields so that the Lagrangian has an overall factor of $1/g_{YM}^2$ in front \cite{DiVecchia:2004jw}.
This different normalization affects the anomalous dimension of composite operators for $N_f \neq 2 N_c$, which acquire
an extra contribution due to the beta function, but it is of course  immaterial for $N_f = 2N_c$. It is in this
normalization that the chiral operator ${\rm Tr} \phi^\ell$  has vanishing anomalous dimension for all $N_f$.

 We find\footnote{\label{M3foot}The spin chain with this nearest-neighbor Hamiltonian reproduces the one-loop anomalous
 dimension of all operators with $L >2$, where $L$ is the number of sites. The $L=2$ case is special:
 the double-trace terms in the scalar potential, which give 
 subleading contributions (at large $N$) for $L>2$, become important for $L=2$ and must be added separately. 
This special case  plays a role in the protection of ${\rm Tr} {\cal M}_{\bf 3}$, see section  \ref{sec:BPS-Spectrum}.
}
{\scriptsize{
 \begin{eqnarray}\label{HQCD}
 & H_{k,k+1}= 
 \\
& \nonumber \\
& \nonumber \\
 &  \bordermatrix{
 &  \phi^{\mathfrak{p}} \phi^{\mathfrak{q}} & Q_{\mathcal{I}} \bar{Q}^{\mathcal{J}} &  \bar{Q}^{\KK}Q_{\LL} & \bar{Q}^{\II} \phi^{\mathfrak{p}} & \phi^{\mathfrak{p}} {Q}_{\II} \cr
 &&&&  \cr
 \phi_{\mathfrak{p}^{\prime}} \phi_{\mathfrak{q}^{\prime}} & 2 \delta_{\mathfrak{p}^{\prime}}^{\mathfrak{p}} \delta_{\mathfrak{q}^{\prime}}^{\mathfrak{q}}+g^{\mathfrak{p} \mathfrak{q}}g_{\mathfrak{p}^{\prime} \mathfrak{q}^{\prime}}-2 \delta_{\mathfrak{q}^{\prime}}^{\mathfrak{p}} \delta_{\mathfrak{p}^{\prime}}^{\mathfrak{q}} &  \sqrt{\frac{N_{f}}{N_c}}g_{\mathfrak{p}^{\prime} \mathfrak{q}^{\prime}} \delta_{\mathcal{I}}^{\mathcal{J}} & 0 & 0 & 0
 \cr &&&&
 \cr
 \bar{Q}^{\II^{\prime}}Q_{\JJ^{\prime}} &  \sqrt{\frac{N_{f}}{N_c}}g^{\mathfrak{p} \mathfrak{q}} \delta_{\mathcal{J}^{\prime}}^{\mathcal{I}^{\prime}} & (2 \delta_{\mathcal{I}}^{\mathcal{I}^{\prime}} \delta_{\mathcal{J}^{\prime}}^{\mathcal{J}}-  \delta_{\mathcal{I}}^{\mathcal{J}} \delta_{\mathcal{J}^{\prime}}^{\mathcal{I}^{\prime}}) \frac{N_{f}}{N_c} & 0 & 0 &0
\cr
 &  & +\frac{1}{2}(1+\xi)\delta_{\mathcal{I}}^{\mathcal{I}^{\prime}} \delta_{\mathcal{J}^{\prime}}^{\mathcal{J}} &  &  &
\cr
Q_{\KK^{\prime}} \bar{Q}^{\LL^{\prime}} & 0 & 0 & 2 \delta_{\LL}^{\KK} \delta_{\KK^{\prime}}^{\LL^{\prime}} & 0 &0
\cr &&&-\frac{1}{2}(1+\xi)\delta_{\KK^{\prime}}^{\KK} \delta_{\LL}^{\LL^{\prime}}& &
\cr
 Q_{\II^{\prime}} \phi_{\mathfrak{p}^{\prime}} & 0 & 0 & 0 & \frac{1}{4}(7-\xi) \delta^{\mathcal{I}}_{\mathcal{I}^{\prime}} \delta_{\mathfrak{p}^{\prime}}^{\mathfrak{p}} & 0
 \cr
 &&&&&
 \cr
 \phi_{\mathfrak{p}^{\prime}} {\bar Q}^{\II^{\prime}}& 0 & 0 & 0 &0&  \frac{1}{4}(7-\xi) \delta_{\mathcal{I}}^{\mathcal{I}^{\prime}} \delta_{\mathfrak{p}^{\prime}}^{\mathfrak{p}} 
 }\nonumber
 \end{eqnarray}
}} 
The indices $ \mathfrak{p},  \mathfrak{q} =  \pm$ label the $U(1)_r$ charges of $ \phi$ and $ \bar  \phi$, in other terms 
we have defined $ \phi^-  \equiv  \phi$, $ \phi^+  \equiv  \bar  \phi$, and $g_{\mathfrak{p}  \mathfrak{q}} =  \left( \begin{array} {cc}
 0 & 1  \\
 1 & 0  \end{array}
 \right)$. The parameter $\xi$ is  the gauge parameter that appears in the gluon propagator as $\frac{1}{k^2}(g_{\mu \nu}-(1-\xi)\frac{k_\mu k_\nu}{k^2})$. 
Although the form of  nearest-neighbor Hamiltonian depends on gauge choice $\xi$, it is easy to check that $\xi$ dependence drops when $H$ acts on a closed chain.
In the following we will set $\xi=-1$.\footnote{This  choice corresponds  to setting to zero the self-energy of  $Q$ and $\bar Q$. Then
our Hamiltonian can also be used as is to calculate the anomalous dimension of operators with open {\it flavor} indices, of the schematic form $\bar Q^i \dots Q_j$. 
For $\xi\neq -1$ there are extra contributions form the self-energy of the $Q^i$ and $\bar Q_j$ at the edge of the chain.}

 We may rewrite $H_{k k+1}$  more concisely (we have set $\xi =-1$) as
\[   \label{H_SCQCD}
H_{k,k+1}= \bordermatrix{
 &  \f \f & Q \bar{Q} &  \bar{Q}Q & \bar{Q} \f & \f Q \cr
 &&&&&  \cr
 \f \f & 2 \mathbb{I} + \mathbb{K} -2 \mathbb{P} &  \sqrt{\frac{N_{f}}{N_c}}  & 0 & 0 &0\cr
 \bar{Q}Q &  \sqrt{\frac{N_{f}}{N_c}}  & (2 \mathbb{I} - \mathbb{K}) \frac{N_{f}}{N_{c}} & 0 & 0&0 \cr
Q \bar{Q} & 0 & 0 & 2 \mathbb{K} & 0&0 \cr
 Q \f & 0 & 0 & 0 & 2  &0
 \cr
 \f \bar{Q} & 0 & 0 & 0 & 0 &2 
 } 
 \]
The symbols $\mathbb{I}, \mathbb{P}$ and $ \mathbb{K}$ for identity, permutation and trace operators respectively. Their position in the matrix specifies the space in which they act. For example, the operator $ \mathbb P$ that appears in the matrix element of $ \langle  \f_{\pp^{\prime}}   \f_{\qq^{\prime}} |  \f^{\pp}  \f^{\qq}  \rangle$ is $ \delta ^{\pp}_{{\qq}^ \prime}  \delta^{\qq}_{{\pp}^ \prime}$,
 the operator $ \mathbb{K}$ that appears in the matrix element $ \langle  {\bar Q}^{\II^\prime} Q_{\JJ^\prime} | Q_{\II} {\bar Q}^{\JJ} \rangle$ stands for the operator $\delta^{\II^\prime}_{\JJ^\prime} \delta_{\mathcal{I}}^{\mathcal{J}}$ and so on. The entries where no symbols appear have an unambiguous index structure.
 In appendix B we present an equivalent from of the Hamiltonian in terms of composite (dimeric) impurities.

Although not immediately obvious from the form (\ref{H_SCQCD}), the Hamiltonian of the SCQCD spin chain preserves parity, once
the constraints on the states allowed by the index structure are taken into account. Parity is  in fact a symmetry of the spin chain
for the whole interpolating theory, the transformation rules are given  below in (\ref{parityrules}).

For $N_f =0$, the Hamiltonian can be consistently truncated to the space of
 $\phi$ (and $ \fbar$): it reduces $2 \mathbb{I}_{\f \f}+ \mathbb{K}_{\f \f}-2 \mathbb{P}_{\f \f}$, which is Hamiltonian of the XXZ spin chain, 
confirming the result  found in  \cite{DiVecchia:2004jw}  for pure $ \NN=2$ SYM. The $N_f \neq 0$ the
$ \f$ sector  is not closed in our case due to the leading order glueball-meson mixing. 

 \subsection{\label{SCQCDexcite} Magnons in the SCQCD spin chain}

The chiral operator Tr $\, \f^{\ell}$ and the antichiral operator Tr $\, \bar \f^{\ell}$
are zero-energy eigenstates (in particular the mixing element that is responsible for $ \f  \f  \rightarrow QQ$ is proportional to $ \mathbb K$ in $ \f$ space,
and thus vanishes when
two neighboring $ \f$ fields have the same $U(1)_r$ index).
They correspond to the two ferromagnetic ground states of the spin chain (all spins up or all down). We choose for definiteness the chiral vacuum Tr $\, \f^{\ell}$.
Recall that in our conventions the $U(1)_r$ charge of $\phi$ is $r= -1$, so the ground state obeys $\Delta + r = 0$, where $\Delta$ is the total conformal dimension.
Both $Q$ and $\bar Q$ have $\Delta + r = 1$, but the index structure forbids the insertion of only one of them. The simplest impurities that 
 can be excited on the ground state are $ \bar  \phi$, ${\cal M}_{\bf 1}$ and ${\cal M}_{\bf 3}$, where the last two are ``dimeric'' impurities 
which occupy two sites (recall (\ref{mesonic})). All  of them have $\Delta + r = 2$, and should be viewed in this sense as 
double excitations, though they are the most elementary we can find in the spin chain for ${\cal N}=2$ SCQCD.
  We will see that they can be viewed
as bound states of the elementary impurities of the interpolating theory with $\check g \neq 0$.
 This hidden compositeness makes the scattering problem somewhat harder than usual.

 In the map from the (generalized) single-trace operators to the states of the spin chain, cyclycity of the trace gives periodic boundary conditions on the chain, along with the constraint that the total momentum of all the impurities in the spin be zero.
 As usual, it is convenient to first consider the chain to be infinite, and impose later the zero-momentum constraint on multi-impurity states.
We now proceed to diagonalize the Hamiltonian on the space of states containing a single impurity (which in the present context means a single $\bar \phi$ or ${\cal M}_{\bf 1}$
or ${\cal M}_{\bf 3}$).
The action of $H$ on single impurities in position space is
\begin{eqnarray}
H[\fbar(x)] &=& 6\fbar(x)-\fbar(x+1)-\fbar(x-1)\\
& & + \sqrt{ \frac {2 N_f}{N_c} } \MM_{\bf 1}(x)+\sqrt{ \frac {2 N_f}{N_c} } \MM_{\bf 1}(x-1)\\
H[\MM_{\bf 1} (x)]&=& 4\MM_{\bf 1} (x)+ \sqrt{ \frac {2 N_f}{N_c} } \fbar (x)+ \sqrt{ \frac {2 N_f}{N_c} } \fbar (x+1)\nonumber \\
H[\MM_{\bf 3} (x)]&=&  8\MM_{\bf 3} (x) \, ,
\end{eqnarray}
where the coordinate $x$ denotes the site of the impurity on the chain; for the  dimeric impurities  $\MM_{\bf 1}$ and  $\MM_{\bf 3}$ 
we use the coordinate of the first site.  
To diagonalize the Hamiltonian on the $\bar \phi$/ $\MM_{\bf 1}$  sector, we
go to momentum space,
\begin{eqnarray}
\fbar (p) & \equiv  & \sum_x \fbar (x) e^{ipx} \, ,\quad \MM_{\bf 1}(p)\equiv \sum_x \MM_{\bf 1}(x) e^{ipx}\\
H \left ( 
\begin{array}{c}
\fbar (p) \\
\MM_{\bf 1}
\end{array} 
\right )
 & = & 
\left (
\begin{array}{cc}
6-e^{ip}-e^{-ip} & (1+e^{-ip})\sqrt{\frac {2N_f}{N_c}}  \\
(1+e^{ip})\sqrt{\frac {2N_f}{N_c}} & 4
\end{array}
\right)
 \left ( 
\begin{array}{c}
\fbar (p) \\
\MM_{\bf 1}
\end{array} 
\right ) \,.
\end{eqnarray}
The expressions for the eigenvalues and  eigenvectors are not very illuminating for generic values of the ratio $N_f/N_c$. For the conformal case of $N_f=2 N_c$, however, they simplify. 
The eigenstates for $N_f = 2 N_c$ are
\begin{eqnarray}
\label{Twave}
T(p) & \equiv &  -\frac {1}{2} (1+e^{-ip}) \fbar (p) +\MM_{\bf 1} (p)
 =  \sum_x e^{ipx}[-\frac{1}{2} (\fbar(x)+\fbar(x+1))+\MM_{\bf 1}(x)]\\
 \label{Ttildewave}
\widetilde T (p) & \equiv &   \fbar (p) + \frac {1}{2} (1+e^{ip}) \MM_{\bf 1} (p)
 =  \sum_x e^{ipx}[ \fbar(x)+\frac{1}{2}(\MM_{\bf 1}(x)+\MM_{\bf 1}(x-1)) ] \, ,
 \end{eqnarray}
 with eigenvalues
 \begin{eqnarray}
H  T(p)  & = &  4\sin^2(\frac{p}{2})\, T(p) \,  \label{T(p)}\\
H  \widetilde T(p)  & = & 8 \, \widetilde T(p)  \label{Ttilde(p)}\,.
\end{eqnarray}
Interestingly, precisely at the conformal point
 $N_f=2 N_c$  the magnon excitation $T(p)$ becomes gapless: in general the gap of $T(p)$ is $4-2 \sqrt{2 N_f/N_c}$.
 From now on we will only consider the superconformal case and set $N_f \equiv 2 N_c$.
 Besides $T(p)$ and $\widetilde T(p)$, we have of course also
  the ${\cal M}_{\bf 3}$ momentum eigenstate,
  \be
\MM_{\bf 3}(p)\equiv \sum_x \MM_{\bf 3}(x) e^{ipx} \, ,
\ee
which has the same momentum-independent energy as $\widetilde T(p)$,
\be
H   {\cal M}_{\bf 3 }(p)   = 8  \, {\cal M}_{\bf 3 }(p)\,.
\ee

 \subsection{\label{orbifold-Hamiltonian}Hamiltonian for the interpolating SCFT}

We have generalized the calculation of the one-loop dilation operator  to  the full interpolating family of ${\cal N} = 2$ SCFTs, in the scalar sector. We find 
\begin{eqnarray}
H & = & 
\bordermatrix{
& \phi^{\mathfrak{p}}\phi^{\mathfrak{q}} & Q_{\mathcal{I}\IIh}\bar{Q}^{\JJh\mathcal{J}}  \cr
\phi_{\mathfrak{p}^\prime}\phi_{\mathfrak{q}^\prime} & (2\delta_{\pp^{\prime}}^{\pp}\delta_{\qq^{\prime}}^{\qq}+g^{\pp\qq}g_{\pp^{\prime}\qq^{\prime}}-2\delta_{\qq^{\prime}}^{\pp}	        \delta_{\pp^{\prime}}^{\qq}) & \delta_{\II}^{\JJ}\delta_{\hat{\II}}^{\hat{\JJ}}g_{\pp^{\prime}\qq^{\prime}}\cr
\bar{Q}^{\IIh^\prime \mathcal{I}^\prime} Q_{\mathcal{J}^\prime\JJh^\prime} & \delta_{\JJ^{\prime}}^{\II^{\prime}}\delta_{\hat{\JJ}^{\prime}}^{\hat{\II}^{\prime}}g^{\pp\qq} & (2\delta_{\II}^{\II^{\prime}}\delta_{\JJ^{\prime}}^{\JJ}-\delta_{\II}^{\JJ}\delta_{\JJ^{\prime}}^{\II^{\prime}})\delta_{\hat{\II}}^{\hat{\JJ}}\delta_{\hat{\JJ}^{\prime}}^{\hat{\II}^{\prime}}+2\kappa^{2}\delta_{\II}^{\JJ}\delta_{\JJ^{\prime}}^{\II^{\prime}}\delta_{\IIh}^{\IIh^{\prime}}\delta_{\JJh^{\prime}}^{\JJh}
}
\nonumber
\\
\nonumber
\\
\nonumber
\\
\nonumber
& \oplus & 
\bordermatrix{
& \fh^{\mathfrak{p}}\fh^{\mathfrak{q}} & \bar{Q}^{\JJh \mathcal{J}}Q_{\mathcal{I}\IIh}  \cr
 \fh_{\mathfrak{p}^\prime}\fh_{\mathfrak{q}^\prime} & \kappa^{2}(2\delta_{\pp^{\prime}}^{\pp}\delta_{\qq^{\prime}}^{\qq}+g^{\pp\qq}g_{\pp^{\prime}\qq^{\prime}}-2\delta_{\qq^{\prime}}^{\pp}\delta_{\pp^{\prime}}^{\qq}) & \kappa^{2}\delta_{\II}^{\JJ}\delta_{\hat{\II}}^{\hat{\JJ}}g_{\pp^{\prime}\qq^{\prime}}\cr
Q_{\mathcal{J}^\prime \JJh^\prime}\bar{Q}^{\IIh^\prime \mathcal{I}^\prime} & \kappa^{2}\delta_{\JJ^{\prime}}^{\II^{\prime}}\delta_{\hat{\JJ}^{\prime}}^{\hat{\II}^{\prime}}g^{\pp\qq} & \kappa^{2}(2\delta_{\II}^{\II^{\prime}}\delta_{\JJ^{\prime}}^{\JJ}-\delta_{\II}^{\JJ}\delta_{\JJ^{\prime}}^{\II^{\prime}})\delta_{\hat{\II}}^{\hat{\JJ}}\delta_{\hat{\JJ}^{\prime}}^{\hat{\II}^{\prime}}+2\delta_{\II}^{\JJ}\delta_{\JJ^{\prime}}^{\II^{\prime}}\delta_{\IIh}^{\IIh^{\prime}}\delta_{\JJh^{\prime}}^{\JJh}
}
\nonumber
\\
\nonumber
\\
\nonumber
\\ 
\nonumber
& \oplus & 
\bordermatrix{
& \phi^{\mathfrak{p}}Q_{\mathcal{I}\IIh} & Q_{\mathcal{I}\IIh} \fh^{\mathfrak{p}} \cr
\phi_{\mathfrak{p}^\prime}\bar{Q}^{\IIh^\prime \mathcal{I}^\prime} & 2\delta_{\II}^{\II^{\prime}}\delta_{\IIh}^{\IIh^{\prime}}\delta_{\pp^{\prime}}^{\pp} & -2\kappa\delta_{\II}^{\II^{\prime}}\delta_{\IIh}^{\IIh^{\prime}}\delta_{\pp^{\prime}}^{\pp}\cr
\bar{Q}^{\IIh^\prime \mathcal{I}^\prime} \fh_{\mathfrak{p}^\prime} & -2\kappa\delta_{\II}^{\II^{\prime}}\delta_{\IIh}^{\IIh^{\prime}}\delta_{\pp^{\prime}}^{\pp} & 2\kappa^{2}\delta_{\II}^{\II^{\prime}}\delta_{\IIh}^{\IIh^{\prime}}\delta_{\pp^{\prime}}^{\pp}
}
\nonumber
\\
\nonumber
\\
\nonumber
\\ 
\label{explicitH}
& \oplus & 
\bordermatrix{
& \fh^{\mathfrak{p}}\bar Q^{\JJh \mathcal{J}} & \bar Q^{\JJh \mathcal{J}} \f^{\mathfrak{p}} \cr
\fh_{\mathfrak{p}^\prime} Q_{\mathcal{J}^\prime \JJh^\prime} & 2\kappa^{2}\delta_{\JJ^{\prime}}^{\JJ}\delta_{\JJh^{\prime}}^{\JJh}\delta_{\pp^{\prime}}^{\pp} & -2\kappa\delta_{\JJ^{\prime}}^{\JJ}\delta_{\JJh^{\prime}}^{\JJh}\delta_{\pp^{\prime}}^{\pp}\cr
Q_{\mathcal{J}^\prime \JJh^\prime} \f_{\mathfrak{p}^\prime} & -2\kappa\delta_{\JJ^{\prime}}^{\JJ}\delta_{\JJh^{\prime}}^{\JJh}\delta_{\pp^{\prime}}^{\pp} & 2\delta_{\JJ^{\prime}}^{\JJ}\delta_{\JJh^{\prime}}^{\JJh}\delta_{\pp^{\prime}}^{\pp}
}
\end{eqnarray}
\newpage
In concise form,\footnote{The meaning of the different operators can be read off by comparing with the explicit form above.
Note in particular that to avoid cluttering we have dropped the identity symbol $\mathbb{I}$. 
Also in the subspaces $Q \bar Q$, $\bar Q Q$ we use the notation $\mathbb{K}$  for the trace operator acting on $SU(2)_R$ indices and
 $\hat {\mathbb{K}}$ that acts on the $SU(2)_L$  indices. }
{\scriptsize{
\begin{eqnarray*}
 & H_{k,k+1}=\\
 & \bordermatrix{
& \f\f & Q\bar{Q} & \fh\fh & \bar{Q}Q & \f Q & Q\fh & \fh\bar{Q} & \bar{Q}\f \cr
&&&& \cr
\f\f & (2+\mathbb{K}-2\mathbb{P}) & \mathbb{K} & 0 & 0 & 0 & 0 & 0 & 0\cr
Q\bar{Q} & \mathbb{K}  & (2-\mathbb{K})\hat{\mathbb{K}}+2\kappa^{2}\mathbb{K} & 0 & 0 & 0 & 0 & 0 & 0\cr
\fh\fh & 0 & 0 & \kappa^{2}(2+\mathbb{K}-2\mathbb{P}) & \kappa^{2}\mathbb{K}  & 0 & 0 & 0 & 0\cr
\bar{Q}Q & 0 & 0 & \kappa^{2}\mathbb{K}  & \kappa^{2}(2-\mathbb{K})\hat{\mathbb{K}}+2\mathbb{K} & 0 & 0 & 0 & 0\cr
\f Q & 0 & 0 & 0 & 0 & 2 & -2\kappa & 0 & 0\cr
Q\fh & 0 & 0 & 0 & 0 & -2\kappa & 2\kappa^{2} & 0 & 0\cr
\fh\bar{Q} & 0 & 0 & 0 & 0 & 0 & 0 & 2\kappa^{2} & -2\kappa\cr
\bar{Q}\f & 0 & 0 & 0 & 0 & 0 & 0 & -2\kappa & 2} \nonumber
\end{eqnarray*}
}}
where 
\be
\kappa\equiv \frac{\check g}{g} \, ,\quad g^2 \equiv \frac{g_{YM}^2 N}{8 \pi^2} \, ,\quad \check g^2 \equiv \frac{\check g_{YM}^2 N}{8 \pi^2}\,.
\ee
It is easy to check that in the limit $\kappa \to 0$  this Hamiltonian reduces to that of the SCQCD spin chain, as it should.\footnote{In the comparison, it is important to take
into account the factors that arise by normalizing to one the tree-level two-point function. Recall that in SCQCD $\bar Q_i Q^i$ is contracted summing
over the $N_f = 2 N_c$ flavors, while in the interpolating SCFT $\bar Q_{\check a} Q^{\check a}$ is contracted summing over the $N_c$ colors (leaving
open the $SU(2)_L$ indices).}  

The Hamiltonian can also be compactly written in terms of the  $\mathbb{Z}_2$-projected $SU(2N_c)$ adjoint fields  Z and $\cal X$, 
\begin{equation}
Z=\left(\begin{array}{cc} \phi & 0\\
0 & \check{\phi}\end{array}\right),\qquad 
\mathcal{X}_{\II \IIh} =\left( \begin{array}{cc} 0 & Q_{\II \IIh} \\
- \epsilon_{\II \JJ} \epsilon_{\hat{\II} \hat{\JJ}} \bar{Q}^{\hat{\JJ} \JJ} & 0 \end{array} \right) \,.
\end{equation}
In this notation,
\[
g^2 H=\left(\begin{array}{cccc}
ZZ & {\cal XX} & Z{\cal X} & {\cal X}Z\\
(\gbar+\twist \gdiff)^{2}(2+\mathbb{K}-2\mathbb{P}) & (\gbar+\twist \gdiff)^{2}\mathbb{K}\hat{\mathbb{K}} & 0 & 0\\
\\(\gbar+\twist \gdiff)^{2}\mathbb{K}\hat{\mathbb{K}} & (\gbar+\twist \gdiff)^{2}(2\hat{\mathbb{K}}-\mathbb{K}\hat{\mathbb{K}}) & 0 & 0\\
 & +2(\gbar-\twist \gdiff)^{2}\mathbb{K}\\
\\0 & 0 & 2(\gbar+\twist \gdiff)^{2} & -2(\gbar^{2}-\gdiff^{2})\\
\\0 & 0 & -2(\gbar^{2}-\gdiff^{2}) & 2(\gbar-\twist \gdiff)^{2}\end{array}\right)\, ,
\]
where $\twist$ is  the twist operator (\ref{tau}), and
we have defined $g_\pm \equiv  (g \pm \check g)/2$. 
The Hamiltonian is invariant under the parity operation\footnote{We are indebted to Pedro Liendo for this observation,
which corrects the contrary claim made in v1  of  the arXiv submission of this paper.}
\be \label{parityrules}
Z^a_{\; b} \to - Z^b_{\;a}  \, \qquad  {\cal X}^a_{\; b} \to -  {\cal X}^b_{\; a} \,.
\ee
where here $a, b = 1, ... 2 N_c$. As it is an exact invariance of the Lagrangian,
this parity symmetry is expected to persists to all loops.

\subsection{\label{orbifold-excite}Magnons in the interpolating spin chain}

The spin chain of the interpolating SCFT admits two degenerate
chiral vacua with $\Delta + r = 0$, namely Tr $\,\phi^\ell$ and  Tr $\,\check \phi^\ell$.
The elementary  impurities are $Q$ and $\bar Q$,
which have $\Delta + r = 1$. In the infinite chain it makes sense to consider states with a single impurity. A single $Q$ impurity separates the $\phi$ vacuum
to its left from the $\check \phi$ vacuum on its right; viceversa for a $\bar Q$ impurity.

The action of the Hamiltonian on a single $Q$ impurity in position space is \[
g^2 HQ_{\II \IIh}(x)=2(g^{2}+ \check{g}^{2})Q_{\II \IIh}(x)-2g \check{g}[Q_{\II \IIh}(x-1)+Q_{\II \IIh}(x+1)] \]
 Fourier transforming as $Q(p)= \sum_{x}e^{ipx}Q(x)$ we have 
 \begin{eqnarray}
g^2 HQ_{\II \IIh}(p) & = & 2(g^{2}+ \check{g}^{2}-2g \check{g} \cos p)Q_{\II \IIh}(p) \nonumber  \\
 & = & [2(g- \check{g})^{2}+4g \check{g}(1- \cos p)]Q_{\II \IIh}(p) \nonumber  \\
 & = & [2(g- \check{g})^{2}+8g \check{g} \sin^{2}( \frac{p}{2})]Q_{\II \IIh}(p) \end{eqnarray}
Hence the dispersion relation for $Q_{\II \IIh}(p)$ is, \[
 E(p;\kappa)= 2 (1-\kappa)^2 +8 \kappa\left(  \sin^{2} \frac{p}{2}\right) \,.
 \]
The magnon is gapless at the orbifold point $\kappa =1$, and it develops a gap as we move
towards SCQCD. Precisely at the SCQCD point, the single impurity state ceases to be meaningful and its
dispersion relation trivializes.
An identical analysis holds for the $\bar{Q}$
impurity, leading to the same dispersion relation.

\section{\label{sec:BPS-Spectrum}Protected Spectrum}

In this section we put to use the one-loop Hamiltonian 
to study the protected spectrum of ${\cal N}=2$ SCQCD and of the interpolating SCFT. 
The results presented here were quoted without proof and used in our previous paper 
 \cite{Gadde:2009dj}. The remainder of the present paper is independent of this section, and
  readers mainly interested in dynamics and integrability of the spin chain may proceed directly to section 5.
 
We are going to determine all the generalized single-trace operators in the scalar sector of SCQCD having vanishing one-loop anomalous dimension. 
 We find the complete list of such operators to be:\footnote{As explained in \cite{Gadde:2009dj}, ${\cal N}=2$ SCQCD has
  a second class of protected operators, which are outside the scalar sector.}
 \be
\Tr\, \phi^{k+2},\qquad \Tr [T\phi^k],\qquad \Tr\MM_{\bf 3}.
\label{list}
\ee
Here, $T\equiv \f \fbar-\MM_{\bf 1}$ and $k\geq0$. 
We are first led to (\ref{list})  by an educated guess. In section \ref{rep-theory} we list all
operators in the scalar sector that obey any of the
the shortening or semi-shortening conditions of the ${\cal N} = 2$
superconformal algebra, which have been completely classified \cite{Dobrev:1985vh, Dobrev:1985qz, Dobrev:1985qv, Dolan:2002zh, Dobrev:2004tk}.
Using the spin-chain Hamiltonian,
we compute the one-loop anomalous dimension of these candidate protected
states, and find that only (\ref{list}) have zero anomalous dimension. Even though here  we only perform a one-loop analysis, the operators (\ref{list}) can be seen to be protected at full quantum level using the superconformal index \cite{Gadde:2009dj}.

In section \ref{orbifoldprotected}, we list the protected operators of the orbifold theory (they can be exhaustively enumerated by a variety of methods \cite{Gadde:2009dj}) and follow their 
evolution along the exactly marginal line $\kappa$.

\subsection{\label{rep-theory}Protected spectrum in ${\cal N} = 2$ SCQCD}

A generic long multiplet $\AA_{R,r(j,\bar{j})}^{\Delta}$ of the $\NN=2$
superconformal algebra is generated by the action of the $8$ Poincar\'e supercharges
$\QQ$ and $\bar{\QQ}$ on a superconformal primary, which by definition is
 annihilated by all  conformal supercharges $\SS$. If  some combination of
the  $\QQ$'s  also annihilates the primary, the corresponding multiplet
is shorter and the conformal dimensions of all its members are protected against quantum corrections.
We follow the conventions of \cite{Dolan:2002zh} for the possible shortening 
conditions for the $\NN=2$ superconformal algebra, see table \ref{shortening}.

\begin{table}
\begin{centering}
\begin{tabular}{|c|l|l|l|l|}
\hline 
\multicolumn{4}{|c|}{Shortening Conditions} & Multiplet\tabularnewline
\hline
\hline 
$\BB_{1}$  & $\QQ_{\alpha}^{1}|R,r\rangle^{h.w.}=0$  & $j=0$ & $\Delta=2R+r$  & $\BB_{R,r(0,\bar{j})}$\tabularnewline
\hline 
$\bar{\BB}_{2}$  & $\bar{\QQ}_{2 \dot{\alpha}}|R,r\rangle^{h.w.}=0$  & $\bar j=0$ & $\Delta=2R-r$  & $\bar{\BB}_{R,r(j,0)}$\tabularnewline

\hline 
$\EE$  & $\BB_{1}\cap\BB_{2}$  & $R=0$  & $\Delta=r$  & $\EE_{r(0,\bar{j})}$\tabularnewline
\hline 
$\bar \EE$  & $\bar \BB_{1}\cap \bar \BB_{2}$  & $R=0$  & $\Delta=-r$  & $\bar \EE_{r(j,0)}$\tabularnewline
\hline 
$\hat{\BB}$  & $\BB_{1}\cap\bar{B}_{2}$  & $r=0$, $j,\bar{j}=0$  & $\Delta=2R$  & $\hat{\BB}_{R}$\tabularnewline
\hline
\hline 
$\CC_{1}$  & $\e^{\alpha\beta}\QQ_{\beta}^{1}|R,r\rangle_{\alpha}^{h.w.}=0$  &  & $\Delta=2+2j+2R+r$  & $\CC_{R,r(j,\bar{j})}$\tabularnewline
 & $(\QQ^{1})^{2}|R,r\rangle^{h.w.}=0$ for $j=0$  &  & $\Delta=2+2R+r$  & $\CC_{R,r(0,\bar{j})}$\tabularnewline
\hline 
$\bar \CC_{2}$  & $\e^{\dot\alpha\dot\beta}\bar\QQ_{2\dot\beta}|R,r\rangle_{\dot\alpha}^{h.w.}=0$  &  & $\Delta=2+2\bar j+2R-r$  & $\bar\CC_{R,r(j,\bar{j})}$\tabularnewline
 & $(\bar\QQ_{2})^{2}|R,r\rangle^{h.w.}=0$ for $\bar j=0$  &  & $\Delta=2+2R-r$  & $\bar\CC_{R,r(j,0)}$\tabularnewline
\hline 
$\mathcal{F}$  & $\CC_{1}\cap\CC_{2}$  & $R=0$  & $\Delta=2+2j+r$  & $\CC_{0,r(j,\bar{j})}$\tabularnewline
\hline 
$\bar{\mathcal{F}}$  & $\bar\CC_{1}\cap\bar\CC_{2}$  & $R=0$  & $\Delta=2+2\bar j-r$  & $\bar\CC_{0,r(j,\bar{j})}$\tabularnewline
\hline 

$\hat{\CC}$  & $\CC_{1}\cap\bar{\CC}_{2}$  & $r=\bar{j}-j$  & $\Delta=2+2R+j+\bar{j}$  & $\hat{\CC}_{R(j,\bar{j})}$\tabularnewline
\hline 
$\hat{\mathcal{F}}$  & $\CC_{1}\cap\CC_{2}\cap\bar{\CC}_{1}\cap\bar{\CC}_{2}$  & $R=0, r=\bar{j}-j$ & $\Delta=2+j+\bar{j}$  & $\hat{\CC}_{0(j,\bar{j})}$\tabularnewline
\hline
\hline 
$\DD$  & $\BB_{1}\cap\bar{\CC_{2}}$  & $r=\bar{j}+1$  & $\Delta=1+2R+\bar{j}$  & $\DD_{R(0,\bar{j})}$\tabularnewline
\hline 
$\bar\DD$  & $\bar\BB_{2}\cap{\CC_{1}}$  & $-r=j+1$  & $\Delta=1+2R+j$  & $\bar\DD_{R(j,0)}$\tabularnewline
\hline 
$\mathcal{G}$  & $\EE\cap\bar{\CC_{2}}$  & $r=\bar{j}+1,R=0$  & $\Delta=r=1+\bar{j}$  & $\DD_{0(0,\bar{j})}$\tabularnewline
\hline
$\bar{\mathcal{G}}$  & $\bar\EE\cap{\CC_{1}}$  & $-r=j+1,R=0$  & $\Delta=-r=1+j$  & $\bar\DD_{0(j,0)}$\tabularnewline
\hline
\end{tabular}
\par\end{centering}

\caption{\label{shortening}Shortening conditions
and short multiplets for the  $\NN=2$ superconformal algebra.}

\end{table}

In table \ref{SCQCD-protected} we list all the generalized single-trace
operators of ${\cal N} = 2$ SCQCD made out of scalar fields, 
which obey any of the possible shortening conditions. 
Using the spin-chain Hamiltonian of section \ref{SCQCDHamiltonian}, we find that the only operators
with zero anomalous dimension are the one listed in (\ref{list})\footnote{Together of course with their conjugates. 
Note that since in our conventions $\phi$ has $r = -1$,  the multiplet $\bar \EE_{-\ell(0,0)}$, $\ell > 0$, is represented
by ${\rm Tr} \phi^\ell$. The conjugate multiplet   $ {\cal E}_{\ell(0,0)}$ is represented by   ${\rm Tr} \bar \phi^\ell$ and
 is of course also protected.}.
 The operators ${\rm Tr}  \, \phi^\ell$ correspond to the vacuum of the spin chain, while
the operators  ${\rm Tr} \, T \,  \phi^\ell$ correspond to the zero-momentum limit of the gapless excitation $T(p)$, eq.  (\ref{T(p)}) .
There is one more protected operator, which is ``exceptional''  in not belonging to an infinite sequence:
$\mbox{Tr}\, \MM_{{\bf 3}}$. Its anomalous dimension is zero for gauge group
$SU(N_{c})$ but not for gauge group $U(N_c)$: 
 the double-trace  terms  in the Lagrangian 
 that arise from the removal of the $U(1)$
are crucial for the protection of this operator (see footnote at page \pageref{M3foot}).

\begin{table}
\begin{centering}
\begin{tabular}{|l|l|l|}
\hline 
Scalar Multiplets  & SCQCD operators  & Protected\tabularnewline
\hline
\hline 
$\bar \BB_{R,-\ell(0,0)}$  & Tr$[\f^{\ell}\MM_{3}^{R}]$  & \tabularnewline
\hline 
$\bar\EE_{-\ell(0,0)}$  & Tr$[\f^{\ell}]$  & $\checked$\tabularnewline
\hline 
$\hat{\BB}_{R}$  & Tr$[\MM_{3}^{R}]$  & $\checked$ for $R=1$\tabularnewline
\hline 
$\bar\CC_{R,-\ell(0,0)}$  & Tr$[T\MM_{3}^{R}\f^{\ell}]$  & \tabularnewline
\hline 
$\bar\CC_{0,-\ell(0,0)}$  & Tr$[T\f^{\ell}]$  & $\checked$\tabularnewline
\hline 
$\hat{\CC}_{R(0,0)}$  & Tr$[T\MM_{3}^{R}]$  & \tabularnewline
\hline 
$\hat{\CC}_{0(0,0)}$  & Tr$[T]$  & $\checked$\tabularnewline
\hline 
$\bar\DD_{R(0,0)}$  & Tr$[\MM_{3}^{R}\f]$  & \tabularnewline
\hline
\end{tabular}
\par\end{centering}

\caption{\label{SCQCD-protected}$\NN=2$ SCQCD protected operators at
one loop}

\end{table}

\subsection{\label{orbifoldprotected}Protected spectrum in the orbifold theory}

As we have reviewed in section 2.2,  ${\cal N} = 2$ SCQCD can be obtained as the $\check g_{YM} \to 0$ limit
of a family of ${\cal N} = 2$ superconformal field theories, which reduces for  $g_{YM} = \check g_{YM}$ to 
  the ${\cal N}=2$  $\mathbb{Z}_2$ orbifold of ${\cal N} = 4$ SYM. 
 In this section we find the protected spectrum of single-trace operators of the interpolating family.
We start at the orbifold point, where the protected states are easy to determine, and  follow  their fate along the exactly marginal line towards ${\cal N} = 2$ SCQCD.

At  the orbifold point, operators fall into two classes: untwisted and twisted.  In the untwisted sector, the protected states 
are well-known, since they are inherited from ${\cal N} = 4$ SYM. 
The protected operators in the twisted sector are chiral with respect to $\NN=1$ subalgebra and could be obtained by analyzing the chiral ring \cite{Gukov:1998kk}.
\footnote{We confirm the spectrum in \cite{Gadde:2009dj} up to one operator that was missed in the analysis of \cite{Gukov:1998kk}.}
Both the classes of operators can be rigorously checked to be protected by computing the superconformal index.\footnote{The calculation
for the orbifold was carried out already in \cite{Nakayama:2005mf}, and confirmed in \cite{Gadde:2009dj} up to a minor emendation.}
Using the index one can also argue
that the protected multiplets found at the orbifold point {\it cannot} recombine into long multiplets
as we vary $\check g$ \cite{Gadde:2009dj}, so in particular  taking $\check g \to 0$ they must evolve into the protected multiplets of
the theory
\be \label{decoupledtheory}
\{ \NN = 2 \; {\rm SCQCD} \; \oplus \; {\rm decoupled} \; SU(N_{\hat c}) \; {\rm vector \; multiplet} \} \, .
\ee
In section \ref{evolution} we follow this evolution in detail. We find that the $SU(2)_L$-singlet protected states
of the interpolating theory evolve into the list (\ref{list}) of protected states of SCQCD, plus some extra states made purely from the decoupled
vector multiplet.  On the other hand, the interpolating theory has also many single-trace protected states with non-trivial $SU(2)_L$ spin,
which are of course absent from the list (\ref{list}): we see that
 in the limit $\check g \to 0$, a  state with $SU(2)_L$ spin $L$ can be  interpreted as a ``multiparticle state'',
 obtained by linking together  $L$  short ``open''  spin chains of SCQCD and decoupled fields $\check \phi$.
 By this route we confirm  that (\ref{list}) is the correct and complete
list of protected single-traces in the scalar sector for $\NN =2$ SCQCD. The results are also suggestive of
a dual string theory interpretation: as $\check g \to 0$,  single closed string states carrying $SU(2)_L$ quantum numbers disintegrate into multiple open strings. 
The above argument, however, doesn't imply that all the protected operators of SCQCD are obtained as degenerations of protected operators of the interpolating theory. Indeed, they aren't. In \cite{Gadde:2009dj}, we discuss an alternative mechanism that brings about more protected SCQCD operators from the decomposition of long multiplets of the interpolating theory as $\check g \to 0$.

In summary, the degeneracy of protected states is independent of the exactly marginal deformation that changes $\check g_{YM}$
and is thus the same for the orbifold  theory 
and for the theory (\ref{decoupledtheory}). 
 At $\check g_{YM} = 0$ there is a symmetry enhancement, $SU(2)_L \times SU(N_{\check c}) \to U(N_f = 2 N_c)$,
 and  we can  consistently truncate  the spectrum of generalized single trace operators
  to singlets of the flavor group $U(N_f)$ -- which in particular do not contain any of the decoupled states $\check \phi$.
  This is the flavor singlet spectrum of ${\cal N} = 2$ SCQCD that we have analyzed in the previous section.

\begin{table}
\begin{centering}
\begin{tabular}{|l|l|}
\hline 
Multiplet  & Orbifold operator ($R,\ell\geq0,\,n\geq2$) \tabularnewline
\hline
\hline 
$\hat{\BB}_{R+1}$  & $\mbox{Tr}[(Q^{+\hat{+}}\bar{Q}^{+\hat{+}})^{R+1}]$ \tabularnewline
\hline 
$\bar \EE_{-(\ell+2)(0,0)}$  & $\mbox{Tr}[\f^{\ell+2}+\fh^{\ell+2}]$  \tabularnewline
\hline 
$\hat{\CC}_{R(0,0)}$  & $\mbox{Tr}[\sum \Torb (Q^{+\hat{+}}\bar{Q}^{+\hat{+}})^{R}]$  \tabularnewline
\hline 
$\bar \DD_{R+1(0,0)}$  & $\mbox{Tr}[\sum(Q^{+\hat{+}}\bar{Q}^{+\hat{+}})^{R+1}(\f^+\fh)]$  \tabularnewline
\hline 
$\bar \BB_{R+1,-(\ell+2)(0,0)}$  & $\mbox{Tr}[\sum_{i}(Q^{+\hat{+}}\bar{Q}^{+\hat{+}})^{R+1}\f^{i}\fh^{\ell+2-i}]$ \tabularnewline
\hline 
$\bar \CC_{R,-(\ell+1)(0,0)}$  & $\mbox{Tr}[\sum_{i} \Torb(Q^{+\hat{+}}\bar{Q}^{+\hat{+}})^{R}\f^{i}\fh^{\ell+1-i}]$  \tabularnewline
\hline
\hline $\AA_{R,-\ell(0,0)}^{\Delta=2R+\ell+2n}$ & $\mbox{Tr}[\sum_{i} \Torb^n(Q^{+\hat{+}}\bar{Q}^{+\hat{+}})^{R}\f^{i}\fh^{\ell-i}]$ \tabularnewline
\hline
\end{tabular}
\par\end{centering}
\caption{\label{orbifold-protected} Superconformal primary
operators in the untwisted sector of the orbifold theory that descend from the $\frac{1}{2}$ BPS primary of ${\cal N} =4$.
 The symbol $\sum$ indicates  summation over all ``symmetric traceless''
permutations of the component fields allowed by the index structure.}
\end{table}

\begin{table}
\begin{centering}
\begin{tabular}{|l|l|}
\hline 
Multiplet  & Orbifold operator $(\ell\geq0)$ \tabularnewline
\hline
\hline 
$\hat{\BB}_{1}$  & $\mbox{Tr}[(Q^{+\hat{+}}\bar{Q}^{+\hat{-}}-Q^{+\hat{-}}\bar{Q}^{+\hat{+}})]=\Tr\,\MM_{\bf 3}$ \tabularnewline
\hline 
$\bar \EE_{-(\ell+2)(0,0)}$  & $\mbox{Tr}[\f^{\ell+2}-\fh^{\ell+2}]$  \tabularnewline
\hline 
\end{tabular}
\par\end{centering}
\caption{\label{twistedtable} Superconformal primary
operators in the twisted section of the orbifold theory.}
\end{table}

\bigskip

\subsection{\label{evolution}Away from the orbifold point: matching with ${\cal N} = 2$ SCQCD}

In the limit $\check g \to 0$, we must be able to match the protected states of the interpolating family
with  protected states of $\{ {\cal N} = 2$ SCQCD $\oplus$ decoupled vector multiplet$\}$.
 For the purpose of this discussion, the protected states naturally splits into two sets:
$SU(2)_L$ singlets and $SU(2)_L$ non-singlets. It is clear that all the
(generalized) single-trace operators of ${\cal N} = 2$ SCQCD must arise from the
$SU(2)_L$ singlets.

\newpage

\subsubsection{$\mathbf{SU(2)_L}$ singlets}

They are: 
\begin{enumerate}
\item[(i)] One  $\hat {\cal B}_1$ multiplet, corresponding to the primary $ \Tr  [Q_{   \IIh \{ \II  }\bar{Q}_{\JJ  \} }^{ \hat{\II}}] = \Tr \,{\cal M}_{\bf 3}$. Since this is the only
operator with these quantum numbers, it cannot mix with anything and its form is independent of $\check g$.
\item[(ii)] Two $\bar \EE_{-\ell (0, 0)}$ multiplets for each $\ell \geq 2$,  corresponding
to  the primaries $\Tr\, [ \phi^\ell \pm {\check \phi}^\ell ]$.

  For each $\ell$, there is a two-dimensional space
of protected operators, and we may choose whichever basis is more convenient.
For $g = \check g$, the natural basis vectors are the untwisted and twisted  combinations
(respectively even and odd under $\phi \leftrightarrow \check \phi$), while for $\check g = 0$ the natural basis vectors
are $\Tr\, \phi^\ell$ (which is an operator of ${\cal N} =2$ SCQCD) and $\Tr\,  {\check \phi}^\ell$
(which belongs to the decoupled sector).
\item[(iii)]  One $\hat {\cal C}_{0 (0,0)}$ multiplet (the stress-tensor multiplet), corresponding to the primary $\Tr \, \Torb = \Tr \, [T +  \check \phi \bar {\check \phi} ]$.
We have checked that this combination is an eigenstate with zero eigenvalue for all $\check g$.

For $\check g = 0$,  we may trivially subtract out the decoupled piece $\Tr  \, \check \phi \bar {\check \phi}$ and recover
 $\Tr \, T$, the stress-tensor multiplet of ${\cal N} = 2$ SCQCD.
 \item[(iv)] One $\bar{\cal C}_{0, -\ell (0,0)}$ multiplet for each $\ell \geq 1$. In the limit $\check g \to 0$, we expect this multiplet to evolve to the  $\Tr\, T \phi^\ell$
multiplet of ${\cal N} = 2$ SCQCD. Let us check this in detail. 

The primary of $\bar {\cal C}_{0, -\ell (0,0)}$ has  $R =0$, $r = -\ell$ and $\Delta = \ell +2$.
 The space of operators which classically have these quantum numbers is spanned by
 \[
| a \rangle=\mbox{Tr}[{\fh}^{\ell+1}\bar{\topp{\f}}],\quad|b _{i}\rangle\equiv\frac{1}{2}\mbox{Tr}[{\f}^{i}Q_{\II\IIh}{\fh}^{\ell-i}\bar{Q}^{\IIh\II}]\quad\mbox{for}\quad0\leq i\leq \ell \quad\mbox{and}\quad|c_{\ell}\rangle\equiv\mbox{Tr}[{\f}^{\ell+1}\bar\f]\]
Diagonalizing the Hamiltonian in Fourier space, we find the protected
operator to be
\[
|\bar\CC_{0,-\ell(0,0)}\rangle_{\kappa} =  \kappa^{\ell}| a \rangle-\sum_{i=0}^{\ell}\kappa^{\ell-i}|b_{i}\rangle+|c_{\ell} \rangle \,  
\]
where $\kappa \equiv \check g/g$.
 In
the limit $\kappa\rightarrow0$,
\[
|\bar\CC_{0, -\ell(0,0)}\rangle_{\kappa\rightarrow0} =\mbox{Tr}[(\f\bar{\f}-\frac{1}{2}Q_{\II\IIh}\bar{Q}^{\II\IIh}){\f}^{\ell}]  = {\rm Tr} [T\f^{\ell}] \, ,
\]
as claimed.
\end{enumerate}
All in all, we see that this list reproduces the list (\ref{list}) of one-loop protected scalar operators of ${\cal N} = 2$ SCQCD, {\it plus} the extra states $\Tr \check \phi^\ell$ which decouple for $\check g = 0$.
This concludes the argument that that the operators (\ref{list}) are protected at the full quantum level,
and that they are the {\it complete} set of  protected 
generalized single-trace primaries of ${\cal N} = 2$ SCQCD.

\subsubsection{$\mathbf{SU(2)_L}$ non-singlets}

 The basic protected primary of ${\cal N} = 2$ {SCQCD}
 which
   is charged under
 $SU(2)_L$    is the $SU(2)_L$ triplet contained in the mesonic operator ${\cal O}^i_{{\bf 3_R } \, j} = (\bar Q^i_a Q^a_j)_{\bf3_R}$.
Indeed writing the $U(N_f = 2 N_c)$ flavor 
indices $i$ as $i = (\check a, \hat \II)$, with $\check a =1, \dots N_f/2 = N_c$ ``half'' flavor indices and $\II = \hat \pm$ $SU(2)_L$ indices,
we can decompose
\be
{\cal O}^i_{{\bf 3_R } \, j}  \rightarrow {\cal O}^{\check a}_{{\bf 3_R  3_L} \, \check b}  \, , \quad {\cal O}^{\check a}_{{\bf 3_R  1_L} \, \check b}  \, .
\ee
In particular we may consider the highest weight combination for both $SU(2)_L$ and $SU(2)_R$,
\be
(\bar{Q}^{+\hat{+}}Q^{+\hat{+}})^{\check a}_{\; \check b} \, .
\ee
States with higher $SU(2)_L$ spin can be built by taking products of   ${\cal O}_{{\bf 3_R  3_L} }$ with $SU(2)_L$ and $SU(2)_R$
indices separately symmetrized  -- and  this is the only way
to obtain protected states of ${\cal N} =2$ SCQCD charged under $SU(2)_L$ which have finite conformal dimension in the Veneziano limit.
 It  is  then a priori clear that a protected primary of the interpolating theory with $SU(2)_L$ spin $L$
  must evolve as $\check g \to 0$
  into a   {product} of $L$ copies of $(\bar{Q}^{+\hat{+}}Q^{+\hat{+}})$ and of as many additional decoupled scalars $\check \phi$ and $\bar {\check \phi}$
as needed to make up for the correct $U(1)_r$ charge and conformal dimension.
It is amusing to follow in more detail this evolution for the various multiplets:
\begin{enumerate}
\item[(i)]  $\hat {\BB}_{R}$ multiplet. 

This is a trivial case, since for each $R$ there is only one operator
with the correct quantum numbers, namely
\[
|\hat{\BB}_{R}\rangle_{\kappa}\equiv\mbox{Tr}[(Q^{+\hat{+}}\bar{Q}^{+\hat{+}})^{R}] \, ,
\]
for all $g$ and $\check g$. 
We have checked that it is indeed an eigenstate of zero eigenvalue for all couplings.

\item[(ii)]   $\bar\DD_{R(0,0)}$ multiplet.

The primary of $\bar\DD_{R(0,0)}$ has $SU(2)_R$ spin equal $R$, $U(1)_r$ charge $r=-1$ and 
 $\Delta=2R+1$. 
The space of operators which classically have 
these quantum numbers is two-dimensional, spanned by Tr$[(Q^{+\hat{+}}\bar{Q}^{+\hat{+}})^{R}{\f}]$
and Tr$[(\bar{Q}^{+\hat{+}}Q^{+\hat{+}})^{R}{\topp{\f}}]$. The spin chain
Hamiltonian  in this subspace reads
\[
g^{2}H_{\bar\DD}=\left(\begin{array}{cc}
4g^{2} & -4g\topp g\\
-4g\topp g & 4\topp g^{2}\end{array}\right)\]
The protected operator (eigenvector with zero eigenvalue) is 
\[\label{schematic sum}
|\bar\DD_{R(0,0)}\rangle_{\kappa}\equiv\mbox{Tr}[\kappa(Q^{+\hat{+}}\bar{Q}^{+\hat{+}})^{R}{\f}+(\bar{Q}^{+\hat{+}}Q^{+\hat{+}})^{R}{\topp{\f}}] \,.
\] 
For $\kappa = 0$, the protected operator
 is interpreted as a ``multi-particle state'' of $R$ open chains of SCQCD and one decoupled scalar $\check \phi$.
For example for $R=2$,
the operator will be broken
into the following gauge-invariant pieces, 
\[
(\bar{Q}^{+\hat{+}}Q^{+\hat{+}})_{\:\:\topp b}^{\topp a}\:,\quad(\bar{Q}^{+\hat{+}}Q^{+\hat{+}})_{\:\:\topp c}^{\topp b}\quad\mbox{and}\quad\topp{\f}_{\:\:\topp a}^{\topp c} \,.
\]
In the limit $\check  g \to 0$, the ``closed chain'' of the interpolating theory effectively breaks into ``open chains'' of  $\{$${\cal N} = 2$ SCQCD $\oplus$ decoupled multiplet$\}$,
with the rupture points at the contractions of the ``half-flavor'' indices $\check a$, $\check b$, $\check c$.

\item[(iii)]  
$\bar \BB_{R,r(0,0)}$ multiplet.

Finding the protected multiplet for arbitrary coupling amounts to diagonalizing
the spin-chain Hamiltonian of the interpolating theory
in the space of operators with quantum numbers $R$, $r$ and $\Delta=2R-r$.
The dimension of this space increases rapidly with $R$ and $r$. Let us focus on
two simple cases.

\emph{case $1$: $R=1$, $r\equiv-\ell<0$} 

In this case,  the space is $\ell+1$ dimensional, spanned by
\[
|\psi_{i}\rangle\equiv\mbox{Tr}[{\f}^{i}Q^{+\hat{+}}{\fh}^{\ell-i}\bar{Q}^{+\hat{+}}] \, , \quad i =0, \dots \ell \,.
\]
The protected operator is found to be
\begin{equation}
|\bar\BB_{1,-\ell(0,0)}\rangle_{\kappa}\equiv\sum_{i=0}^{\ell}\kappa^{i}|\psi_{i}\rangle\label{eq:B multiplet}\end{equation}
In our schematic notation of $\sum$, introduced earlier, the same  operator would read
\[
|\bar\BB_{1,-\ell(0,0)}\rangle_{\kappa}=\mbox{Tr}[\sum_{i}\kappa^{i}(Q^{+\hat{+}}\bar{Q}^{+\hat{+}}){\f}^{i}{\fh}^{\ell-i}] \, .
\] 
Note that at $\kappa=0$, the $U(1)_r$ charge of the operator  is all carried by the decoupled scalars ${\topp{\f}}$ -- there are
no ${\f}$. This is  again consistent with the picture of the closed chain disintegrating
into open pieces. 

\emph{case 2: $r=-2$, $R=2$}

The relevant vector space is spanned by the operators
\[
\begin{array}{cccccc}
|0\rangle & = & \mbox{Tr}[\f\f Q^{+\hat{+}}\bar{Q}^{+\hat{+}}Q^{+\hat{+}}\bar{Q}^{+\hat{+}}] & \quad\quad|\topp 0\rangle & = & \mbox{Tr}[Q^{+\hat{+}}{\topp{\f}}{\topp{\f}}\bar{Q}^{+\hat{+}}Q^{+\hat{+}}\bar{Q}^{+\hat{+}}]\\
|1\rangle & = & \mbox{Tr}[\f Q^{+\hat{+}}{\topp{\f}}\bar{Q}^{+\hat{+}}Q^{+\hat{+}}\bar{Q}^{+\hat{+}}] & \quad\quad|\topp 1\rangle & = & \mbox{Tr}[Q^{+\hat{+}}{\topp{\f}}\bar{Q}^{+\hat{+}}\f Q^{+\hat{+}}\bar{Q}^{+\hat{+}}]\\
|2\rangle & = & \mbox{Tr}[\f Q^{+\hat{+}}\bar{Q}^{+\hat{+}}\f Q^{+\hat{+}}\bar{Q}^{+\hat{+}}] & \quad\quad|\topp 2\rangle & = & \mbox{Tr}[Q^{+\hat{+}}{\topp{\f}}\bar{Q}^{+\hat{+}}Q^{+\hat{+}}{\topp{\f}}\bar{Q}^{+\hat{+}}]\end{array}\]
The Hamiltonian in this subspace is (the basis vectors are read in the sequence $|0 \rangle$, $|\check 0 \rangle$, $|1 \rangle$, $\dots$) 
\[
g^{2}H_{\bar\BB_{2,-2(0,0)}}=\left(\begin{array}{cccccc}
4g^{2} & 0 & -2g\topp g & -2g\topp g & 0 & 0\\
0 & 4\topp g^{2} & -2g\topp g & -2g\topp g & 0 & 0\\
-2g\topp g & -2g\topp g & 4g^{2}+4\topp g^{2} & 0 & -2g\topp g & -2g\topp g\\
-2g\topp g & -2g\topp g & 0 & 4g^{2}+4\topp g^{2} & -2g\topp g & -2g\topp g\\
0 & 0 & -2g\topp g & -2g\topp g & 4g^{2} & 0\\
0 & 0 & -2g\topp g & -2g\topp g & 0 & 4\topp g^{2}\end{array}\right)\]
There is an eigenvector with zero eigenvalue for all $\kappa$, namely
\begin{eqnarray*}
|\bar\BB_{2,-2(0,0)}\rangle_{\kappa} & \equiv & \kappa^{2}|0\rangle+|\topp 0\rangle+\kappa|1\rangle+\kappa|\topp 1\rangle+\kappa^{2}|2\rangle+|\topp 2\rangle\\
 & = & \mbox{Tr}[\sum_{i}\kappa^{i}(Q^{+\hat{+}}\bar{Q}^{+\hat{+}})^{2}{\f}^{i}{\fh}^{2-i}]
 \end{eqnarray*}
As expected, for $\kappa = 0$ the operator contains  ${\topp{\f}}$ and no $\phi$.

Extrapolating  from these cases, we make an educated guess for the form for general protected operator,
\[
|\bar\BB_{R,-\ell(0,0)}\rangle_{\kappa}=\mbox{Tr}[\sum_{i}\kappa^{i}(Q^{+\hat{+}}\bar{Q}^{+\hat{+}})^{R}{\f}^{i}{\fh}^{\ell-i}] \,.
\] 
In the limit $\kappa\rightarrow 0$, this operator breaks into $R$ mesons $(\bar Q Q)^{\check a} _{\; \check b}$
of ${\cal N} = 2$ SCQCD and $\ell$ decoupled scalars ${\check \phi}^{\check a}_{\; \check b}$.

\item[(iv)]$\hat \CC_{R(0,0)}$  and $\bar\CC_{R,-\ell(0,0)}$ multiplets.

We have not studied these cases in detail since they are technically quite involved. It is clear
however that for $\check g \to 0$ the protected primaries must evolve into states of the schematic
form
\be \label{schematicjunk}
\Tr \left[ \, {\cal O}_{\bf 3_R 3_L }^R  {\check \phi}^{\ell + n} \bar{\check \phi}^n \right] \, ,
\ee
with $\ell=0$, $n=1$ for $\hat \CC_{R(0,0)}$ and $n=1$ for $\bar\CC_{R,-\ell(0,0)}$.
\end{enumerate}

 \section{Two-body scattering}

In this section we study the scattering of  two magnons in the spin chain for the  interpolating SCFT.
We take the chain to be infinite. Because of the index structure of the impurities, one of the asymptotic magnons must be a
 $Q$  and the other a $\bar Q$, and their ordering is fixed --
 we can have a $Q$ impurity always to the left
 of a $\bar Q$ impurity, or viceversa. The scattering is thus pure reflection.
 For the case of $Q$  to the {\it left} of $ \bar{Q}$, and suppressing momentarily the  $SU(2)_L \times SU(2)_R$ quantum numbers,
   the asymptotic form of the eigenstates of the Hamiltonian is
\[
 \sum_{x_1  \ll x_2}\left(  e^{i p_1 x_1 + i p_2 x_2} + S(p_2,p_1)  e^{i p_2 x_1 + i p_1 x_2}\right)
 | \ldots \f Q(x_{1}) \fh \ldots \fh \bar{Q}(x_{2}) \f \ldots \rangle \, .\]
This is the definition  of the two-body $S$-matrix. 
In fact, thanks to the nearest-neighbor nature
of the spin chain,  if the impurities are not adjacent we are already
  in the ``asymptotic'' region, so $x_1 \ll x_2$ should be interpreted as $x_1 < x_2 -1$.
Similarly, for the case where  $Q$  to the {\it right} of $ \bar{Q}$  the asymptotic form of the two-magnon state is
   \[
 \sum_{x_1  \ll x_2}\left(  e^{i p_1 x_1 + i p_2 x_2} + \check S(p_2,p_1)  e^{i p_2 x_1 + i p_1 x_2}\right)
 | \ldots \fh \bar Q(x_{1}) \f \ldots \f {Q}(x_{2}) \fh \ldots \rangle \, ,\]
 which defines $\check S$.  The two-body S-matrices $S$ and $\check S$ are 
related by exchanging $g \leftrightarrow \check g$,
  \be
  S (p_1, p_2 ; g, \check g)= \check S (p_1, p_2 ; \check g, g) \, .
  \ee
 The total energy 
 of a two-magnon state is just the sum of the energy of the individual magnons,
  \be \label{totalE}
  E(p_1, p_2 ; \kappa)
  =\left( 2(1- \kappa)^{2}+ 8 \kappa( \sin^{2} \frac{p_{1}}{2} ) \right) +\left(  2(1- \kappa)^{2}  +8 \kappa ( \sin^{2} \frac{p_{2}}{2}) \right)\,.
  \ee
Besides the continuum of states with real momenta $p_1$ and $p_2$, there can be  bound and ``anti-bound'' states
for special complex values of the momenta. 
A bound state occurs when
\be \label{bound}
S(p_1, p_2) =\infty\,,\quad   {\rm with} \quad p_1 = \frac{P}{2} - i q \,, \quad  p_2=  \frac{P}{2} + i q \, , 
\quad q >0 \, .
\ee
Since $S(p_2, p_1)= 1/S(p_1, p_2) \to 0$,
the asymptotic wave-function is \be \label{boundwave}
e^{i P \frac{x_1 + x_2}{2} -q (x_2 - x_1)}\, ,
\ee which
 is indeed normalizable (since $x_2 > x_1$ in our  conventions). A bound state has  smaller energy
than   any state in the two-particle continuum  with the same total momentum $P$.
 An anti-bound state occurs when
\be \label{antibound}
S(p_1, p_2) =\infty\,,\quad   {\rm with} \quad p_1 = \frac{P}{2} - i q + \pi \, ,\quad  p_2=  \frac{P}{2} + i q - \pi \, , 
\quad q >0 \, .
\ee
The asymptotic wave-function is now 
\be \label{antiboundwave}
(-1)^{x_2-x_1} e^{i P \frac{x_1 + x_2}{2} -q (x_2 - x_1)}\,. 
\ee
The energy of an anti-bound state is strictly bigger  than the
two-particle continuum.
It is easy to see
that  (\ref{bound}) and (\ref{antibound}) are the only allowed possibilities for complex  $p_1$ and $p_2$ such
that the total momentum and the total energy  are real.

The analysis of two-body scattering proceeds independently in four different sectors, corresponding
the choice of the triplet or singlet combinations for $SU(2)_L$ and $SU(2)_R$.  
In each sector, we will compute the S-matrix and look for the (anti)bound states  associated to its poles.

 \subsection{$3_{L} \otimes 3_{R}$ Sector}

In the  $3_{L} \otimes3_{R}$ sector, we write the general
two-impurity state with $Q$ to the left of $\bar Q$
as
\[ \label{two-body}
| \Psi_{3 \otimes3} \rangle= \sum_{x_{1}<x_{2}} \Psi_{3 \otimes3}(x_{1},x_{2})| \ldots \f Q(x_{1}) \fh \ldots \fh \bar{Q}(x_{2}) \f \ldots \rangle _{3 \otimes 3} \,.
\]
There is no mixing with states containing $\bar \phi$ and $\bar {\check \phi}$ since they have different $SU(2)_L \times SU(2)_R \times U(1)_r$
quantum numbers. Acting with the Hamiltonian, one finds:
\begin{itemize}
\item
For $x_{2}>x_{1}+1$,
\begin{eqnarray} \label{bulkH}
g^2 H \cdot \Psi_{3 \otimes3}(x_{1},x_{2}) & = & 4({g}^{2}+ \check{g}^{2}) \Psi_{3 \otimes3}(x_{1},x_{2})-2{g} \check{g} \Psi_{3 \otimes3}(x_{1}+1,x_{2})-2{g} \check{g} \Psi_{3 \otimes3}(x_{1}-1,x_{2})
\nonumber \\
 &  &
 -2{g} \check{g} \Psi_{3 \otimes3}(x_{1},x_{2}+1)-2{g} \check{g} \Psi_{3 \otimes3}(x_{1},x_{2}-1)\,. \end{eqnarray}
\item For $x_{2}=x_{1}+1$, \[
\label{contactH}
g^2 H \cdot \Psi_{3 \otimes3}(x_{1},x_{2})=4{g}^{2} \Psi_{3 \otimes3}(x_{1},x_{2})-2{g} \check{g} \Psi_{3 \otimes3}(x_{1}-1,x_{2})-2{g} \check{g} \Psi_{3 \otimes3}(x_{1},x_{2}+1) \,.\]
\end{itemize}
The  plane wave states $e^{i(p_{1}x_{1}+p_{2}x_{2})}$ and $e^{i(p_{1}x_{2}+p_{2}x_{1})}$
are separately eigenstates for the ``bulk'' action of the Hamiltonian (\ref{bulkH}), with eigenvalue (\ref{totalE}).
The action
of the Hamiltonian on the state with adjacent impurities, equ.(\ref{contactH}), provides the boundary condition that fixes the exact
eigenstate of asymptotic momenta $p_1$, $p_2$, 
 \[
 \Psi_{3 \otimes3}(x_{1},x_{2})=e^{i(p_{1}x_{1}+p_{2}x_{2})}+S_{3 \otimes3}(p_{1},p_{2})e^{i(p_{1}x_{2}+p_{2}x_{1})} \, ,\]
where 
\[
S_{3 \otimes3}(p_{1},p_{2})=- \frac{1+e^{ip_{1}+ip_{2}}-2 \kappa e^{ip_{1}}}{1+e^{ip_{1}+ip_{2}}-2 \kappa e^{ip_{2}}}\, , \qquad \kappa \equiv \frac{\check g}{g}\,. \]
In this sector, the S-matrix coincides with the familiar S-matrix 
of the XXZ chain, with the identification $\Delta_{XXZ} = \kappa$.  
   The pole of the S-matrix,
 \[
e^{ip_{2}}= \frac{1+e^{i(p_{1}+p_{2})}}{2 \kappa}  \, ,
\]
is associated to a bound state. Writing  $p_1 = P/2 - i q$, $p_2= P/2 + i q$, we have
\be
e^{-q} = \frac{\cos(\frac{P}{2}) }{\kappa}\,.
\ee
The wave-function is normalizable provided $q >0$, which implies $2\, {\rm arccos}\, \kappa < |P| < \pi$. 
Substituting $p_1$ and $p_2$ into the expression for the total energy (\ref{totalE}), we find that the
dispersion relation of the bound state is simply
\be
\left[ Q\bar Q\right]^{bound}_{3_L \, 3_R}: \qquad E= 4 \sin^{2}( \frac{P}{2}) \, ,\quad   2 \,{\rm arccos}\, \kappa < |P| < \pi\,.
\ee
This dispersion relation is plotted as the dotted (orange) curve  in the left column of figure~\ref{DispersionPlots}.
When the total momentum $P$ is smaller than $2 \arccos \kappa$ the bound state dissolves into the two-particle continuum.
The bound state exists for the full range of $P$ at the orbifold point $\kappa =1$, but the allowed range of $P$ shrinks as $\kappa$ is decreased,
and the bound state  disappears in the SCQCD limit $\kappa \to 0$.

The S-matrix in the  $3_{L} \otimes3_{R}$ sector 
with $Q$ to the {\it right} of $\bar Q$ is obtained by switching $g \leftrightarrow \check g$,
\be
\check S_{3 \otimes3}(p_{1},p_{2}; \kappa)= S_{3 \otimes3}(p_{1},p_{2}; 1/\kappa)=- \frac{1+e^{ip_{1}+ip_{2}}-\frac{2}{ \kappa} e^{ip_{1}}}{1+e^{ip_{1}+ip_{2}}-\frac{2}{ \kappa} e^{ip_{2}}}\, .
\ee
 Now the pole of the S-matrix is associated to a bound state with 
 \be
 e^{-q} = \kappa \cos(\frac{P}{2}) \,.
 \ee
The bound state exists for all $P$ in  the whole range of $\kappa \in (0,1]$. Its dispersion relation is
\be
\left[\bar Q  Q\right]^{bound}_{3_L \, 3_R}:  \qquad E= 4 \kappa^2 \sin^{2}( \frac{P}{2})\, ,
\ee
plotted as the dotted (orange) curve in the right column of figure~\ref{DispersionPlots}.
The existence of this bound state is consistent with our analysis of the protected spectrum in section 4.

 \subsection{$1_{L} \otimes3_{R}$ Sector}

 The general two-body state with $Q$ to the left of $\bar Q$ 
is 
\[
| \Psi_{1 \otimes3} \rangle= \sum_{x_{1}<x_{2}} \Psi_{1 \otimes3}(x_{1},x_{2})| \ldots \f Q(x_{1}) \fh \ldots \fh \bar{Q}(x_{2}) \f \ldots \rangle_{1 \otimes 3}
\]
The action of the Hamiltonian for $x_{2}=x_{1}+1$ is now \[
g^2 H \cdot \Psi_{1 \otimes3}(x,x+1)=8{g}^{2} \Psi_{1 \otimes3}(x,x+1)-2{g} \check{g} \Psi_{1 \otimes3}(x-1,x+1)-2{g} \check{g} \Psi_{1 \otimes3}(x,x+2) \,.\]
Writing 
\[
 \Psi_{1 \otimes3}(x_{1},x_{2})=e^{i(p_{1}x_{1}+p_{2}x_{2})}+S_{1 \otimes3}(p_{2},p_{1})e^{i(p_{1}x_{2}+p_{2}x_{1})} \,,
 \]
we find 
 \[
S_{1 \otimes3}(p_{1},p_{2}; \kappa)=- \frac{1+e^{ip_{1}+ip_{2}}-2( \kappa- \frac{1}{\kappa})e^{ip_{1}}}{1+e^{ip_{1}+ip_{2}}-2( \kappa- \frac{1}{\kappa})e^{ip_{2}}} \, ,
 \]
 which is again the S-matrix of the XXZ chain, now with $\Delta = \kappa - \frac{1}{\kappa}$.
The S-matrix blows up for
\be
e^{ip_{2}} =   \frac{1+e^{i(p_{1}+p_{2})}}{2( \kappa- \frac{1}{\kappa})} \, .
\ee
This pole is associated to an  {\it anti}-bound state. Parametrizing  $p_1 = P/2 -iq +\pi$, $p_2 = P/2 -iq -\pi$,  
the location of the pole is given by
\be \label{M3bound}
e^{-q} = \frac{\cos  (\frac{P}{2})}{\frac{1}{\kappa}-\kappa} \,. 
\ee
Normalizability of the wave-function requires $q >0$, which occurs for a restricted range of $P$
for $\kappa_{*} < \kappa <1$, and  for the full range of $P$ for $\kappa < \kappa_{*}$, 
\begin{eqnarray} \label{twocases}
 2 \, {\rm arccos}(\frac{1}{\kappa}-\kappa) < |P| < \pi  \quad {\rm for} \;\frac{\sqrt{5}-1}{2} <\kappa <1 \, \; \\
 0 < |P| < \pi \quad {\rm for} \; 0< \kappa < \frac{\sqrt{5}-1}{2}  \,. \nonumber
\end{eqnarray}
Substituting in $E(p_1, p_2; \kappa)$ we find the dispersion relation for the anti-bound state,
\be
\left[ Q \bar Q\right]^{antibound}_{1_L \, 3_R}: \qquad E = \frac{4(2- \kappa^{2})}{1- \kappa^{2}}- \frac{4 \kappa^{2}}{1- \kappa^{2}} \sin^{2} \frac{P}{2}\, ,
\]
 which is plotted  as the solid (red) curve in the left column of figure~\ref{DispersionPlots}.
The anti-bound state is absent at the orbifold point $\kappa =1$. For $\kappa \to 0$, $q \to + \infty$, so that
 the wave-function (\ref{antiboundwave}) localizes to two neighboring sites and in fact coincides with   the dimeric excitation  ${\cal M}_{\bf 3} = (Q \bar Q)_{\bf 3} $ of $\NN=2$ SCQCD;
 in the limit we smoothly recover the ${\cal M}_{\bf 3}$ dispersion relation $E(P) = 8$.

For $\bar Q Q$ scattering, we have
\be
\check S_{1 \otimes3}(p_{1},p_{2}; \kappa) = S_{1 \otimes3}(p_{1},p_{2}; 1/\kappa)= - \frac{1+e^{ip_{1}+ip_{2}}-2( \frac{1}{\kappa}- \kappa)e^{ip_{1}}}{1+e^{ip_{1}+ip_{2}}-2( \frac{1}{\kappa}- \kappa)
e^{ip_{2}}} \, .
\ee
Now the pole corresponds to a bound state, indeed it occurs for $p_1= P/2 -iq$, $p_2= P/2+ iq$ with $q$ and $P$ related as in (\ref{M3bound}). Clearly the allowed range of $P$
 is as in (\ref{twocases}).
We find the dispersion relation 
\be
\left[ Q \bar Q\right]^{bound}_{1_L \, 3_R}: \qquad E  = \frac{4\kappa^{2}}{(1-\kappa^{2})}(1-2\kappa^{2}+\sin^{2}\frac{P}{2})\, ,
\ee
which is plotted as the solid (red) curve in the right column of figure~\ref{DispersionPlots}.

 \subsection{$3_{L} \otimes1_{R}$ Sector}

The scattering problem in the $3_{L} \otimes1_{R}$ sector is solved by the same technique. We find 
\[
S_{3 \otimes1}(p_{1},p_{2})=\check S_{3 \otimes1}(p_{1},p_{2})=-1\, , \]
which coincides with the scattering matrix of  free
fermions, or with the $\Delta_{XXZ} \to \infty$ limit of the S-matrix for the XXZ chain.  Clearly there are no (anti)bound states.

 \subsection{$1_{L} \otimes1_{R}$ Sector}

The analysis for the $1_{L} \otimes1_{R}$
sector is slightly more involved because a two-impurity state 
is not closed under the action of Hamiltonian: when $Q$ and $\bar Q$ are
next to each other they can transform into $\phi \bar \phi$.
 The general state for $Q \bar Q$ scattering in the singlet sector
 is \begin{eqnarray}
| \Psi_{1 \otimes1} \rangle & = &  \sum_{x_{1}<x_{2}} \Psi_{1 \otimes1}(x_{1},x_{2})| \ldots \f Q(x_{1}) \fh \ldots \fh \bar{Q}(x_{2}) \f \ldots \rangle_{1 \otimes 1} \\
 &  & + \sum_{x} \Psi_{\fbar}(x)| \ldots \f \fbar(x) \f \ldots \rangle \,. \nonumber
 \end{eqnarray}
 The first term is an eigenstate for ``bulk'' action of the Hamiltonian ($x_2 > x_1 +1$)  with  the usual eigenvalue $E(p_1, p_2 ; \kappa)$ of equ.(\ref{totalE}). The
action of the Hamiltonian for $x_2 = x_1 +1$ is
\begin{eqnarray}
g^2 H \cdot \Psi_{1 \otimes1}(x,x +1 ) & = & 4({g}^{2}+ \check{g}^{2}) \Psi_{1 \otimes1}(x ,x+1)-2{g} \check{g} \Psi_{1 \otimes1}(x-1,x+1)-2{g} \check{g} \Psi_{1 \otimes1}(x,x+2) \nonumber \\
 &  & +2{g}^{2} \Psi_{\fbar}(x)+2{g}^{2} \Psi_{\fbar}(x+1) \,.
 \end{eqnarray}
Furthermore, 
\begin{eqnarray}
g^2 H \cdot \Psi_{\fbar}(x) & = & 6{g}^{2} \Psi_{\fbar}(x)-{g}^{2} \Psi_{\fbar}(x+1)-{g}^{2} \Psi_{\fbar}(x-1) \\
 &  & +2{g}^{2} \Psi_{1 \otimes1}(x,x+1)+2{g}^{2} \Psi_{1 \otimes1}(x-1,x) \, . \nonumber
 \end{eqnarray}
We take the ansatz 
\begin{eqnarray}
 \Psi_{1 \otimes1}(x_{1},x_{2}) & = & e^{i(p_{1}x_{1}+p_{2}x_{2})}+S_{1 \otimes1}(p_{2},p_{1})e^{i(p_{1}x_{2}+p_{2}x_{1})} \\
 \Psi_{\fbar}(x) & = & S_{\fbar}(p_{2},p_{1})e^{i(p_{1}+p_{2})x} \,.
 \end{eqnarray}
Note that $S_{1 \otimes1}(p_{1},p_{2})$ still has the interpretation
of the scattering matrix of the magnons $Q$ and $ \bar{Q}$, which are the asymptotic excitations,
while $ \fbar$ is an ``unstable'' excitations created
during the collision of $Q$ and $ \bar{Q}$. 
We find 
\begin{eqnarray}
S_{1 \otimes1}(p_{1},p_{2}) & = & - \left( \frac{1+e^{ip_{1}+ip_{2}}-2( \kappa- \frac{1}{\kappa})e^{ip_{1}}}{1+e^{ip_{1}+ip_{2}}-2( \kappa- \frac{1}{\kappa})e^{ip_{2}}} \right) \left( \frac{1+e^{ip_{1}+ip_{2}}-2 \kappa e^{ip_{1}}}{1+e^{ip_{1}+ip_{2}}-2 \kappa e^{ip_{2}}} \right)^{-1} \\
S_{\fbar}(p_{1},p_{2}) & = &  \frac{4ie^{i(p_{1}+p_{2})}( \sin p_{1}- \sin p_{2})}{(1+e^{ip_{1}+ip_{2}}-2 \kappa e^{ip_{1}})(1+e^{ip_{1}+ip_{2}}-2( \kappa- \frac{1}{\kappa})e^{ip_{2}})} \,.
\end{eqnarray}
$S_{1 \otimes  1}$ is the product of two factors, and it admits two poles. The first factor coincides with $S_{1 \otimes 3}$, so its pole is associated to
an anti-bound state entirely analogous to the anti-bound state in the $1_L \otimes 3_R$ sector. The pole is located at $p_1 = P/2 -i q  +\pi$, $p_2 = P/2 + i q - \pi$,
with 
\be
e^{-q}=\frac{\cos(P/2)}{\frac{1}{\kappa}-\kappa} \, .
\ee
The dispersion relation is again
\be
 \left[ Q  \bar Q\right]^{antibound}_{1_L \, 1_R}: \qquad E= \frac{4(2- \kappa^{2})}{1- \kappa^{2}}- \frac{4 \kappa^{2}}{1- \kappa^{2}} \sin^{2} \frac{P}{2}\, ,
\ee
and the range of $P$ for which the wave-function is normalizable is as in (\ref{twocases}) -- see the solid
 (red) curve in the left column of figure~\ref{DispersionPlots}.
It is interesting to analyze the explicit form of the wave-function in the $\kappa \to 0$ limit. The  $Q \bar Q$ piece has the form
\be
\Psi_{1\otimes1}(x_{1},x_{2})  =  (-1)^{x_2 - x_1} e^{iP(\frac{x_{1}+x_{2}}{2})}e^{-q(x_{2}-x_{1})} \, ,\quad q \to \infty
\ee
so only the $x_2 = x_1 +1$ term survives in the  limit, and we recover
  the dimeric impurity ${\cal M}_{\bf 1}$ of SCQCD. A short calculation gives 
\be
\frac{\Psi_{\bar{\phi}}(x)}{\Psi(x,x+1)}|_{\kappa\to0} =\frac{2}{(1+e^{iP})} \,.
\ee
Comparison with (\ref{Ttildewave}) shows that that in the $\kappa \to 0$ limit the  antibound state in the $Q \bar Q$ singlet sector becomes precisely 
the dimeric excitation $\widetilde T$ of SCQCD.

The pole in the second factor of $S_{1 \otimes  1}$ 
 corresponds instead to a bound state, with
\be
e^q = \frac{\cos(P/2)}{\kappa}\,.
\ee
The dispersion relation and range of existence are
\be
\left[Q  \bar Q\right]^{bound}_{1_L \, 1_R}: \qquad E =  4 \sin^{2} \frac{q}{2} \, ,\quad  0 < |P| < 2 \, {\rm arccos} \kappa \, ,
\ee
which are shown as the dashed (green) curve on the left column of figure~\ref{DispersionPlots}. This bound state is absent at the orbifold point and
comes into full existence (for any $P$) in the SCQCD limit $\kappa \to 0$. The natural guess is  that in this limit
it reduces to the gapless $T(p)$ magnon of SCQCD, and it does:
\be
\frac{\Psi_{\bar{\phi}}(x)}{\Psi(x,x+1)}|_{\kappa\to0} =-\frac{1 + e^{-i P}}{2} \, ,
\ee
in agreement with (\ref{Twave}).

The S-matrix in the  $\bar Q Q$ channel is obtained as usual by $\kappa \to 1/\kappa$,
\begin{eqnarray*}
\check S_{1 \otimes1}(p_{1},p_{2};\kappa) & = & - \left( \frac{1+e^{ip_{1}+ip_{2}}+ 2( \kappa- \frac{1}{\kappa})e^{ip_{1}}}{1+e^{ip_{1}+ip_{2}}+2( \kappa- \frac{1}{\kappa})e^{ip_{2}}} \right) \left( \frac{1+e^{ip_{1}+ip_{2}}-\frac{2}{ \kappa} e^{ip_{1}}}{1+e^{ip_{1}+ip_{2}}-\frac{2}{ \kappa} e^{ip_{2}}} \right)^{-1} \\
\check S_{\fbar}(p_{1},p_{2};\kappa) & = &  \frac{4ie^{i(p_{1}+p_{2})}( \sin p_{1}- \sin p_{2})}{(1+e^{ip_{1}+ip_{2}}-\frac{2}{ \kappa} e^{ip_{1}})(1+e^{ip_{1}+ip_{2}}+ 2( \kappa- \frac{1}{\kappa})e^{ip_{2}})} \,.
\end{eqnarray*}
The pole in the first factor of $\check S_{1 \otimes1}$ corresponds to a bound state, with
\be
\left[\bar Q   Q\right]^{bound}_{1_L \, 1_R}: \qquad E(P)=\frac{4 \kappa^2}{1-\kappa^2} \left(1 - 2 \kappa^2 + \sin^2 \frac{P}{2} \right)\, ,
\ee
with the range of existence given by (\ref{twocases}).
Finally, the pole in the second factor of $\check S_{1 \otimes1}$ never corresponds to a normalizable solution.

\subsection{Summary}
\begin{figure}[t!]
\begin{centering}
\begin{tabular}{|l|c|c|}
\hline 
 & $Q\bar{Q}$ scattering channel & $\bar{Q}Q$ scattering channel\tabularnewline
\hline
\hline 
$\kappa=0.999$ & \includegraphics[scale=0.6]{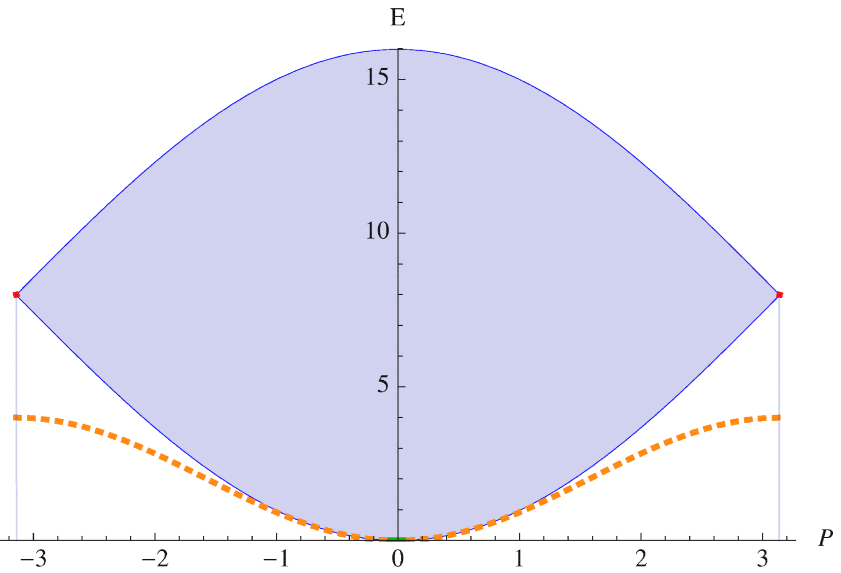} & \includegraphics[scale=0.6]{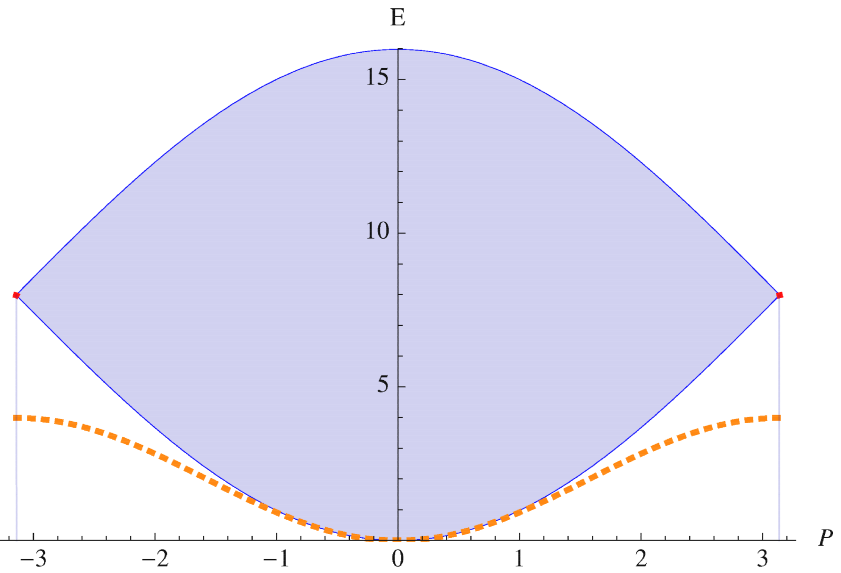}\tabularnewline
\hline 
$\kappa=0.65$ & \includegraphics[scale=0.6]{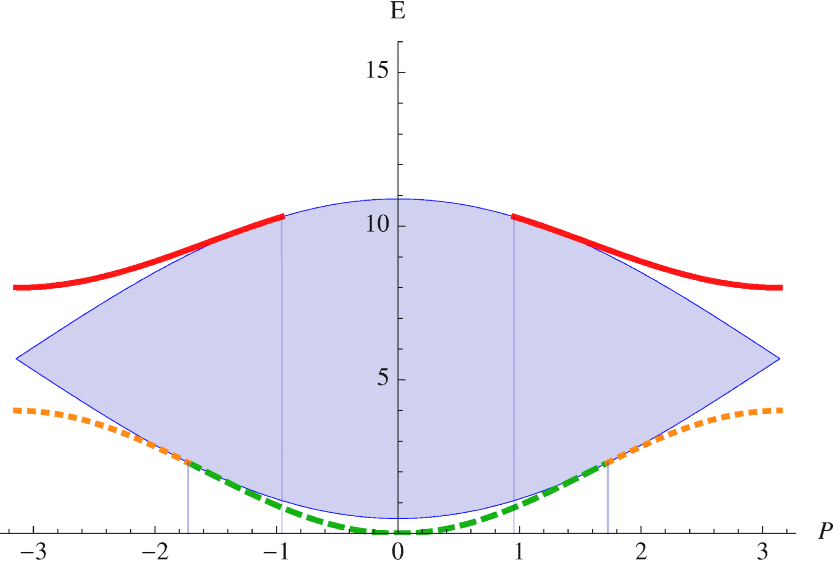} & \includegraphics[scale=0.6]{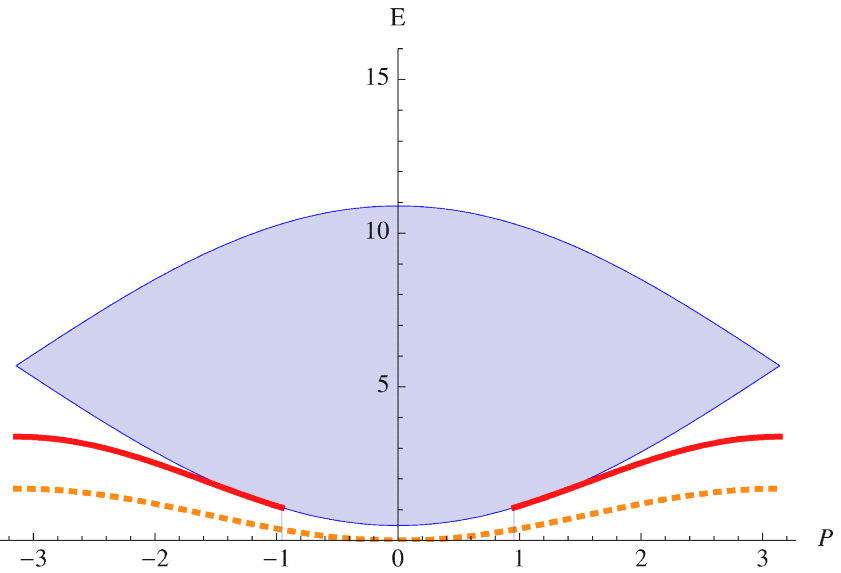}\tabularnewline
\hline 
$\kappa=\frac{-1+\sqrt{5}}{2}$ & \includegraphics[scale=0.6]{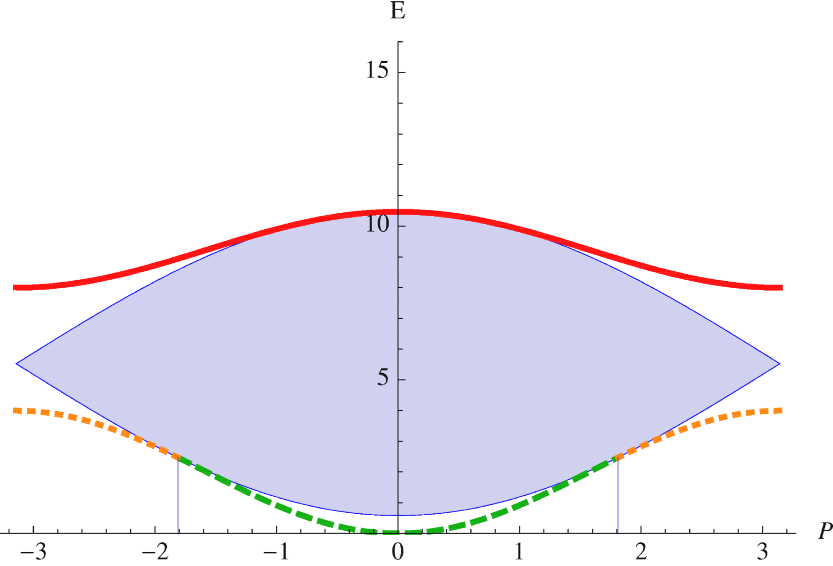} & \includegraphics[scale=0.6]{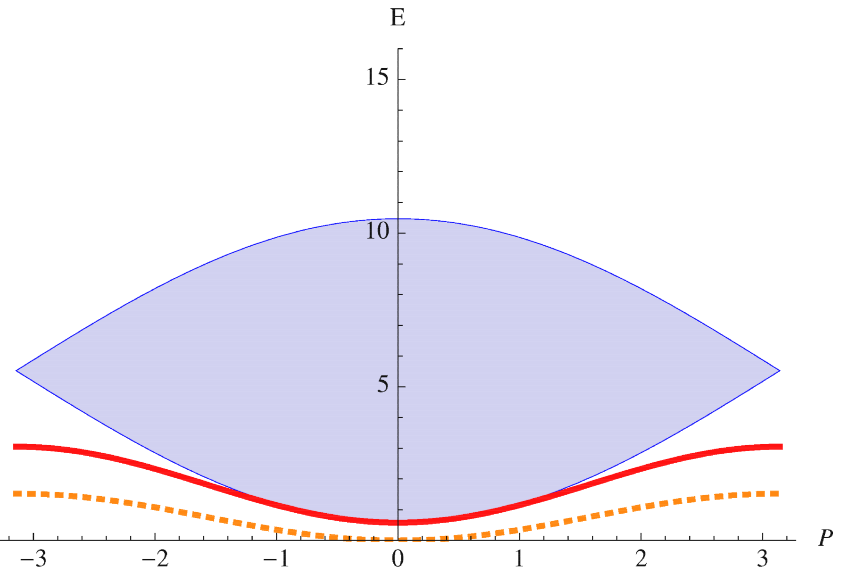}\tabularnewline
\hline 
$\kappa=0.35$ & \includegraphics[scale=0.6]{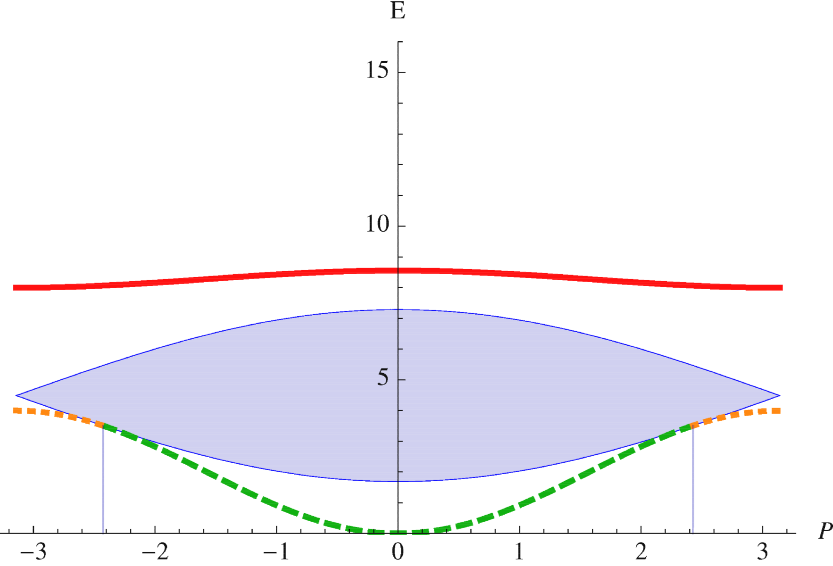} & \includegraphics[scale=0.6]{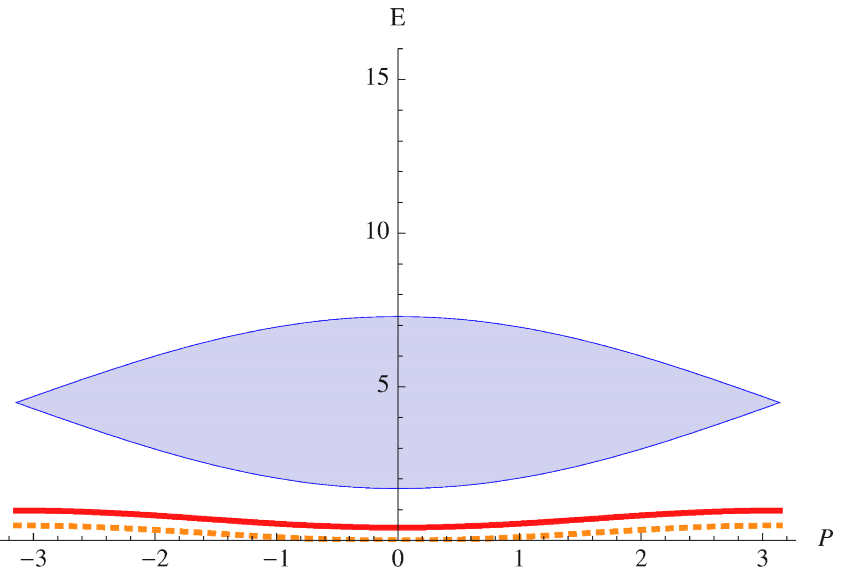}\tabularnewline
\hline 
$\kappa=0.001$ & \includegraphics[scale=0.6]{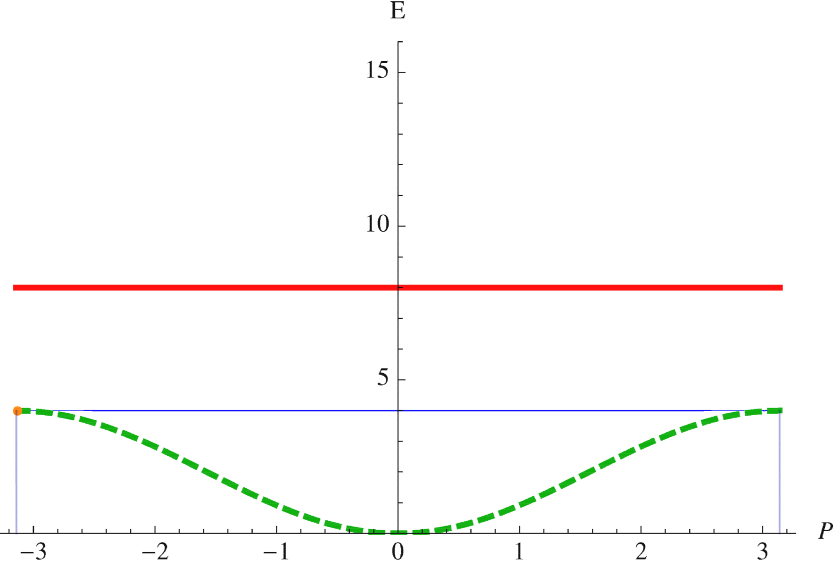} & \includegraphics[scale=0.6]{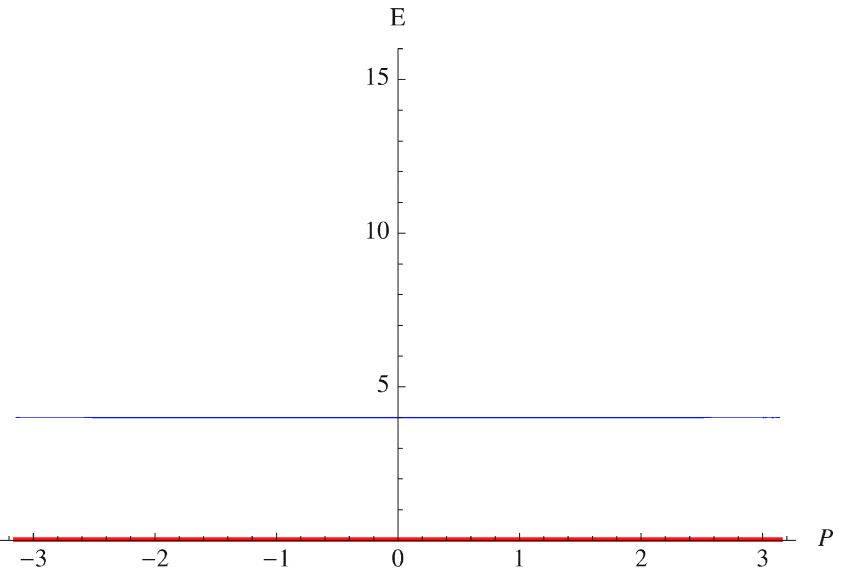}\tabularnewline
\hline
\end{tabular}
\par\end{centering}
\caption{Plots of the dispersion relations of the (anti)bound states for different values of $\kappa$. \label{DispersionPlots}
The shaded region represents the two-particle continuum.  }
\end{figure}

The two-body scattering problem in the spin chain of the interpolating SCFT
admits a rich spectrum of bound and anti-bound states. The results are summarized in table \ref{BoundTable} and figure~\ref{DispersionPlots}.
The $Q \bar Q$ scattering channel (that is, the channel with $Q$ to the left of $\bar Q$, and the $\phi$ vacuum on the outside) is the one relevant to make contact
with ${\cal N}=2$ SCQCD, which is obtained in the $\kappa \to 0$ limit. Remarkably, the magnon excitations of 
SCQCD are recovered as the smooth  limits of  the $Q \bar Q$ (anti)bound states: as $\kappa \to 0$ the wavefunctions of the (anti)bound states
 localize to two neighboring sites and reproduce the ``dimeric'' magnons $T(p)$,  $\widetilde T(p)$ and ${\cal M}_{ \bf 3}(p)$
of SCQCD.

\begin{table}
\begin{centering}
\begin{tabular}{|l|l|l|l|}
\hline 
 & Pole of the S-matrix & Range of existence & Dispersion relation $E(P)$ \tabularnewline
\hline
\hline 
${\cal M}_{33}$ & $e^{-q}=\cos(\frac{P}{2})/\kappa$ & $2 \arccos\kappa< |P| < \pi$ & $4\sin^{2}(\frac{P}{2})$\tabularnewline
\hline 
$T$ & $e^{q}=\cos(\frac{P}{2})/\kappa$ & $0<|P|<2\arccos\kappa$ & $4\sin^{2}(\frac{P}{2})$\tabularnewline
\hline 
\textcolor{red}{$\widetilde{T}$ and ${\cal M}_{3}$} & \textcolor{red}{$e^{-q}=\cos(\frac{P}{2})/(\kappa-\frac{1}{\kappa})$} &
\textcolor{red}{See  equ.(\ref{twocases})}
 & \textcolor{red}{$\frac{4\kappa^{2}}{(1-\kappa^{2})}(\frac{2}{\kappa^{2}}-1-\sin^{2}\frac{P}{2})$}\tabularnewline
\hline
\hline 
$\check{{\cal M}}_{33}$ & $e^{-q}=\kappa\cos(\frac{P}{2})$ & $0<|P|<\pi$ & $4\kappa^{2}\sin^{2}(\frac{P}{2})$\tabularnewline
\hline 
$\check{T}$ & $e^{q}=\kappa\cos(\frac{P}{2})$ & No solution & \tabularnewline
\hline 
$\check{\widetilde{T}}$ and $\check{{\cal M}}_{3}$ & $e^{-q}=\cos(\frac{P}{2})/(\frac{1}{\kappa}-\kappa)$ & See  equ.(\ref{twocases}) & $\frac{4\kappa^{2}}{(1-\kappa^{2})}(1-2\kappa^{2}+\sin^{2}\frac{P}{2})$\tabularnewline
\hline
\end{tabular}
\par\end{centering}
\caption{\label{BoundTable}
Dispersion relations and range of existence of the various (anti)bound  states 
in  two-body scattering.
The first three entries correspond to the $Q \bar Q$ channel and the last three entries
to the $\bar{Q} Q$ channel. The color-coding of the third entry is a reminder that these are {\it anti-}bound states with energy above the two-particle continuum.}
\end{table}

 \subsection{Left/right factorization of the two-body S-matrix }

As is well-known, the magnon  excitations of the ${\cal N}=4$ SYM spin chain transform in the fundamental representation of $SU(2|2) \times SU(2|2)$,
and their two-body S-matrix factorizes into the product of the S-matrices for the ``left'' and ``right'' $SU(2|2)$. The $\mathbb{Z}_2$ orbifold preserves
this factorization. Remarkably, this left/right factorization persists even away from the orbifold point, for the full interpolation SCFT --
or at least this is what happens at one-loop in the scalar sector. 
Our results for the S-matrix in the $Q \bar Q$ channel in the four different $SU(2)_L \times SU(2)_R$ sectors are summarized
 in table \ref{summaryS}, where we have defined
 \[
 \SS(p_{1},p_{2}, \kappa) \equiv- \frac{1-2 \kappa e^{ip_{1}}+e^{i(p_{1}+p_{2})}}{1-2 \kappa e^{ip_{2}}+e^{i(p_{1}+p_{2})}} \, ,
 \]
 {\it i.e.} the standard S-matrix of the XXZ chain, with $\Delta_{XXZ}= \kappa$. 
 \begin{table}[H]
 \begin{centering}
 \begin{tabular}{|r|l|}
 \hline 
$L \otimes R$  & $S(p_{1},p_{2}, \kappa)$ \tabularnewline
 \hline
 \hline 
$1 \otimes1$  & $- \SS(p_{1},p_{2}, \kappa- \frac{1}{\kappa}) \SS^{-1}(p_{1},p_{2}, \kappa)$ \tabularnewline
 \hline 
$1 \otimes3$  & $ \SS(p_{1},p_{2}, \kappa- \frac{1}{\kappa})$ \tabularnewline
 \hline 
$3 \otimes1$  & $-1$ \tabularnewline
 \hline 
$3 \otimes3$  & $ \SS(p_{1},p_{2}, \kappa)$ \tabularnewline
 \hline
 \end{tabular}
 \par \end{centering}
 \caption{\label{summaryS}
 The S-matrix in the $Q \bar Q$ scattering channel. }
 \end{table}
We see that we can write
\[
S(p_{1},p_{2};\kappa)=\frac{S_{L}(p_{1},p_{2}; \kappa)S_{R}(p_{1},p_{2}; \kappa) }{S_{3 \otimes 3}(p_1, p_2; \kappa)}
\]
 where $S_{L}$ and $S_{R}$ are defined in table \ref{factorization-S}.

 \begin{table}[H]
 \begin{centering}
 \begin{tabular}{|c|l||c|l|}
 \hline 
$SU(2)_{L}$  & $S_{L}(p_{1},p_{2}; \kappa)$  & $SU(2)_{R}$  & $S_{R}(p_{1},p_{2}; \kappa)$ \tabularnewline
 \hline
 \hline 
${\bf 1}$  & $ \SS(p_{1},p_{2}; \kappa- \frac{1}{\kappa})$  & ${\bf 1}$  & $ -1$ \tabularnewline
 \hline 
${\bf 3}$ & $ \SS(p_{1},p_{2}; \kappa)$  & ${\bf 3}$  & $   \SS(p_{1},p_{2}; \kappa)$ \tabularnewline
 \hline
 \end{tabular}
 \par \end{centering}
 \caption{Definitions of the $SU(2)_L$ and $SU(2)_R$ S-matrices.}
 \label{factorization-S}
 \end{table}

In the analysis of the Yang-Baxter equation, it will be useful to write the S-matrices in both  the $SU(2)_L$ and $SU(2)_R$ sectors using
the identity $(\mathbb{I})$ and trace $(\mathbb{K})$ tensorial structures,
\begin{eqnarray}
S_L (p_1, p_2; \kappa) & =& A_L (p_1, p_2; \kappa) \, \mathbb{I} +B_L (p_1, p_2; \kappa) \, \mathbb{K}\\
S_R (p_1, p_2; \kappa) & =& A_R(p_1, p_2; \kappa)  \, \mathbb{I} +B_R (p_1, p_2; \kappa) \, \mathbb{K} \,.
\end{eqnarray}
Writing the indices explicitly,
 \be
(S_R)_{\II \JJ}^{\MM \NN}   =  A_R \,\, \delta_{\II}^{\MM} \delta_{\JJ}^{\NN}+B_R \,\, \epsilon_{\II \JJ}\epsilon^{\MM \NN} \, ,
 \ee
 Recalling that eigenvalue of  $\mathbb{K}$ on the triplet is zero while it is two on the singlet, we see that
\begin{eqnarray}
A & = & S_{\bf 3}\\
B & = & \frac{1}{2}(S_{\bf 1} - S_{\bf 3}) \, .
\end{eqnarray}
The values of $S_{\bf 1}$ and $S_{\bf 3}$ in both the $SU(2)_L$ and $SU(2)_R$ sectors
can be read off from table \ref{factorization-S},
  \begin{eqnarray}
A_L(p_{1},p_{2}, \kappa ) & =  &\SS(p_{1},p_{2}, \kappa ) \\
B_L(p_{1},p_{2}, \kappa ) & = &\frac{1}{2}\left( \SS(p_{1},p_{2}, \kappa - \frac{1}{\kappa })- \SS(p_{1},p_{2}, \kappa )\right)\\
A_R(p_{1},p_{2}, \kappa ) & = & \SS(p_{1},p_{2}, \kappa ) \\
B_R(p_{1},p_{2}, \kappa ) &   =& - \frac{1}{2}(1+ \SS(p_{1},p_{2}, \kappa ))\, .
 \end{eqnarray}
In complete analogy, in the $ \bar{Q}Q$  channel we have the factorization
\be
\check S(p_{1},p_{2};\kappa)=\frac{\check S_{L}(p_{1},p_{2}; \kappa)\check S_{R}(p_{1},p_{2}; \kappa) }{\check S_{3 \otimes 3}(p_1, p_2; \kappa)} \, ,
\ee
and we can write
\begin{eqnarray}
\check S_L(p_1, p_2; \kappa) & =& \check A_L (p_1, p_2; \kappa)\, \mathbb{I} +\check B_L(p_1, p_2; \kappa) \, \mathbb{K}\\
\check S_R(p_1, p_2; \kappa) & =& \check A_R (p_1, p_2; \kappa)\, \mathbb{I} +\check B_R (p_1, p_2; \kappa)\, \mathbb{K} \,.
\end{eqnarray}
As always, each ``checked'' quantity is obtained from the corresponding unchecked one by sending $\kappa \to 1/\kappa$.

\section{Yang-Baxter Equation}

The  one-loop spin chain of the $\mathbb{Z}_2$ orbifold of ${\cal N}=4$ SYM is known to be integrable \cite{Beisert:2005he,Solovyov:2007pw}.
A natural question is whether integrability persists for the $\check g \neq g$. 
We can explore the integrability of the spin chain for the interpolating SCFT by checking the Yang-Baxter equation
for the two-body S-matrix.
Integrability of the spin chain amounts to the existence of higher conserved quantities beyond the momentum and the Hamiltonian, which 
would imply  exact  factorization of many-body scattering  into a sequence of two-body scatterings.
 For this to happen it is necessary  that different ways to factorize three-body scattering into two-body scatterings
 should commute: the Yang-Baxter equation expresses this consistency condition.

The two-body S-matrix  of our theory factorizes   into the S-matrix for the $SU(2)_L$ sector  times the S-matrix for the  $SU(2)_R$ sector.
 The Yang-Baxter equation must be satisfied separately in each sector. Clearly this is a
sufficient condition for the full Yang-Baxter equation to hold; it is also a necessary condition since we can always restrict the asymptotic states to one sector
by setting their quantum numbers in the other sector to be highest weights.
  \begin{figure}[t]
 \begin{centering}
 \includegraphics[scale=0.4]{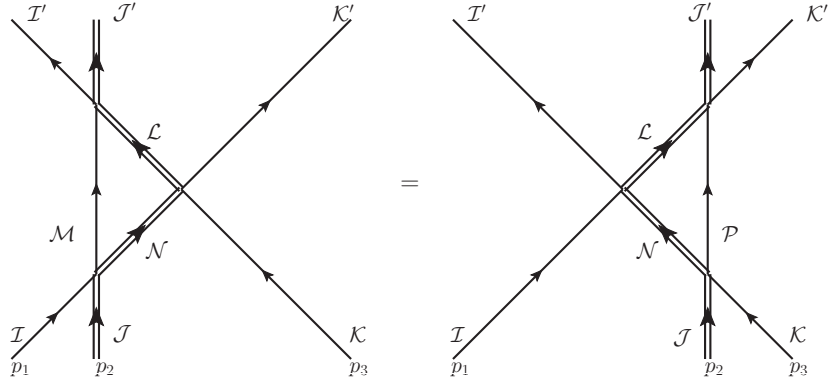}
 \par \end{centering}
 \caption{\label{YB}Yang-Baxter equation in each $SU(2)$ sector. Simple lines represent $Q$ impurities, double lines $\bar Q$ impurities.}
 \end{figure}
In each sector, the Yang-Baxter equation is represented by the diagram of figure \ref{YB}, and reads explicitly
\[
S_{\II \JJ}^{\MM \NN}(p_{1},p_{2}) \topp S_{\NN \KK}^{\LL \KK^{\prime}}(p_{1},p_{3})S_{\MM \LL}^{\II^{\prime} \JJ^{\prime}}(p_{2},p_{3})= \topp S_{\LL \PP}^{\JJ^{\prime} \KK^{\prime}}(p_{1},p_{2})S_{\II \NN}^{\II^{\prime} \LL}(p_{1},p_{3}) \topp S_{\JJ \KK}^{\NN \PP}(p_{2},p_{3}) \, 
 \]
Using the decomposition introduced in the previous section, we can write the left-hand side as
 \begin{eqnarray}
 &  & S_{\II \JJ}^{\MM \NN}(p_{1},p_{2}) \topp S_{\NN \KK}^{\LL \KK^{\prime}}(p_{1},p_{3})S_{\MM \LL}^{\II^{\prime} \JJ^{\prime}}(p_{2},p_{3}) \nonumber  \\
 & = & A \topp AA \delta_{\KK}^{\KK^{\prime}} \delta_{\II}^{\II^{\prime}} \delta_{\JJ}^{\JJ^{\prime}} + A \topp BBg_{\JJ \KK} \delta_{\II}^{\KK^{\prime}}g^{\II^{\prime} \JJ^{\prime}}+B \topp BAg_{\II \JJ} \delta_{\KK}^{\II^{\prime}}g^{\JJ^{\prime} \KK^{\prime}} 
\nonumber  \\
 & + & (A \topp AB+B \topp AA+2B \topp AB+B \topp BB) \delta_{\KK}^{\KK^{\prime}}g_{\II \JJ}g^{\II^{\prime} \JJ^{\prime}}+A \topp BAg_{\JJ \KK}g^{\JJ^{\prime} \KK^{\prime}} \delta_{\II}^{\II^{\prime}} \nonumber
 \end{eqnarray}
We have suppressed the momentum arguments with the convention that the first symbol in each term is a function of $(p_1,p_2)$, the second is function of $(p_1,p_3)$ and the third $(p_2,p_3)$. Similarly, for the right-hand side
\begin{eqnarray*}
 &  &  \topp S_{\LL \PP}^{\JJ^{\prime} \KK^{\prime}}(p_{1},p_{2})S_{\II \NN}^{\II^{\prime} \LL}(p_{1},p_{3}) \topp S_{\JJ \KK}^{\NN \PP}(p_{2},p_{3}) \\
 & = &  \topp AA \topp A \delta_{\II}^{\II^{\prime}} \delta_{\JJ}^{\JJ^{\prime}} \delta_{\KK}^{\KK^{\prime}}+ \topp AB \topp Bg^{\II^{\prime} \JJ^{\prime}}g_{\JJ \KK} \delta_{\II}^{\KK^{\prime}}+ \topp BB \topp Ag^{\JJ^{\prime} \KK^{\prime}}g_{\II \JJ} \delta_{\KK}^{\II^{\prime}}  \\
 & + &  \topp AB \topp Ag_{\II \JJ}g^{\II^{\prime} \JJ^{\prime}} \delta_{\KK}^{\KK^{\prime}}+( \topp AA \topp B+ \topp BA \topp A+2 \topp BA \topp B+ \topp BB \topp B)g^{\JJ^{\prime} \KK^{\prime}} \delta_{\II}^{\II^{\prime}}g_{\JJ \KK} 
 \end{eqnarray*}
Collecting the terms with the same index structure, the Yang-Baxter equation in each $SU(2)$ sector reduces to the following five equations:
\begin{eqnarray}
A \topp AA & = &  \topp AA \topp A \label{YB1} \\
A \topp BB & = &  \topp AB \topp B \\
B \topp BA & = &  \topp BB \topp A \\
2B \topp AB+A \topp AB+B \topp AA+B \topp BB & = &  \topp AB \topp A \label{4th}  \\
A \topp BA & = &  2 \topp BA \topp B+\topp AA \topp B+ \topp BA \topp A+ \topp BB \topp B \,.\label{YB5}
\end{eqnarray}

\begin{figure}[t]
\begin{centering}
\includegraphics[scale=0.7]{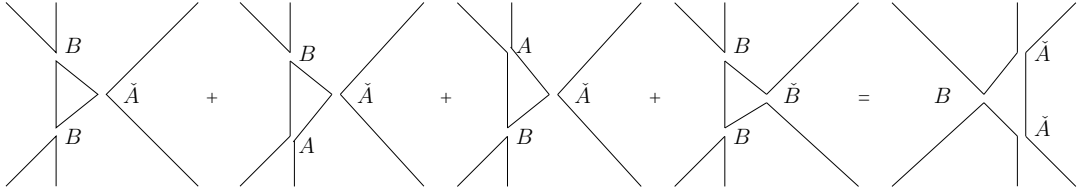} \qquad\qquad\qquad\qquad 
\par \end{centering}
\caption{\label{YBexample}Example of $SU(2)$ index flow. Collection of the terms with index structure $g_{\II \JJ}\delta_{\KK}^{\KK^\prime}g^{{\II^\prime}{\JJ^\prime}}$ gives rise to the 4th YB equation. Similarly, other Yang-Baxter equations also can be understood in graphical manner.}
\end{figure}
At the orbifold point, $\kappa = 1/\kappa =1$ and thus $A = \check A$, $B = \check B$: the first three equations are trivial; the forth and fifth become equivalent.
In both the $SU(2)_L$ and $SU(2)_R$ sectors  (which are in fact equivalent for $\kappa =1$), the remaining equation is easily checked.
Thus as expected, the Yang-Baxter equation is satisfied at the orbifold point. 
 We then find that YB is {\it violated} as we move away from the orbifold
point, for all $\kappa \in (0, 1)$,
showing conclusively that the spin chain of the interpolating theory is {\it not} integrable for general $\kappa$. 
To our surprise however,
YB holds again in the SCQCD limit $\kappa \to 0$! 
We take this as a hint that planar ${\cal N}=2$ SCQCD might be integrable, at least at one loop.

\section{Discussion}

\begin{figure}[t!]
\begin{centering}
\begin{tabular}{c|c}
\includegraphics[scale=0.8]{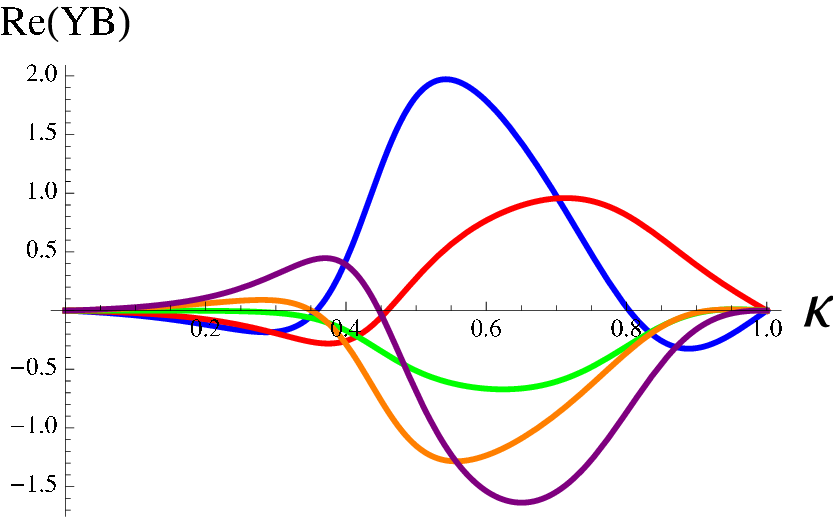} & \includegraphics[scale=0.8]{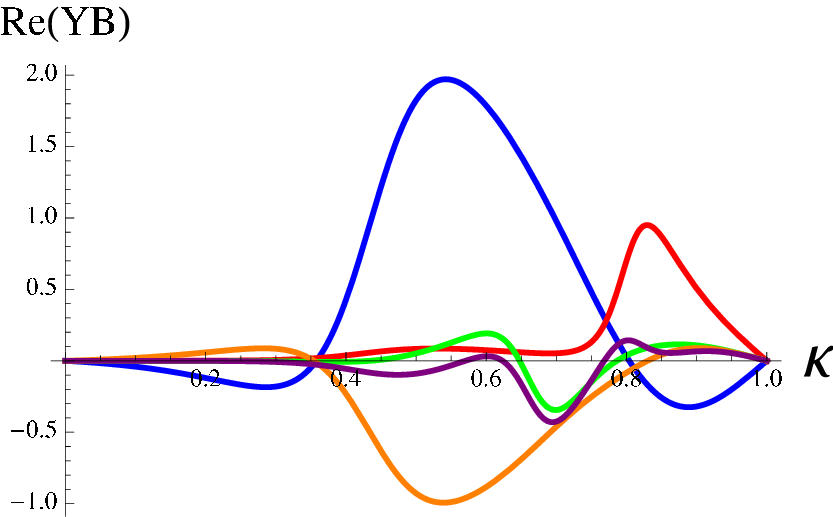}\tabularnewline
\hline 
\includegraphics[scale=0.8]{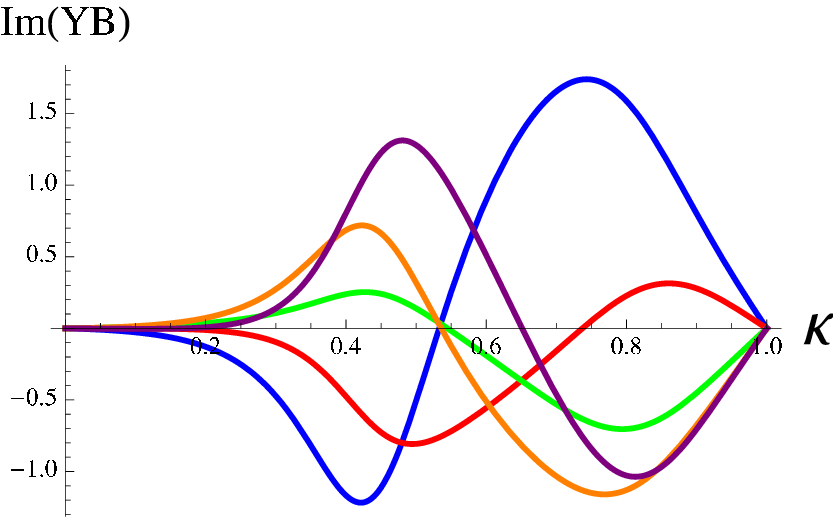} & \includegraphics[scale=0.8]{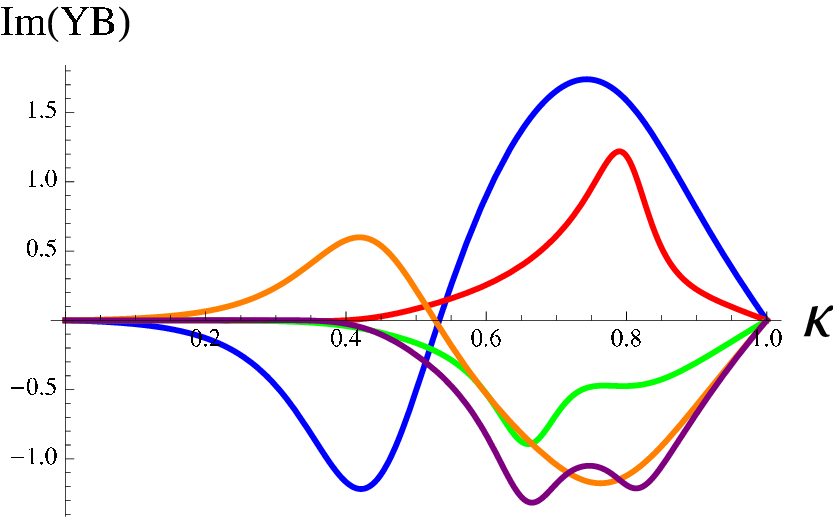}\tabularnewline
\hline 
$SU(2)_{R}$ sector  & $SU(2)_{L}$ sector\tabularnewline
\end{tabular}
\par\end{centering}
\caption{
\label{YBfailure} 
The figure shows the differences between the left and right-hand sides
of the five Yang-Baxter equations, as a function of $\kappa$, for the specific  choice of momenta $p_{1}=0.3$, $p_{2}=0.8$ and $p_{3}=1.4$.
The blue, red, green, orange and purple curves show {\it(l.h.s)$- $(r.h.s)}
for the the first to fifth equation.}
\end{figure}

Ordinarily, verification of the Yang-Baxter equation for the two-magnon S-matrix
counts as strong evidence for integrability. In our case, however,
for  $\kappa$ {\it strictly} zero, 
the elementary $Q$ impurities  ``freeze'',  and only  $Q\bar Q$ dimers
can propagate on the chain.  Correspondingly, the $Q$ dispersion relation  becomes momentum-independent, 
\be
E_Q(p; \kappa) =2(1-\kappa)^2 + 8 \kappa \, \sin^2 (\frac{p}{2}) \xrightarrow[\kappa \to 0]{}    2 \, ,
\ee
and  the S-matrix also degenerates to a simple expression.
Verification of YB strictly at $\kappa = 0$ may then appear like an accident due to this degenerate limit.
What we find more significant, and non-trivial evidence for integrability,
 is that the integrable point $\kappa =0$ is reached
smoothly, with YB failing infinitesimally for infinitesimal $\kappa$ -- this is clear since the S-matrices are analytic (rational) functions of $\kappa$.
This smooth behavior is illustrated in figure \ref{YBfailure}, where we plot the differences between the left and right hand sides of the five equations (\ref{YB1}--\ref{YB5})
(for some choice of the momenta).

An elegant way to conclusively prove  one-loop integrability at $\kappa =0$
would be to exhibit the algebraic Bethe ansatz for the SCQCD spin chain. 
The simplest guess for the R-matrix does not appear to work \cite{integrability}, but the issue deserves further investigation.

Another possible approach is to {\it assume} integrability 
 to derive Bethe equations for the periodic chain, and then check whether  their (numerical) solutions agree with the solutions obtained by direct
diagonalization  of the Hamiltonian. This is not entirely straightforward, because we cannot
work strictly at $\kappa = 0$.  The naive Bethe equations 
at $\kappa = 0$  have no interesting solutions for finite values of the Bethe roots -- 
the non-trivial dynamics is hidden in  Bethe roots with infinite imaginary parts (in the momentum variable). We saw this phenomenon 
in the evolution of the bound states as $\kappa \to 0$, where the individual magnon momenta behave as $ i \log \kappa$.
Taking the SCQCD limit $\kappa \to 0$ too early we lose information about the bound states.
(It is conceivable that the failure of the (simplest) algebraic Bethe ansatz is also due to this 
order-of-limits issue.)
Nevertheless, it makes sense to write Bethe equations for  small $\kappa$, viewed as a regulator to be removed at the end of the calculation.
We can also calculate the S-matrix of the bound states, by using the fusion procedure
for infinitesimal $\kappa$, and check {\it their} YB equation in the SCQCD limit. 
The consistency of this approach should follow from the smoothness of the $\kappa \to 0$ limit.

A natural extension of our work is the calculation of one-loop dilation operator in the complete theory, including fermions and derivatives \cite{fullH}.
Let us briefly comment on the symmetry structure of the complete spin chain. As is well-known, the symmetry of the ${\cal N}=4$
spin chain in the excitation picture is $PSU(2|2)_L \times PSU(2|2)_R \times \mathbb{R}$, where
the central factor $\mathbb{R}$ is identified with the Hamiltonian. The $\mathbb{Z}_2$ orbifold projection preserves
the $PSU(2|2)_R$ in the ``right'' sector (this is a subgroup of the ${\cal N}=2$ superconformal group $SU(2,2|2)$),
but breaks $PSU(2|2)_L$ to the bosonic subgroup $SU(2)_L \times SU(2)_{\alpha}$, where $SU(2)_\alpha$ denotes the left-handed Lorentz symmetry.
At the orbifold point $\kappa =1$, the breaking is only due to a global twist of the chain, while locally the symmetry is the same as
in ${\cal N}=4$, but for $\kappa \neq 1$ the symmetry is truly broken. All in all,
the symmetry of the spin chain of the interpolating theory is $SU(2)_L \times SU(2)_\alpha \times PSU(2|2)_R \times \mathbb{R}$.
In this paper we have found that in the  two-body S-matrix of $Q$ impurities has a left $\times$ right factorization, and we expect this
feature to persist for the full chain.

An obvious question is whether symmetry is sufficient to fix the form of the S-matrix, 
as it does to all loops  in ${\cal N}=4$ SYM (up to an overall scalar factor). While unlikely for $S_L$, this is likely for $S_R$,
which has a large supergroup symmetry.
In fact, the symmetry in the right sector of the interpolating SCFT
the same as in (either sector of) ${\cal N}=4$ SYM. The $S_R$ matrix
of ${\cal N}=4$ is uniquely fixed, up to an overall scalar factor, from the (centrally extended) $SU(2|2)_R$ symmetry \cite{Beisert:2005tm}.
But our results for $S_R$ in the interpolating theory are definitely different (for $\kappa \neq 1$) from the ${\cal N}=4$ results.
This is clear already in the scalar sector studied in this paper, by inspection of the S-matrix of the
$Q_{{\cal I} \hat +}$ impurities.  This discrepancy is explained by the fact in our case
the magnons transform in a reducible representation of SU(2|2) (two copies of the fundamental representation).
It will  be interesting to see whether these assumptions can be relaxed to
reproduce (and possibly uniquely fix) 
Remarkably, it is still possible to use symmetry to  fix uniquely the form of
the $S_R$ matrix in the interpolating theory, up to a free parameter that can be identified
with $\kappa$. This analysis will be presented elsewhere \cite{twistedmagnons}.

Finally it would be very interesting to evaluate the two-body S-matrix at strong coupling, in the dual string sigma-model, and
 see whether it has the same $\kappa$ dependence as the perturbative S-matrix. 
 Failure of integrability for generic $\kappa$ is not an issue here, since we would not be using
 in any way factorization of $n$-body scattering, but rather focus on the two-body
 S-matrix, which we expect to have a smooth interpolation from weak to strong coupling.
  The sigma-model at the orbifold point is well-known, and moving away from the orbifold point corresponds
to changing the value of a theta angle $\beta$ (the period of the NSNS $B$-field through the collapsed cycle of the orbifold) \cite{Lawrence:1998ja,Klebanov:1999rd}.
The orbifold point corresponds to $\beta =1/2$, while the SCQCD limit corresponds to $\beta \to 0$.
From the dual side, it is natural to expect integrability precisely at the two extrema 0 and 1/2,
but not for generic values of the $B$-field. A toy model for this behavior 
 is the $O(3)$ sigma-model in a magnetic field \cite{Zamolodchikov:1992zr}.

One of our  original motivations was to collect ``bottom-up'' clues about the string dual of ${\cal N}=2$ SCQCD.
While firm conclusions will have to wait a higher-oder (all order?) analysis, we
 can already see a qualitative agreement with the ``top-down'' approach of our previous paper \cite{Gadde:2009dj}. We argued
that ${\cal N}=2$ SCQCD is dual to a non-critical string background, with seven geometric dimensions, containing both an $AdS_5$
and an $S^1$ factor. Rotation in $S^1$ corresponds to the $U(1)_r$ quantum number. In lightcone
quantization of the sigma-model, the lightcone coordinates would be obtained by combining  
this $S^1$ and the timelike direction of $AdS_5$. 
We then expect five bosonic gapless excitations,
four associated to the transverse AdS coordinates and one to the seventh dimension. The vacuum of the lightcone
sigma-model corresponds to chiral  vacuum ${\rm Tr} \, \phi^\ell$ of the spin chain,  while the four AdS excitations correspond to derivative impurities on the chain.
 In the scalar sector that we have studied in this paper,
 {\it one} gapless excitation is then expected, the one corresponding to the seventh dimension:
just what we found, the gapless magnon $T(p)$.  As $\kappa \to 0$, the $Q$ impurities, carriers of the $SU(2)_L \times SU(2)_R$ quantum numbers
associated with the three extra dimensions (the transverse $S^3$, see \cite{Gadde:2009dj} for details),
become non-dynamical, and only their composite bound state $T(p)$ survives as a gapless mode. 
We interpret this phenomenon as the field theory counterpart of the transition from the critical to the non-critical background.

\section*{Acknowledgements}
It is a pleasure thank Vladimir Korepin, Pedro Liendo, Juan Maldacena, Joe Minahan, Joe Polchinski, Soo-Jong Rey, Martin Rocek, Radu Roiban,
Matthias Staudacher, Pedro Vieira, and Kostya Zarembo for useful discussions. 
This work was  supported in part by DOE grant DEFG-0292-ER40697 and by NSF grant PHY-0653351-001. Any
opinions, findings, and  conclusions or recommendations expressed in this 
material are those of the authors and do not necessarily reflect the views of the National 
Science Foundation.

 \appendix
 
\section{\label{trick}Simplified computation of the one-loop dilation operator}

In this appendix we determine the one-loop spin-chain Hamiltonian by a simple shortcut. 
The interactions contributing to $H_{k,k+1}$ at one loop are listed schematically in figure \ref{feyn-dia}.  
The first and second interactions (self-energy and gluon exchange) in figure \ref{feyn-dia} are proportional to the identity operator in $V_k\otimes V_{k+1}$, while the non-trivial tensorial
structures  are contributed only by the third diagram (quartic interaction). The idea is to evaluate explicitly the third diagram, and to fix the terms
proportional to the identity by requiring that the anomalous dimensions of a few protected operators vanish.
\subsection{SCQCD}
\begin{figure}[t]
\subfloat[ \label{fig:quartic1}]{\qquad \includegraphics[scale=0.5]{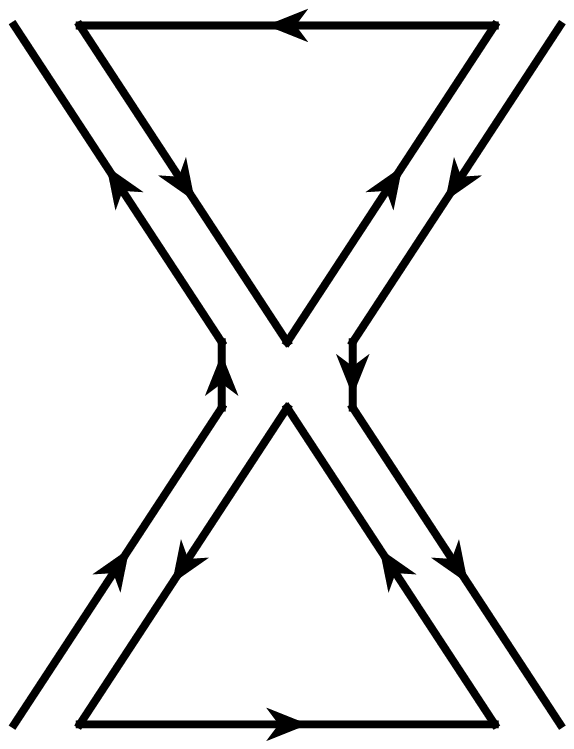}$ \qquad$}\subfloat[ \label{fig:quartic2}]{\includegraphics[scale=0.5]{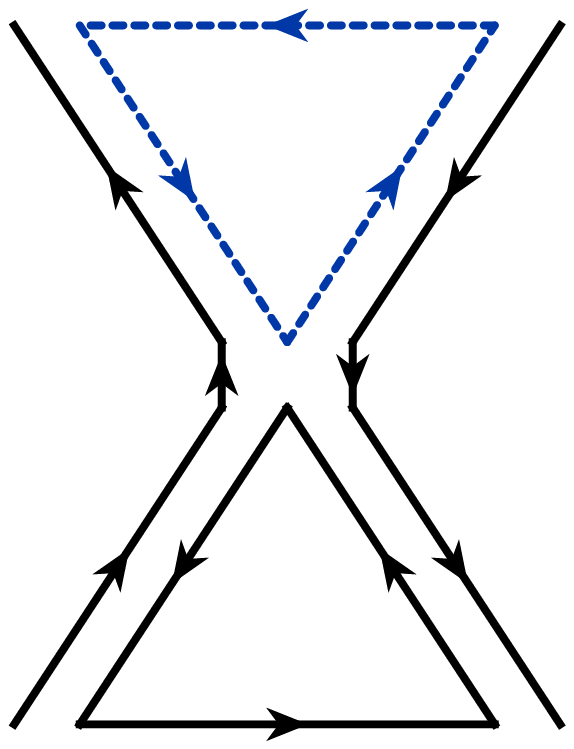}$ \qquad$}\subfloat[ \label{fig:quartic3}]{\includegraphics[scale=0.5]{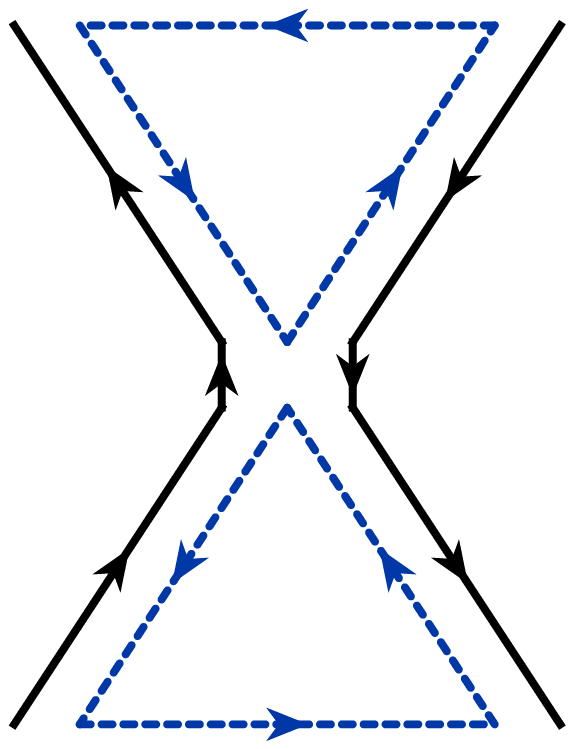}$ \qquad$}\subfloat[ \label{fig:quartic4}]{\includegraphics[scale=0.5]{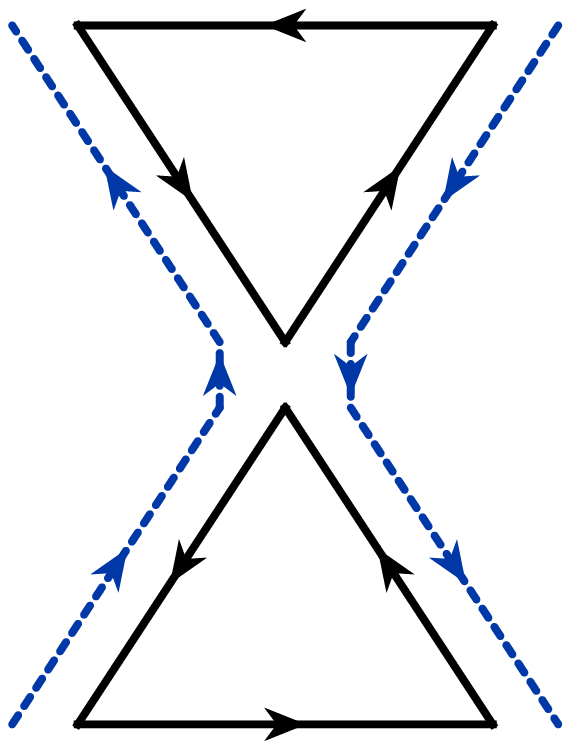}}
\caption{\label{quartic}The color/flavor structure of the quartic vertex. The solid black
line represents the flow of the color index while the dotted blue
line show the flow of the flavor index. Diagram
(a) shows the $ \f^{4}$
interaction vertex, whose contribution is proportional to $N_{c}$
as compared to the tree level. In  (b) the $Q^{2} \f^{2}$
interaction vertex has a factor of $N_{f}/N_c$ compared to (a) because
of the presence of one flavor loop. The $Q^{4}$ vertex in 
(c) has an additional factor of $(N_{f}/N_{c})^{2}$ compared to (a)
due to the presence of two flavor loops. Diagram (d), however,
does not carry any additional $N_{f}/N_{c}$ factors.}
\end{figure}

Let us recall our notations.
The indices $ \mathfrak{p},  \mathfrak{q} =  \pm$ label the $U(1)_r$ charges of $ \phi$ and $ \bar  \phi$, in other terms 
we  define $ \phi^-  \equiv  \phi$, $ \phi^+  \equiv  \bar  \phi$, and $g_{\mathfrak{p}  \mathfrak{q}} =  \left( \begin{array} {cc}
 0 & 1  \\
 1 & 0  \end{array}
 \right)$.
 The elements of the Hamiltonian due to  quartic vertices are:
\bea
\langle \phi_{\mathfrak{p}^{\prime}} \phi_{\mathfrak{q}^{\prime}}|H| \phi^{\mathfrak{p}} \phi^{\mathfrak{q}} \rangle_{\f^{4}}&=& \delta_{\mathfrak{p}^{\prime}}^{\mathfrak{p}} \delta_{\mathfrak{q}^{\prime}}^{\mathfrak{q}}+g^{\mathfrak{p} \mathfrak{q}}g_{\mathfrak{p}^{\prime} \mathfrak{q}^{\prime}}-2 \delta_{\mathfrak{q}^{\prime}}^{\mathfrak{p}} \delta_{\mathfrak{p}^{\prime}}^{\mathfrak{q}} 
\label{quart1}\\
\langle \phi_{\mathfrak{p}^{\prime}} \phi_{\mathfrak{q}^{\prime}}|H|Q_{\mathcal{I}} \bar{Q}^{\mathcal{J}} \rangle_{Q^{2} \f^{2}}&=&  \sqrt{\frac{N_{f}}{N_{c}}}g_{\mathfrak{p}^{\prime} \mathfrak{q}^{\prime}} \delta_{\mathcal{I}}^{\mathcal{J}}
\label{quart2} \\
\langle \bar{Q}^{\II^{\prime}}Q_{\JJ^{\prime}}|H|Q_{\mathcal{I}} \bar{Q}^{\mathcal{J}} \rangle_{Q^{4}}&=&\frac{N_{f}}{N_{c}}(2 \delta_{\mathcal{I}}^{\mathcal{I}^{\prime}} \delta_{\mathcal{J}^{\prime}}^{\mathcal{J}}-  \delta_{\mathcal{I}}^{\mathcal{J}} \delta_{\mathcal{J}^{\prime}}^{\mathcal{I}^{\prime}})
\label{quart3}\\
\langle Q_{\JJ^{\prime}}\bar{Q}^{\II^{\prime}}|H|\bar{Q}^{\mathcal{J}} Q_{\mathcal{I}} \rangle_{Q^{4}}&=& 2  \delta_{\mathcal{I}}^{\mathcal{J}} \delta_{\mathcal{J}^{\prime}}^{\mathcal{I}^{\prime}}-  \delta_{\mathcal{I}}^{\mathcal{I}^{\prime}} \delta_{\mathcal{J}^{\prime}}^{\mathcal{J}}
\label{quart4}
\eea
The factors of $\frac{N_{f}}{N_{c}}$ are explained in figure \ref{quartic}. Figures \ref{fig:quartic1},\ref{fig:quartic2},\ref{fig:quartic3},\ref{fig:quartic4} correspond to equations (\ref{quart1},\ref{quart2},\ref{quart3},\ref{quart4}) respectively.
This fixes the Hamiltonian up to the terms proportional to the identity, 
{\scriptsize{
 \begin{eqnarray*}
 & H_{k,k+1}=\\
 & \nonumber \\
 &  \bordermatrix{
 &  \phi^{\mathfrak{p}} \phi^{\mathfrak{q}} & Q_{\mathcal{I}} \bar{Q}^{\mathcal{J}} &  \bar{Q}^{\KK}Q_{\LL} & Q_{\II} \phi^{\mathfrak{p}} \cr
 &&&&  \cr
 \phi_{\mathfrak{p}^{\prime}} \phi_{\mathfrak{q}^{\prime}} & \alpha \delta_{\mathfrak{p}^{\prime}}^{\mathfrak{p}} \delta_{\mathfrak{q}^{\prime}}^{\mathfrak{q}}+g^{\mathfrak{p} \mathfrak{q}}g_{\mathfrak{p}^{\prime} \mathfrak{q}^{\prime}}-2 \delta_{\mathfrak{q}^{\prime}}^{\mathfrak{p}} \delta_{\mathfrak{p}^{\prime}}^{\mathfrak{q}} &  \sqrt{\frac{N_{f}}{N_c}}g_{\mathfrak{p}^{\prime} \mathfrak{q}^{\prime}} \delta_{\mathcal{I}}^{\mathcal{J}} & 0 & 0 \cr
 \bar{Q}^{\II^{\prime}}Q_{\JJ^{\prime}} &  \sqrt{\frac{N_{f}}{N_c}}g^{\mathfrak{p} \mathfrak{q}} \delta_{\mathcal{J}^{\prime}}^{\mathcal{I}^{\prime}} & \beta \delta_{\mathcal{I}}^{\mathcal{I}^{\prime}} \delta_{\mathcal{J}^{\prime}}^{\mathcal{J}}-  \delta_{\mathcal{I}}^{\mathcal{J}} \delta_{\mathcal{J}^{\prime}}^{\mathcal{I}^{\prime}} \frac{N_{f}}{N_c} & 0 & 0 \cr
Q_{\KK^{\prime}} \bar{Q}^{\LL^{\prime}} & 0 & 0 & \gamma \delta_{\KK^\prime}^{\KK}\delta^{\LL^\prime}_{\LL}+2 \delta_{\LL}^{\KK} \delta_{\KK^{\prime}}^{\LL^{\prime}} & 0 \cr
 \bar{Q}^{\II^{\prime}} \phi_{\mathfrak{p}^{\prime}} & 0 & 0 & 0 & \eta \delta_{\mathcal{I}}^{\mathcal{I}^{\prime}} \delta_{\mathfrak{p}^{\prime}}^{\mathfrak{p}}}
 \end{eqnarray*}
}} 
We can now find the coefficients $\alpha,\beta,\gamma$ and $\eta$ from knowledge of the protected spectrum. Vanishing of the anomalous dimension of  $\Tr \f^k$ gives $\alpha=2$. 
Another protected multiplet is the multiplet containing the stress-energy tensor. Its superconformal primary, called $\mbox{Tr}\,T$, has $R,r=0$ and $ \Delta=2$.
Hence, it is a linear combination of $\mbox{Tr}[Q_{\II} \bar{Q}^{\II}]$ and $ \mbox{Tr}[ \f \fbar]$.
The restriction of the Hamiltonian to this subspace is
\[
H= \quad  \bordermatrix{
& \mbox{Tr}[ \f \fbar] &  &  \mbox{Tr}[\MM_{\bf 1}] \cr
\mbox{Tr}[ \f \fbar] & 4 &  & 2 \sqrt{\frac{2 N_{f}}{N}} \cr
\mbox{Tr}[\MM_{\bf 1} ] & 2 \sqrt{\frac{2 N_{f}}{N}} &  & ( \beta+ \gamma)-2( \frac{N_{f}}{N_{c}}-2)}
\]
This matrix must have a zero at the superconformal
point $N_{f}=2N_{c}$, yielding $\beta+ \gamma=4$.
Finally, the fact that $\mbox{Tr}\,T \f$ is also a protected operator gives the relation $\beta+2 \eta=8 $. We started with four coefficients $\alpha$, $\beta$, $\gamma$, $\eta$
and imposed three relations. The undetermined degrees of freedom
 corresponds to the ``gauge'' freedom of adding to the nearest neighbor Hamiltonian
terms  that vanish upon evaluating the full $H$ on a closed chain. We may solve the constraints by writing
\be
\alpha=2\, ,\qquad \beta=4  +\frac{1}{2}(1 + \xi) \, ,\qquad \gamma=-\frac{1}{2}(1 + \xi)\, ,\qquad \eta=\frac{1}{4}(7-\xi) \, ,
\ee
where $\xi$ is the arbitrary gauge parameter. The resulting Hamiltonian is in perfect agreement (for $N_f = 2 N_c$) with the answer (\ref{HQCD}) obtained by the slightly
lengthier route of
 explicit evaluating
 all relevant one-loop diagrams. All in all, this confirms our understanding of the protected spectrum.

\subsection{Interpolating SCFT}
We can repeat the same exercise for the interpolating SCFT. The quartic vertices give
\bea
\langle \phi_{\mathfrak{p}^{\prime}} \phi_{\mathfrak{q}^{\prime}}| \phi^{\mathfrak{p}} \phi^{\mathfrak{q}} \rangle_{\f^{4}}  &=&  \delta_{\mathfrak{p}^{\prime}}^{\mathfrak{p}} \delta_{\mathfrak{q}^{\prime}}^{\mathfrak{q}}+g^{\mathfrak{p} \mathfrak{q}}g_{\mathfrak{p}^{\prime} \mathfrak{q}^{\prime}}-2 \delta_{\mathfrak{q}^{\prime}}^{\mathfrak{p}} \delta_{\mathfrak{p}^{\prime}}^{\mathfrak{q}} \\
 \langle \fh_{\mathfrak{p}^{\prime}} \fh_{\mathfrak{q}^{\prime}}| \fh^{\mathfrak{p}} \fh^{\mathfrak{q}} \rangle_{\fh^{4}}&=&  \kappa^{2} (\delta_{\mathfrak{p}^{\prime}}^{\mathfrak{p}} \delta_{\mathfrak{q}^{\prime}}^{\mathfrak{q}}+g^{\mathfrak{p} \mathfrak{q}}g_{\mathfrak{p}^{\prime} \mathfrak{q}^{\prime}}-2 \delta_{\mathfrak{q}^{\prime}}^{\mathfrak{p}} \delta_{\mathfrak{p}^{\prime}}^{\mathfrak{q}}) \\
\langle \bar{Q}^{\LLh \LL}Q_{\KK \KKh}|Q_{\II \IIh} \bar{Q}^{\JJh \JJ} \rangle_{Q^4} & = & 2\, \delta_{\IIh}^{\JJh} \delta_{\KK}^{\JJ} \delta_{\KKh}^{\LLh} \delta_{\II}^{\LL}-  \delta_{\II}^{\JJ} \delta_{\IIh}^{\JJh} \delta_{\KK}^{\LL} \delta_{\KKh}^{\LLh} \nonumber  \\
 &+  &  \kappa^{2}( 2\,\delta_{\KKh}^{\JJh} \delta_{\II}^{\JJ} \delta_{\IIh}^{\LLh} \delta_{\KK}^{\LL}-  \delta_{\II}^{\LL} \delta_{\IIh}^{\LLh} \delta_{\KK}^{\JJ} \delta_{\KKh}^{\JJh}) \label{QbarQ}\\
\langle Q_{\II \IIh} \bar{Q}^{\JJh \JJ}| \bar{Q}^{\LLh \LL}Q_{\KK \KKh} \rangle_{Q^4}&=& 
2\, \delta_{\IIh}^{\JJh} \delta_{\KK}^{\JJ} \delta_{\KKh}^{\LLh} \delta_{\II}^{\LL}-  \delta_{\II}^{\JJ} \delta_{\IIh}^{\JJh} \delta_{\KK}^{\LL} \delta_{\KKh}^{\LLh} \nonumber  \\
&+  &  \kappa^2( 2\,\delta_{\KKh}^{\JJh} \delta_{\II}^{\JJ} \delta_{\IIh}^{\LLh} \delta_{\KK}^{\LL}- \delta_{\II}^{\LL} \delta_{\IIh}^{\LLh} \delta_{\KK}^{\JJ} \delta_{\KKh}^{\JJh}) \label{QQbar}\\
\langle \phi_{\mathfrak{p}^{\prime}} \phi_{\mathfrak{q}^{\prime}}|Q_{\mathcal{I} \IIh} \bar{Q}^{\JJh \mathcal{J}} \rangle_{Q^{2} \f^{2}} & = & g_{\mathfrak{p}^{\prime} \mathfrak{q}^{\prime}} \delta_{\mathcal{I}}^{\mathcal{J}} \delta_{\IIh}^{\JJh} \\
\langle \fh_{\mathfrak{p}^{\prime}} \fh_{\mathfrak{q}^{\prime}}| \bar{Q}^{\JJh \mathcal{J}}Q_{\mathcal{I} \IIh} \rangle_{Q^{2} \fh^{2}} & = &  \kappa^{2}g_{\mathfrak{p}^{\prime} \mathfrak{q}^{\prime}} \delta_{\mathcal{I}}^{\mathcal{J}} \delta_{\IIh}^{\JJh} \\
\langle \bar{Q}^{\JJh \JJ} \fh_{\mathfrak{q}}| \phi^{\mathfrak{p}}Q_{\II \IIh} \rangle_{\f Q \fh \bar{Q}} & = & -2\kappa \delta_{\qq}^{\pp} \delta_{\II}^{\JJ} \delta_{\IIh}^{\JJh}   \\
\langle \phi^{\mathfrak{p}} \bar{Q}^{\JJh \JJ}|Q_{\II \IIh} \fh_{\mathfrak{q}} \rangle_{\f Q \fh \bar{Q}} & = & -2\kappa \delta_{\qq}^{\pp} \delta_{\II}^{\JJ} \delta_{\IIh}^{\JJh}
\eea
The first four elements can have  additional identity pieces. They are easily determined by imposing the symmetry under $g \leftrightarrow \check g$, $Q\leftrightarrow\bar Q$ and $\f \leftrightarrow \check \f$ and by requiring the Hamiltonian to reduce to that of SCQCD in the limit $\kappa \to 0$. 
The one loop Hamiltonian (\ref{explicitH}) is precisely reproduced by this method.
 
\section{\label{composite}The Hamiltonian for SCQCD in the Dimer Picture}

In this appendix we 
rewrite the Hamiltonian for SCQCD as acting
on adjoint fields and
 dimers
$Q_{\II} \bar{Q}^{\JJ}$, regarded as basic objects. We define
the singlet combination $ \MM= \frac{1}{\sqrt{2}} \MM_{\II}^{\mbox{ } \JJ} \delta_{\JJ}^{\II}$
and the triplet $ \MM^{i}= \frac{1}{\sqrt{2}} \MM_{\II}^{\mbox{ } \JJ}( \sigma^{i})_{\JJ}^{\II}$,
where $ \sigma^{i}$ are three Pauli matrices. These can be rewritten
in an $SO(4)$ notation as $ \MM^{m}= \frac{1}{\sqrt{2}} \MM_{\II}^{\mbox{ } \JJ}( \sigma^{m})_{\JJ}^{\II}$,
where $m= 0, \ldots,3$ and $ \sigma^{0} \equiv \mathbb{I}_{2 \times2}$.

 Consider the action of $H$ on following sequence in the spin chain, 
\[
 \begin{array}{ccccccc}
 \f^{\pp} &  & Q_{\II} &  &  \bar{Q}^{\JJ} &  &  \f^{\qq} \\
 &  \frac{1}{2}(3+ \frac{\xi}{2})  &  & (5- \frac{\xi}{2}) \mathbb{I}_{QQ}-2 \mathbb{K}_{QQ} &  &  \frac{1}{2}(3+ \frac{\xi}{2})  \\
 &  \downarrow &  &  \downarrow &  &  \downarrow \\
 \f_{\pp^{\prime}} &  &  \bar{Q}^{\II^{\prime}} &  & Q_{\JJ^{\prime}} &  &  \f_{\qq^{\prime}} \end{array} \]
In the new picture, where  $ \MM$ is regarded as a basic impurity, the
 middle term $( 5- \frac{\xi}{2}) \mathbb{I}_{QQ}- 2 \mathbb{K}_{QQ}$
is the ``self energy'' of $ \MM$, and we split it evenly between
the $\phi {\cal M}$ and ${\cal M} \phi$ matrix elements. 
So we write \begin{eqnarray*}
  \langle \ldots \f_{\pp^{\prime}} \bar{\MM}_{\mbox{ } \JJ^{\prime}}^{\II^{\prime}} \ldots|  H | \ldots \f^{\pp} \MM_{\II}^{\mbox{ } \JJ} \ldots \rangle & = & [ \frac{1}{2}(3+ \frac{\xi}{2})+ \frac{1}{2}(5- \frac{\xi}{2})] \delta_{\mathfrak{p}^{\prime}}^{\mathfrak{p}} \delta_{\mathcal{I}}^{\mathcal{I}^{\prime}} \delta_{\JJ^{\prime}}^{\JJ} 
  - \delta_{\mathfrak{p}^{\prime}}^{\mathfrak{p}} \delta_{\mathcal{I}}^{\JJ} \delta_{\JJ^{\prime}}^{\II^{\prime}} \\
 & = & (4 \delta_{\mathcal{I}}^{\mathcal{I}^{\prime}} \delta_{\JJ^{\prime}}^{\JJ}- \delta_{\mathcal{I}}^{\JJ} \delta_{\JJ^{\prime}}^{\II^{\prime}}) \delta_{\mathfrak{p}^{\prime}}^{\mathfrak{p}} \\
  \langle \ldots \f_{\pp^{\prime}} \bar{\MM}^{m^{\prime}} \ldots| H | \ldots \f^{\pp} \MM^{m} \ldots \rangle 
 & = &  \delta_{\mathfrak{p}^{\prime}}^{\mathfrak{p}} \delta^{mm^{\prime}}(4-2 \delta^{m0})\,. 
\end{eqnarray*}
Similarly, to find the action of $H$ on two neighboring ${\cal M}$s, we consider 
 the sequence 
 \[
 \begin{array}{ccccccc}
Q_{\II} &  &  \bar{Q}^{\JJ} &  & Q_{\KK} &  &  \bar{Q}^{\LL} \\
 & (5- \frac{\xi}{2}) \mathbb{I}_{QQ}-2 \mathbb{K}_{QQ} &  & ( \frac{\xi}{2}-1) \mathbb{I}_{QQ}+2 \mathbb{K}_{QQ} &  & (5- \frac{\xi}{2}) \mathbb{I}_{QQ}-2 \mathbb{K}_{QQ} \\
 &  \downarrow &  &  \downarrow &  &  \downarrow \\
 \bar{Q}^{\II^{\prime}} &  & Q_{\JJ^{\prime}} &  &  \bar{Q}^{\KK^{\prime}} &  & Q_{\LL^{\prime}} \end{array} \]
This gives
 \begin{eqnarray*}
 \langle \ldots \bar{\MM}^{m^{\prime}} \bar{\MM}^{n^{\prime}} \ldots| H | \ldots \MM^{m} \MM^{n} \ldots \rangle & = &  \delta^{mm^{\prime}} \delta^{nn^{\prime}}(13-4 \delta^{m0}-4 \delta^{n0}) \\
 &  & + \delta^{mn} \delta^{m^{\prime}n^{\prime}}- \delta^{mn^{\prime}} \delta^{nm^{\prime}}+i \epsilon^{mnn^{\prime}m^{\prime}} \,.\end{eqnarray*}

 \bibliographystyle{JHEP}
 \bibliography{Orbifoldbib}

 \end{document}